\documentclass{aa}  
\usepackage{graphicx}
\usepackage{txfonts}
\usepackage{natbib}
\bibpunct{(}{)}{;}{a}{}{,} 
\usepackage[switch]{lineno}
\begin{document}

% \linenumbers

   \title{Mass-morphology relation of TNG50 galaxies}

   \author{Bruno M. Celiz
          \inst{1,2,3}\fnmsep\thanks{\email{bruno.celiz@mi.unc.edu.ar}}
          \and
          Julio F. Navarro\inst{4}
          \and
          Mario G. Abadi\inst{2,3}
          \and
          Volker Springel\inst{5}
          }

   \institute{Facultad de Matemática, Astronomía, Física y Computación, UNC, Medina Allende s/n, X5000HUA, Córdoba, Argentina
        \and
            Instituto de Astronomía Teórica y Experimental, CONICET--UNC, Laprida 854, X5000BGR, Córdoba, Argentina
        \and
            Observatorio Astronómico de Córdoba, UNC, Laprida 854, X5000BGR, Córdoba, Argentina
        \and
            Department of Physics and Astronomy, University of Victoria, Victoria, BC, V8P 5C2, Canada
        \and
            Max-Planck-Institut für Astrophysik, Karl-Schwarzschild-Straße 1, D-85741 Garching, Germany
        }

   \date{Received XXX; accepted YYY}
 
    \abstract 
    {We used the cosmological hydrodynamical simulation TNG50 to study the galaxy mass-morphology relation, as measured by the rotational support of the stellar component of simulated galaxies. For isolated galaxies with a stellar mass in the range of $8<\log(\mathit{M_{*}/M_{\odot}}) < 11$, rotational support increases with $\mathit{M_*}$, from dispersion-supported spheroidal dwarfs to massive galaxies with prominent, rotationally supported discs. Our results indicate that this correlation arises from the spatial distribution of star formation in TNG50 galaxies, which occurs primarily in two distinct regions: an unresolved, non-rotating central baryonic clump $(r \lesssim 1~\mathrm{kpc})$ and a rotationally supported outer disc, separated by a  quiescent region. The importance of the inner clump increases with decreasing $\mathit{M_*}$; it makes up less than $20\%$ of all stars in the most massive galaxies, but more than $80\%$ in dwarfs. This explains why dwarfs have less rotational support than massive galaxies and why all dwarfs have similar stellar half-mass radii, regardless of $\mathit{M_*}$. It also explains why massive galaxies in TNG50 appear to form inside-out (as the outer disc grows), whereas dwarfs form outside-in, as star formation in the dominant inner clump moves progressively inward. The clump-disc segregation of star formation in TNG50 galaxies is probably numerical in origin. Inner clumps are formed by the accumulation of low-angular-momentum gas supported by the equation of state introduced to prevent artificial fragmentation. The decoupled-wind feedback implementation in TNG50 helps to preserve the clumps, but disrupts disc formation in its immediate surroundings. This hinders the formation of discs in (dwarf) galaxies whose sizes are not substantially larger than the clump, but it has little effect on the larger discs of more massive systems. Our results argue in favour of taking caution when interpreting the dependence on stellar mass of TNG50 galaxy morphologies, or the evolution of galaxy sizes, especially at the dwarf end.}

   \keywords{
   galaxies: dwarf -- galaxies: kinematics and dynamics -- galaxies: star formation
    }

   \maketitle

\section{Introduction} \label{sec:intro}

Galaxy surveys including SDSS\footnote{\url{https://www.sdss.org/}} \citep{York2000} and GAMA\footnote{\url{https://www.gama-survey.org/}} \citep{Driver2011} have shown that the number of low-mass galaxies dominate the galaxy population \citep{Kauffmann2003Mstar,Blanton2005,LiWhite2009, Baldry2012}. Due to their low luminosity, studies of  dwarf galaxies (defined as those with $\log(M_{*}/M_{\odot}) < 9$) are generally limited to nearby regions, such as the Local Volume or the Local Group of galaxies \citep[see e.g.][]{Mateo1998,Walter2008THINGS,Swaters2009,McC2012,SHIELD2016}. 

Existing studies of dwarf galaxy morphology show that most dwarfs are irregular, with few examples of pure disc or spheroidal morphologies \citep[see e.g.][]{Karachentsev2013, Klypin2015}. For example, \citet{GAMA2016} reported that irregular galaxies outnumber disc-like galaxies in the mass range of $\log(M_{*}/M_{\odot}) \lesssim 9.3,$ with the difference increasing at lower masses (see \citealt{vdBSwater2001} and \citealt{Swaters2009} for a study on late-type dwarf galaxies). \citet{Klypin2015}, on the other hand, reported that only $\sim 10\%$ of bright dwarf galaxies are spheroids.

The physical processes driving the morphological diversity of dwarf galaxies are not fully understood. 
Low-luminosity systems are expected to form in the shallow potential wells of low-mass dark-matter halos \citep[see e.g.][]{Moster2013,Adams2014,Oh2015,SHIELD2016,Behroozi2019,Oman2019}. This implies that the impact of baryonic processes such as stellar feedback is enhanced in dwarfs \citep[see e.g.][]{McQuinn2019,Gutcke2021LYRA,OstrikerKim2022}, possibly with strong effects on galaxy morphology. Dwarf galaxies thus provide an ideal test bed for models of star formation and evolution that aim to track the baryon cycle and its impact on galaxy formation and evolution \citep[see e.g.][]{DekelWoo2003,Ferrarotti2006,Christensen2016,El-Badry2017}.

Direct numerical simulations have also struggled to consistently reproduce the diverse morphologies of dwarf galaxies observed in the Local Volume: a  'weak tension' regarding our understanding of dwarf galaxy evolution in the prevailing $\Lambda$CDM cosmological model \citep{Sales2022}. Key to resolving this tension is the study of the spatially resolved star formation activity in dwarfs, which differ from massive galaxies not only in their morphology, but also in the radial gradients of their stellar populations. 

Massive disc galaxies, for example, tend to form inside-out, as later accreting material settles in the outer disc because of its higher angular-momentum content. Conversely, dwarf galaxies typically exhibit positive radial-stellar-age gradients, with their youngest and most metal-rich stars preferentially inhabiting the inner regions, while the oldest populate the outskirts \citep[see e.g.][]{B-L2016,Albers2019,Cheng2024,Fu2024And,Riggs2024,Tau2024Obs}. How these differences in age gradients between dwarf and luminous galaxies arise is still unclear, but the morphological differences hint at the important role that rotational support may play in the process. Indeed, rotational support seems absent in the stellar components of the faintest galaxies known: the dwarf spheroidal companions of the Milky Way and Andromeda galaxies \citep[see e.g.][]{McC2012}.

We seek to find out why rotational support plays a diminished role in the stellar component of dwarfs. It may be because gaseous discs are not established in these systems before stars form in earnest, or because the feedback effects of star formation itself act to disrupt thin, rotationally supported structures in the shallow potential wells of low-mass halos. It could also be the result of more complex, external effects such as the role of the environment through tidal or ram-pressure stripping, or galaxy mergers.

We used a state-of-the-art hydrodynamic cosmological simulation \citep[TNG50, ][]{Nelson2019TNG50, Pillepich2019} to investigate these issues. In particular, we aimed to study how the rotational support of the stellar component depends on galaxy mass. We also looked into how the spatial distribution of star formation drives the evolution of galaxy size and the morphology of isolated galaxies, with the goal of understanding the physical origin of such trends. 

This paper is organised as follows. In Section \ref{SecMethods}, we present details of the simulation used and of the galaxy sample studied. We also introduce the definition of galaxy rotational support adopted throughout this work. In Section \ref{SecResults}, we analyse the spatially resolved morphology of isolated galaxies of different masses and use them to motivate and interpret the observed trends between galaxy mass, size, and rotational support, as well as the evolution of galaxy sizes. 
Finally, in Section \ref{SecConc}, we summarise our results and present our conclusions.

\begin{figure*}
    \centering
    \includegraphics[width=0.495\columnwidth]{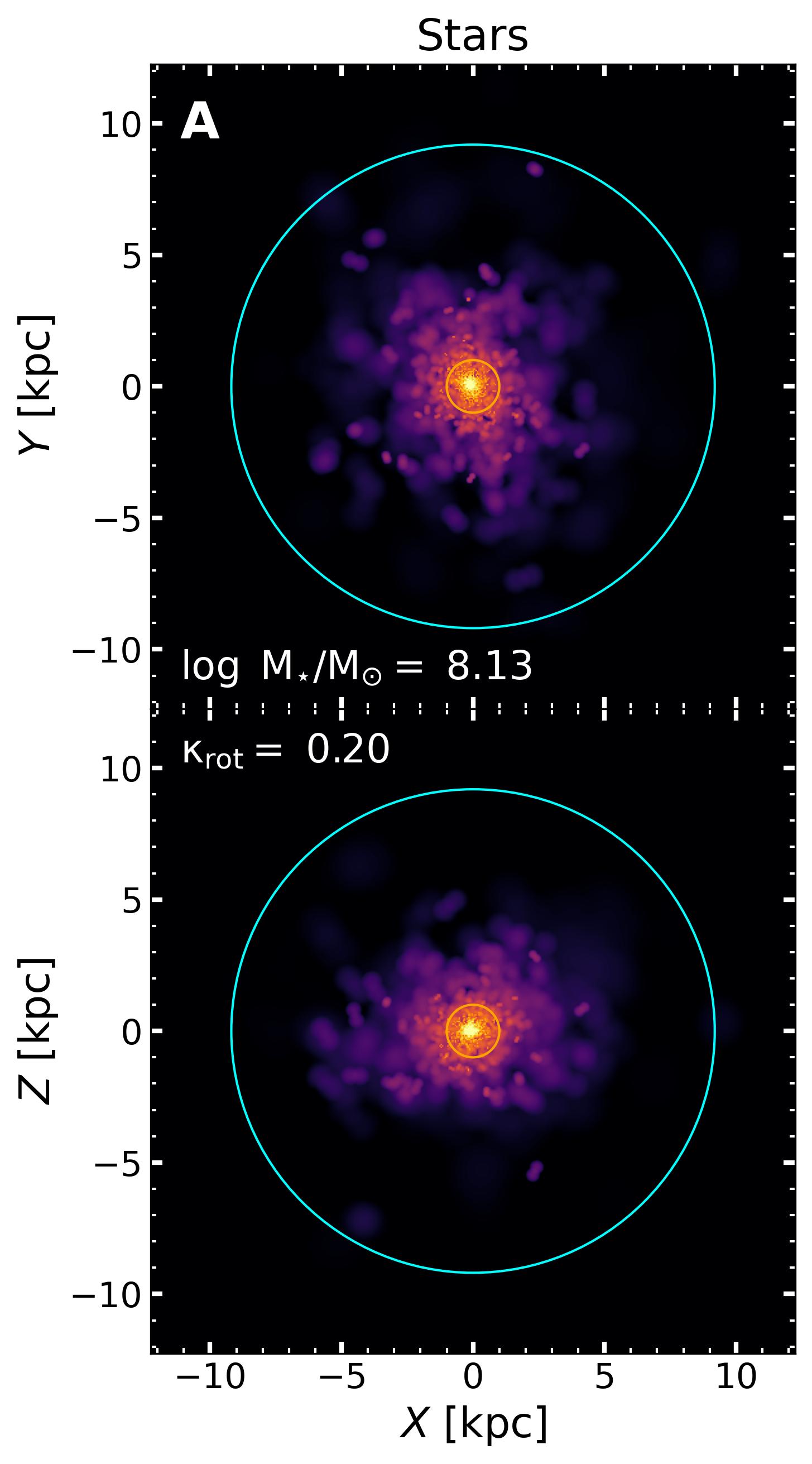}
    \includegraphics[width=0.464\columnwidth]{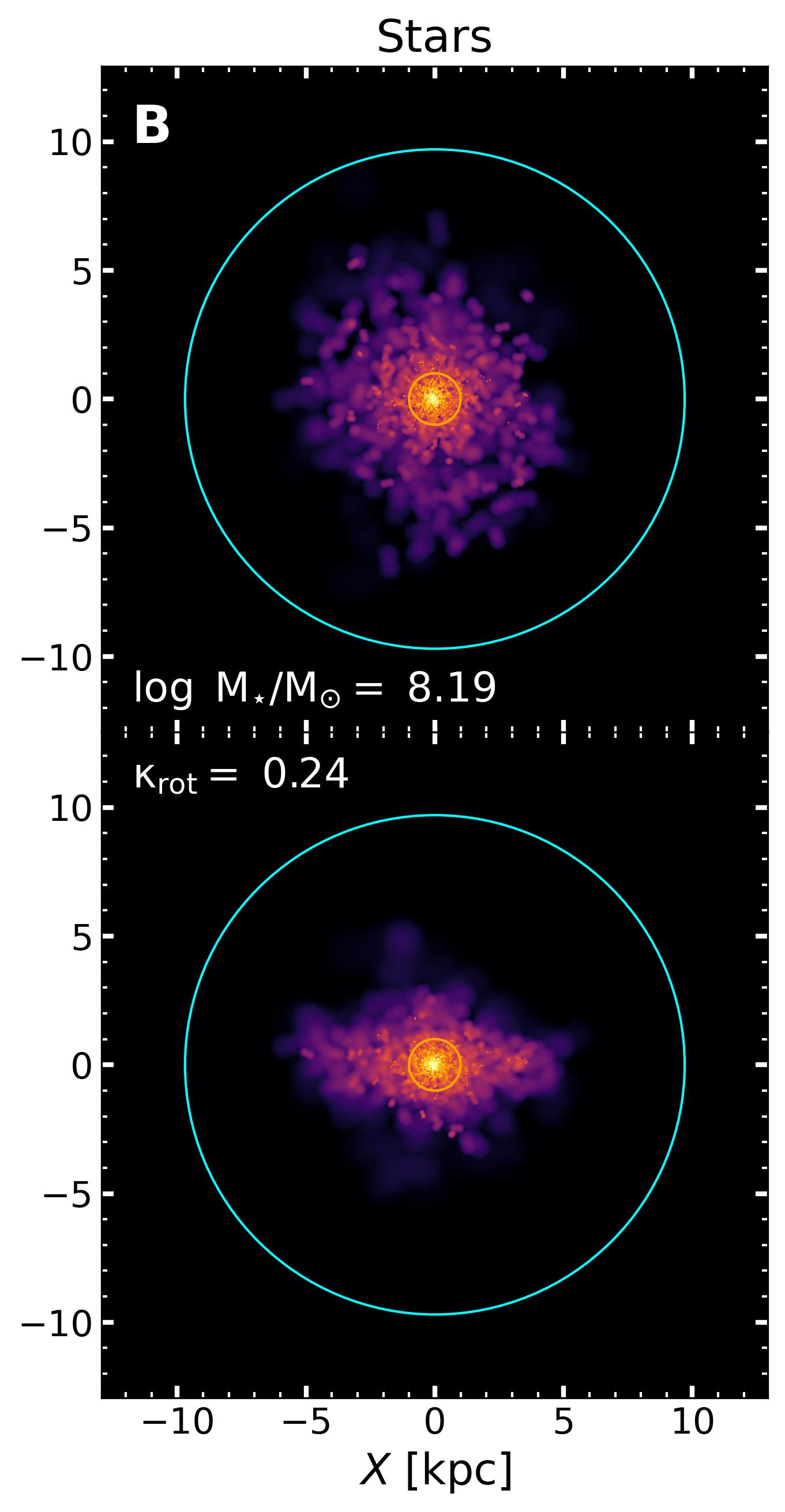}
    \includegraphics[width=0.464\columnwidth]{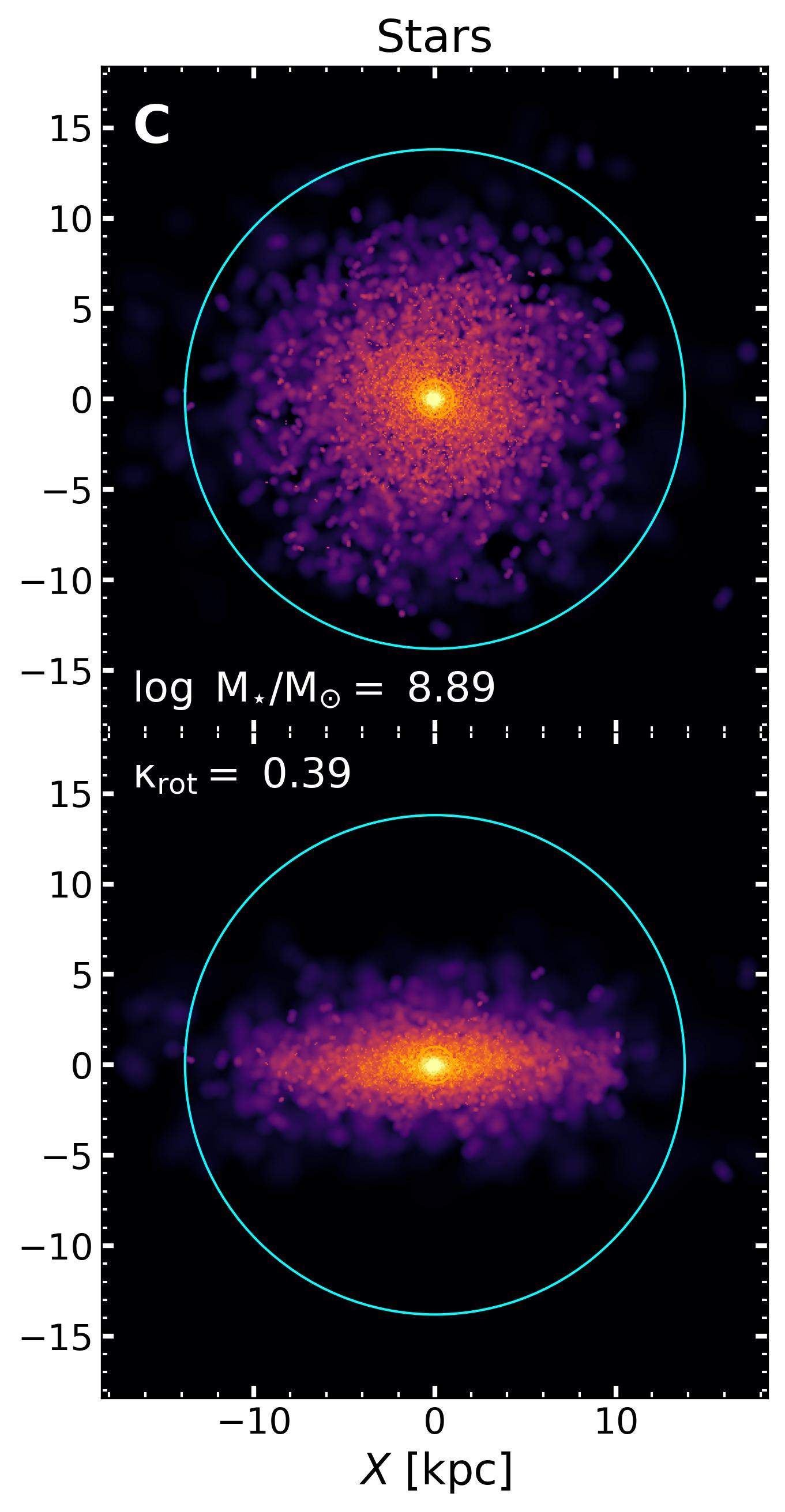}
    \includegraphics[width=0.498\columnwidth]{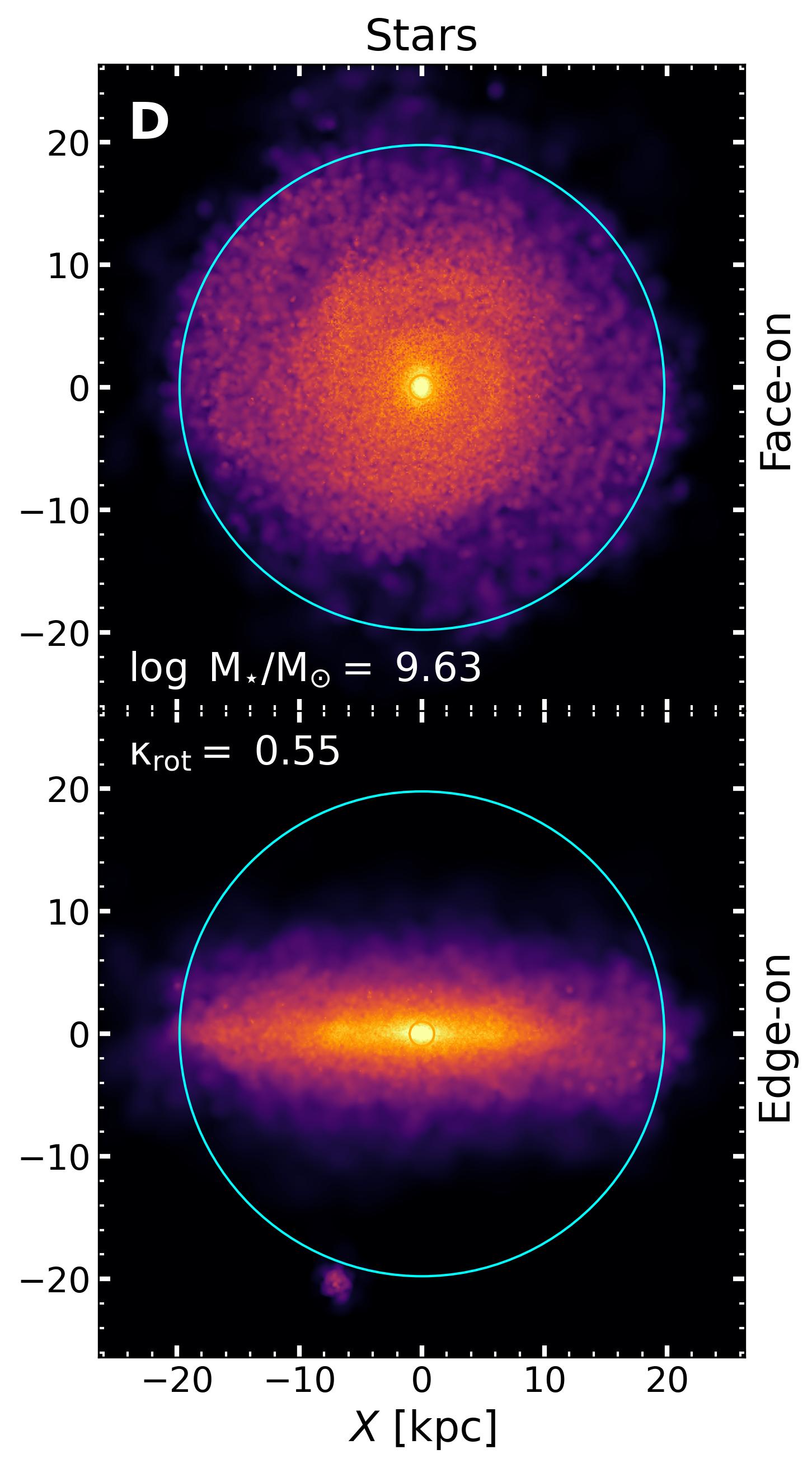}
    \includegraphics[width=0.495\columnwidth]{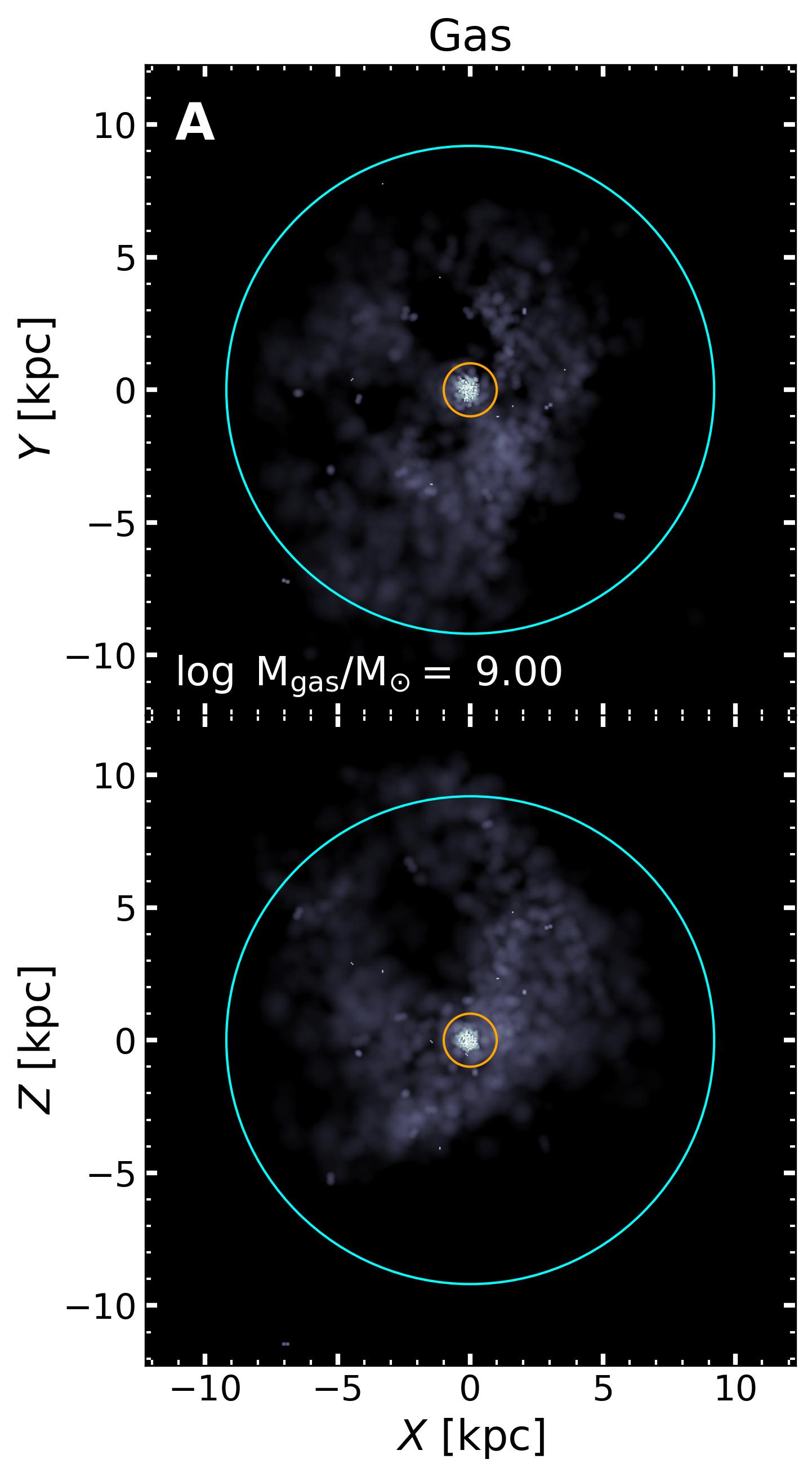}
    \includegraphics[width=0.467\columnwidth]{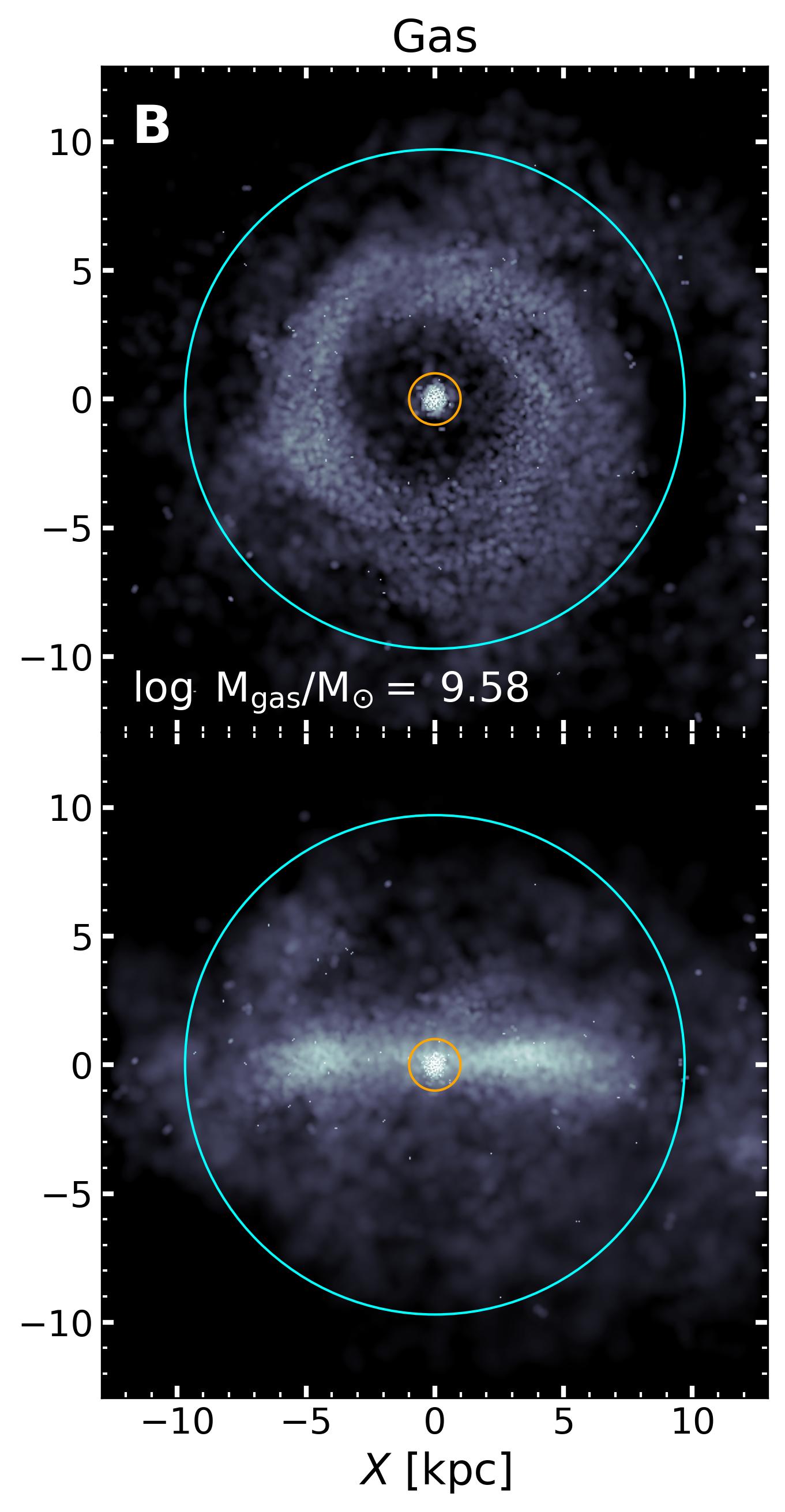}
    \includegraphics[width=0.467\columnwidth]{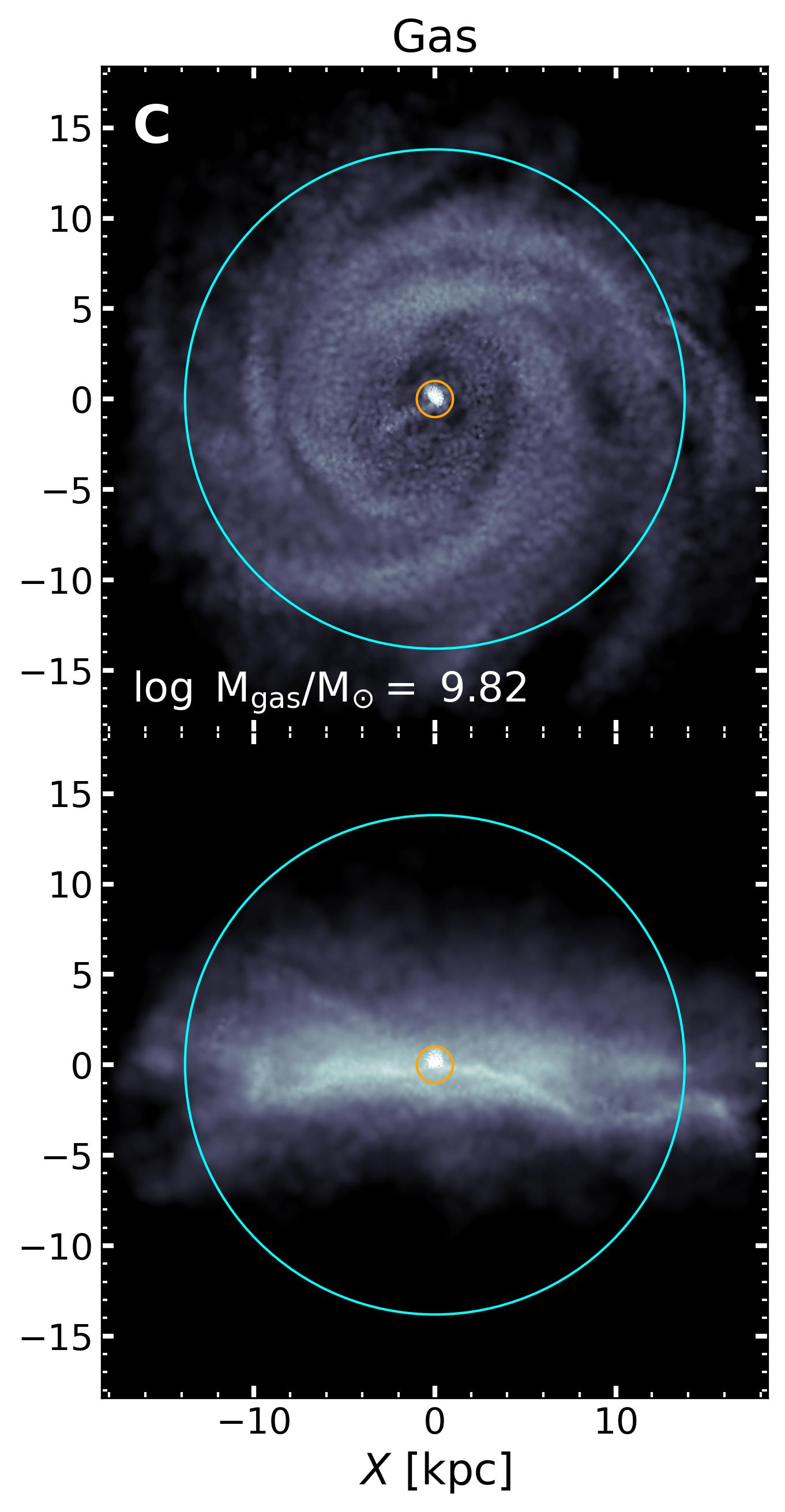}
    \includegraphics[width=0.498\columnwidth]{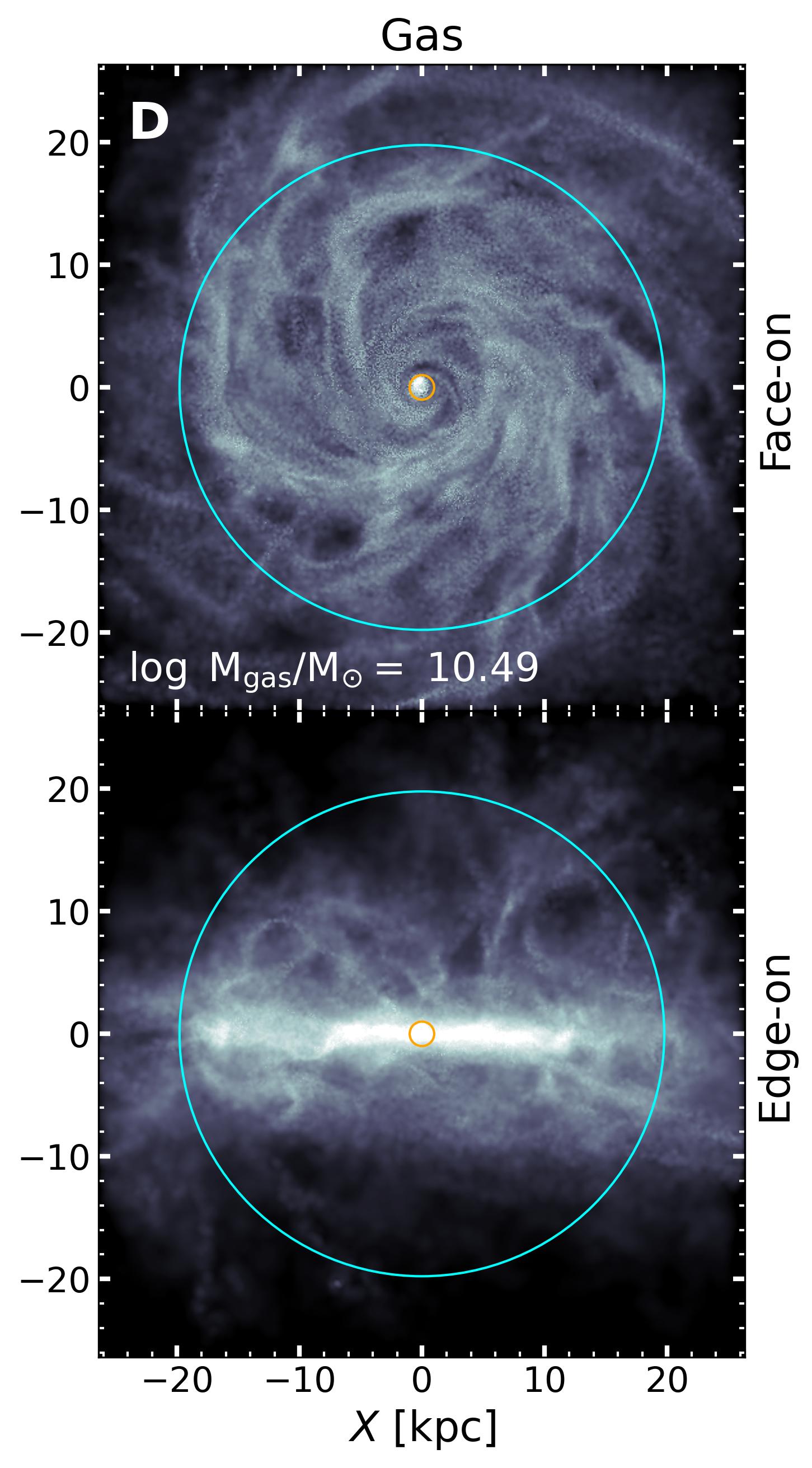}
    
    \caption{Face-on and edge-on projections of stars (top panels) and gas (bottom panels) in four TNG50 central galaxies with varying stellar mass and rotational support. From left to right, galaxies increase in total stellar mass $\mathit{M_{*}}$, total gas mass $M_{\mathrm{gas}}$, and rotational support $\kappa_{\mathrm{rot}}$ (values are shown in white labels). Cyan circles indicate our definition of galaxy size $r_{\mathrm{glx}} = 0.15 ~ r_{200}$; orange circles mark $r = 1$ kpc. Extended gaseous and stellar discs are more clearly evident  in the two most massive galaxies (C and D). All galaxies exhibit a dense baryonic clump (gas and stars) in the central regions of $r\sim 1$ kpc. The TNG50 \texttt{Subhalo\_IDs} for these galaxies are: 830875 (A), 792447 (B), 733632 (C), and 643257 (D). Brighter colours indicate a higher number density of particles. Images were generated with PySPHViewer \citep{BenitezLlambay2017}.}
    \label{FigGxExamples}
\end{figure*}

\section{Methods} \label{SecMethods}

\subsection{Simulation} \label{subsec:simu}

For this study, we used The Next Generation Illustris simulations \citep[IllustrisTNG\footnote{\url{https://www.tng-project.org/}},][]{Miranacci2018, Naiman2018, Pill2018,Springel2018,Nelson2019}, a suite of magneto-hydrodynamic cosmological simulations of a standard $\Lambda$CDM Universe \citep[$h = 0.6774;~\Omega_m = 0.3089;~\sigma_8 = 0.8159 $, consistent with][]{Planck2016}. From initial conditions set at redshift $z=127$, the simulations were evolved with the moving mesh code \texttt{AREPO} \citep{Springel2010, Pakmor2016} forward in time until $z=0$, and the properties of dark-matter and baryon particles are systematically recorded in 100 snapshots at intervals of $\sim$0.15 Gyr.

Our analysis focuses on simulated galaxies identified in the TNG50-1 run \citep[][TNG50 hereafter]{Nelson2019TNG50, Pillepich2019}, a $51.7~\mathrm{Mpc}$ side periodic box containing $2160^3$ particles of dark matter of mass m$_{\mathrm{DM}} = 4.5 \times 10^5~\mathit{M_{\odot}}$, and an equal number of initial gas cells with a `target baryon mass' of m$_{\mathrm{baryon}} = 8.5 \times 10^4~\mathit{M_{\odot}}$. 
The Plummer-equivalent gravitational softening for DM and stars is $\epsilon_{\mathrm{DM,*}} = 0.29$ kpc, and the minimum value of the adaptive gas gravitational softening is $\epsilon_{\mathrm{gas}} = 0.07$ kpc (at redshift $z=0$).

The baryonic treatment included in TNG50 \citep[updated from the previous Illustris project,][]{Vogelsberger2013,Vogelsberger2014,Weinberger2017,Pill2018} allows the gas to cool down to a temperature of $T = 10^4$ K following the cooling and heating rates computed from local density, redshift, and metallicity. The gas above density $n = 0.13~\mathrm{cm^{-3}}$ was modelled using an effective equation of state to describe a dual-phase interstellar medium gas that prevents artificial fragmentation \citep{SpringelHernquist2003}. Star formation occurs in gas cells above this density threshold. Stellar
particles are born through a Chabrier initial mass function \citep{Chabrier2003}, with their subsequent stellar evolution and feedback implemented following the prescriptions described in \citet{Pill2018}.

\subsection{Simulated galaxies}

Galaxy catalogues were generated using the \texttt{friends-of-friends} \citep[FoF,][]{Davies1985} and \texttt{SUBFIND} algorithms \citep{Springel2001, Dolag2009}. We also used the merger tree obtained with the \texttt{SUBLINK} algorithm \citep{R-G2015} to track the temporal evolution of galaxies. 

We restricted our analysis to isolated galaxies (i.e. central galaxies of their own FoF group) with a stellar mass of $8 < \log (\mathit{M_{*}/M_{\odot}}) < 11$. We removed ongoing mergers by excluding galaxies with a satellite of stellar mass higher than $10$\% of its central host. We also excluded backsplash systems that, at some point in their evolution, were part of a massive group or cluster.
We computed galaxy properties, such as stellar mass and size, by defining all particles and cells within $r_{\rm glx}= 0.15 \, r_{200}$ of the halo centre, where $r_{200}$ is the virial\footnote{Throughout this paper, virial quantities are defined at the radius enclosing 200 times the critical density for closure.} radius.

\subsection{Galaxy kinematic morphology}
\label{subsec:kin_morph}

The morphology of a simulated galaxy can be quantified using the kinematics of its stellar component. We used the fraction of kinetic energy invested in ordered rotation, $\kappa_{\mathrm{rot}}$ \citep{Sales2012}, as an estimate of rotational support, which is defined as follows:

\begin{equation}
    \kappa_{\mathrm{rot}} = \frac{K_{\mathrm{rot}}}{K} = \frac{1}{K} \sum \frac{1}{2} m \left( \frac{j_z}{R} \right)^2
\label{eq:kappa_rot_def}
,\end{equation}
where $K = \sum m |{v}|^2/2$ is the sum of the kinetic energy of all stellar particles of the galaxy; $j_z$ is a star particle's specific angular-momentum component perpendicular to the disc plane\footnote{We define the disc plane of a galaxy using the angular momentum of their young stars, with ages under 1 Gyr.} and $R$ its cylindrical galactocentric distance. 

As shown in \cite{Sales2012}, this estimator strongly correlates with the fraction of stars with a high orbital circularity parameter $\epsilon_j = j_z/j_{\mathrm{circ}}(E)$ \citep{Abadi2003}; higher values of $\kappa_{\mathrm{rot}}$ are associated with a more prominent disc-like component. An arbitrary threshold value of $\kappa_{\mathrm{rot}} = 0.5$ is often used to distinguish rotation-dominated galaxies from dispersion-dominated ones \citep[see e.g.][]{R-G2017, Du2021, Du2022}.

\begin{figure}
    \centering
    \includegraphics[width=\columnwidth]{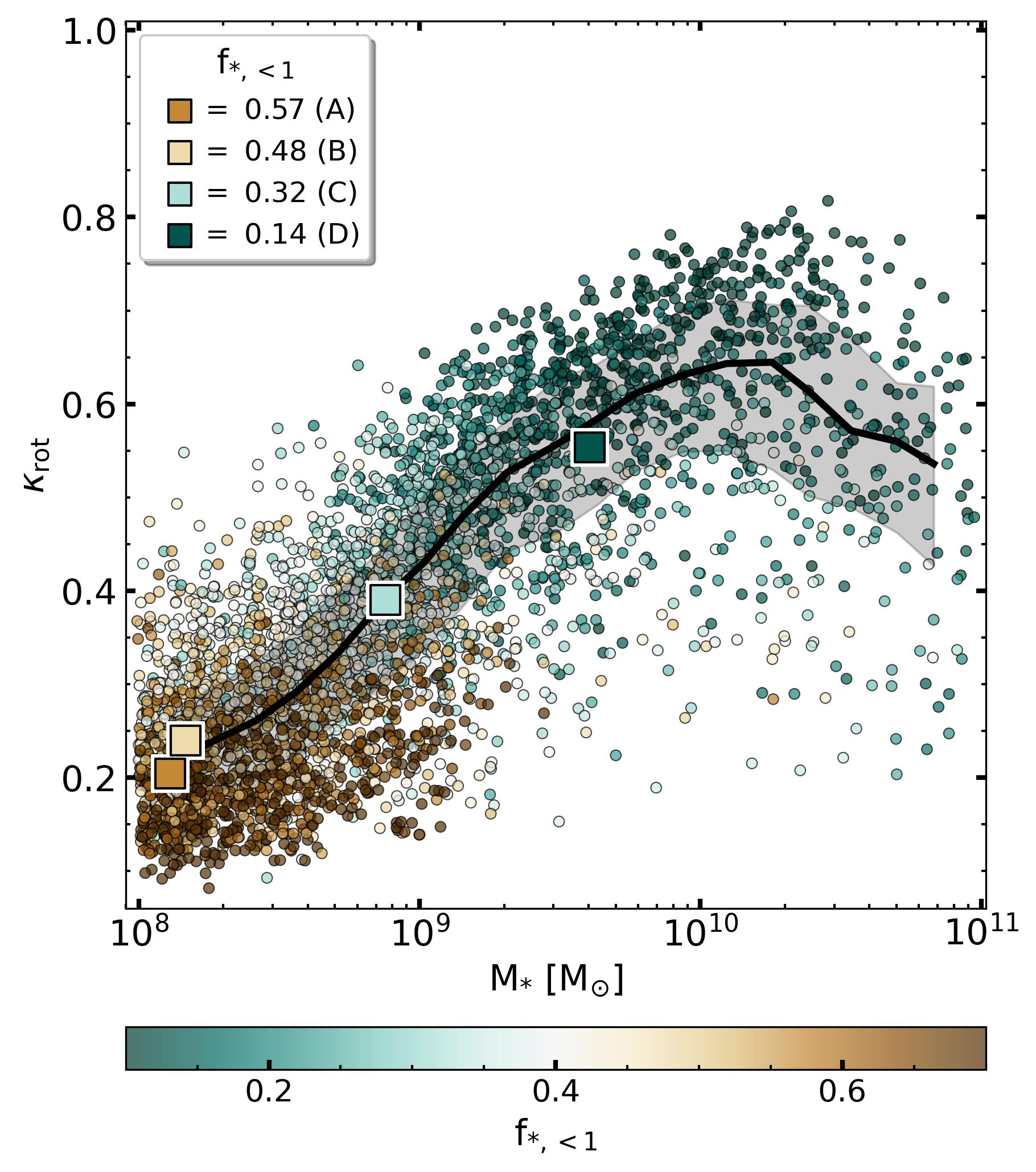}
    
    \caption{Galaxy morphology estimator $\kappa_{\mathrm{rot}}$ as function of stellar mass for 3824 isolated galaxies from TNG50. Each galaxy is represented with a circle coloured according to the fraction of stellar mass enclosed within 1 kpc of its centre: $f_{*,\mathrm{<1}} = M_{*,\mathrm{<1}}/M_{*}$. We indicate the median and the 25th-75th percentiles with a solid black line and shaded region. Filled squares highlight the four example galaxies (A, B, C, and D) shown in Figure \ref{FigGxExamples}. Ordered rotation increases monotonically with stellar mass in the $8.0 < \log(\mathit{M_{*}/M_{\odot}}) \lesssim 10.3$ range, with dwarf galaxies typically having $\kappa_{\mathrm{rot}} \lesssim 0.4$. The colour gradient shows that inner regions become more dominant at lower stellar masses, reaching $f_{*,\mathrm{<1}} > 0.5$, and these galaxies exhibit the lowest rotational support of the sample. At fixed stellar mass, $\kappa_{\mathrm{rot}}$ is anti-correlated with the  stellar-mass fraction in the inner regions $f_{*,\mathrm{<1}}$.}
    \label{FigMstarKrot}
\end{figure}

\section{Results} \label{SecResults}

\subsection{Morphological components}

As illustrative examples, we show face-on and edge-on projections of the stellar (upper panels) and gaseous (lower panels) components of four isolated TNG50 galaxies at redshift $z=0$ in Figure \ref{FigGxExamples}. These galaxies were chosen to have $\kappa_{\mathrm{rot}}$ close to the median value at given stellar mass (see Section \ref{SecKrotMstar}). Stellar mass (and $\kappa_{\mathrm{rot}}$) increases from left to right (galaxies A, B, C, and D). We indicate both $r_{\mathrm{glx}} = 0.15 ~ r_{200}$ (cyan circle) and $r = 1$ kpc (orange circle) in each panel. The former effectively defines the central galaxy as adopted throughout this work; the latter is a fixed physical radius that encloses a dense central baryonic clump found in the innermost region of most TNG50 galaxies at $z=0$. The total stellar mass, total gas mass and $\kappa_{\mathrm{rot}}$ of each galaxy are shown in the labels of Figure \ref{FigGxExamples}.

As anticipated, low values of $\kappa_{\rm rot}$ denote amorphous stellar morphologies without a well-defined disc. High values of $\kappa_{\rm rot}$, on the other hand, indicate systems where a co-planar, rotationally supported disc of stars dominates the morphology of a galaxy.

The spatial distribution of the gas is particularly revealing, especially when seen face-on. Three different regions can be clearly distinguished: (i) a dense $\sim 1\,$kpc inner clump at the centre; (ii) a flattened and extended outer disc-like component; and (iii) a gap between clump and disc. These morphological components are clearly delineated in the gas component of each of these galaxies and are surrounded by a more tenuous, irregularly shaped gaseous envelope. 

Although these gas components can be seen in all four galaxies, their relative prominence varies from galaxy to galaxy. As expected from the values of $\kappa_{\rm rot}$, the disc component grows in importance with increasing stellar mass, from practically non-existent in galaxy A, to clearly dominant in galaxy D. On the other hand, the central clump is seen to have similar size in all galaxies, regardless of mass, while the gap between clump and disc is particularly obvious in galaxy B. These trends suggest a close link between the spatial distribution of the gas and the rotational support of a galaxy, as we discuss next.

\subsection{Rotational support versus galaxy stellar mass}
\label{SecKrotMstar}

We begin by exploring the galaxy-mass dependence on the rotational support of the stellar component (shown in Figure \ref{FigMstarKrot}). Here, each galaxy is represented with a circle and coloured according to the fraction of its stellar mass enclosed within 1 kpc of the centre ($f_{*,\mathrm{<1}} = M_{*,\mathrm{<1}}/M_*$), or, in other words, by the mass fraction attached to the dense inner clump discussed in the previous subsection. The solid black line and the shaded regions indicate the median trend, as well as the 25th and 75th percentiles. The coloured squares correspond to the four example galaxies shown in Figure \ref{FigGxExamples} (galaxies A, B, C, and D).
 
This figure shows a strong positive correlation between $\kappa_{\rm rot}$ and $\mathit{M_*}$, except for the most massive galaxies (i.e. $\log (M_{*}/M_{\odot}) > 10.3$), where there is a hint of a decrease in $\kappa_{\rm rot}$ as $\mathit{M_*}$ keeps increasing. This change in trend is driven by the increasing importance of late major mergers in shaping the most massive galaxies, and by the effects of the energetic feedback from active galactic nuclei \citep[AGNs; see e.g.][]{Naab2014,Genel2015,R-G2017,SotilloRamos2022}.

The main takeaway point of Figure \ref{FigMstarKrot} is, however, the strong stellar-mass dependence of galaxy morphology, as measured by $\kappa_{\rm rot}$. Broadly speaking, most -- if not all -- dwarf (i.e., $\log (M_{*}/M_{\odot}) < 9.0$) galaxies in TNG50 are dominated by a non-rotating spheroidal component, whereas the majority of massive galaxies have rotationally supported discs.

\begin{figure}
    \centering
    \includegraphics[width=0.97\columnwidth]{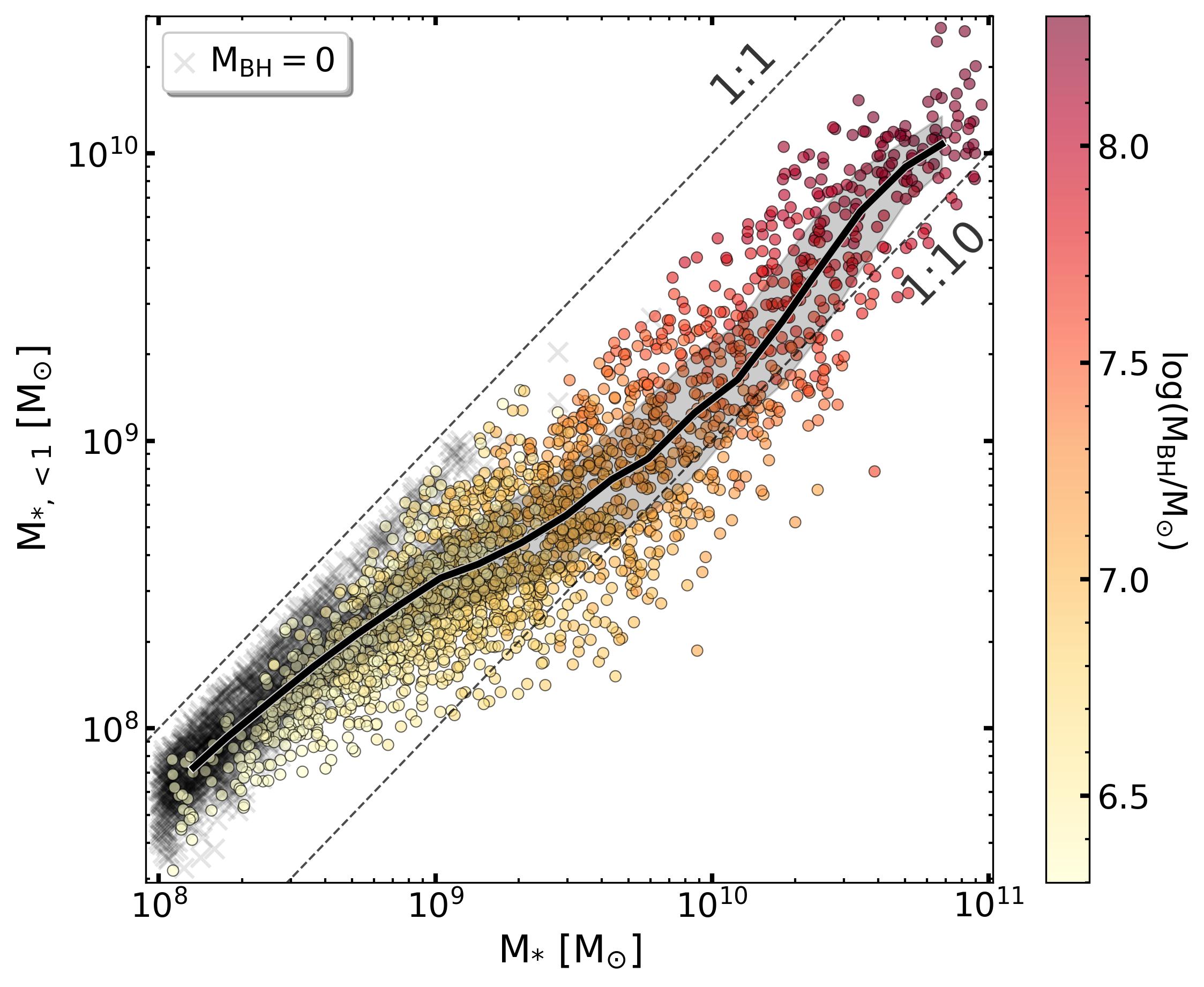}
    \includegraphics[width=0.93\columnwidth]{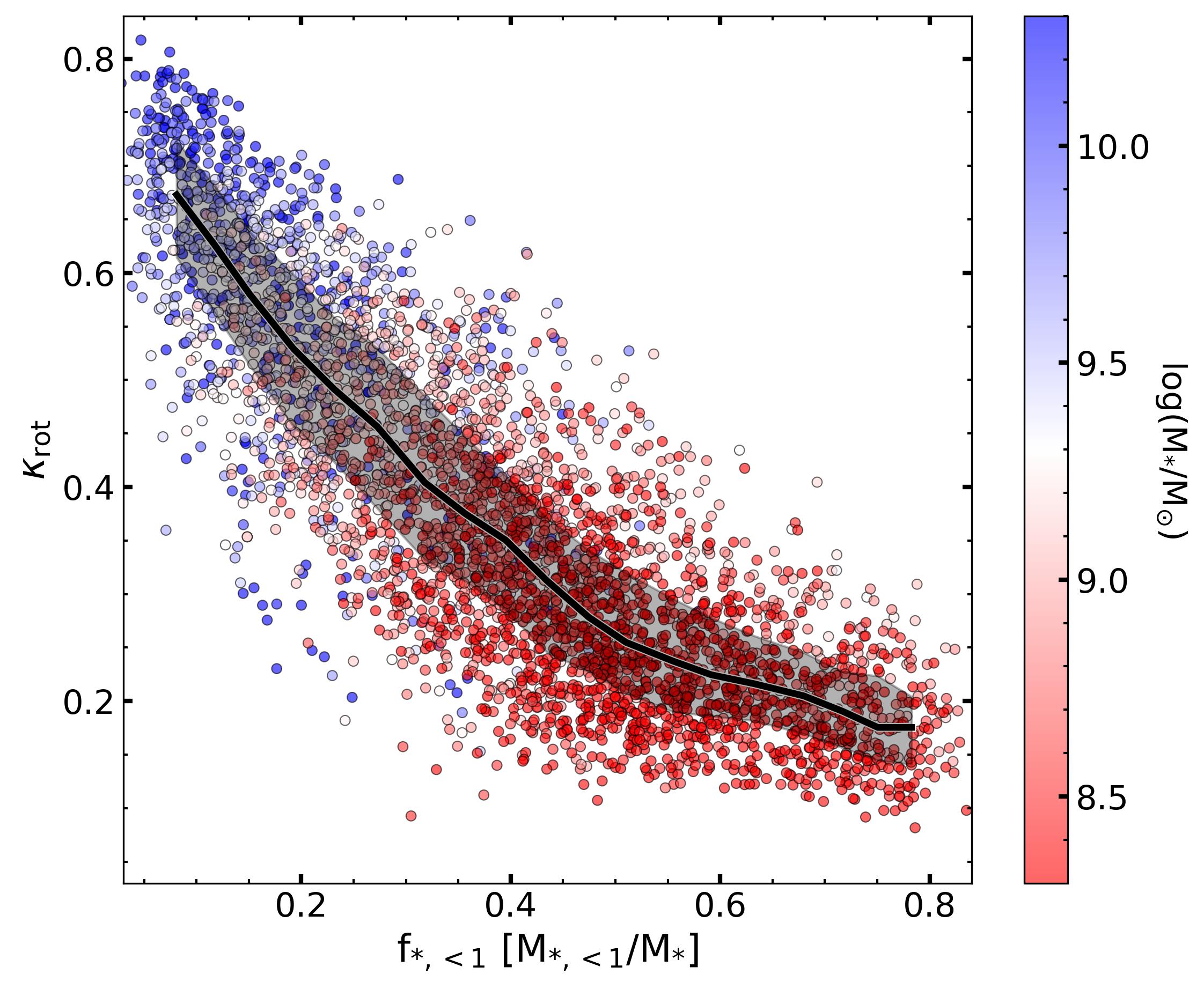}
    
    \caption{Top panel: Stellar mass enclosed within 1 kpc of galaxy centre ($M_{*,\mathrm{<1}}$) as a function of total stellar mass ($\mathit{M_{*}}$), coloured according to the central black-hole mass. In the dwarf regime, most galaxies do not harbour a central massive black hole due to the seeding procedure of TNG50, and we identify them with grey crosses. The dashed grey lines show the 1:1 and 1:10 ratios, indicating that low-mass galaxies have most of their stellar mass enclosed within 1 kpc, while at higher masses the inner region represents roughly 0.1-0.2 of the total mass. Bottom panel: Galaxy morphology parameter $\kappa_\mathrm{{rot}}$ as function of enclosed stellar-mass fraction, $f_{*,\mathrm{<1}}$, coloured according to total stellar mass. The solid black lines and shaded regions indicate the median and the 25th-75th percentiles. At low stellar masses (redder colours), the inner regions become more dominant. Rotational support is anti-correlated with the fraction of stellar mass in the inner regions.}
    \label{fig:M1kpc_f1kpc_wrt_Mstar}
\end{figure}

The origin of this trend is not immediately obvious. One may posit, for example, that gaseous star forming discs are unable to form in dwarfs, and that stars in dwarfs perhaps form in dense clouds before the gas has a chance to coalesce onto a disc. Galaxy B, however, provides a clear counter-example; this system has a distinctive, star forming gaseous disc, yet the stellar component, taken as a whole, barely rotates (i.e. $\kappa_{\rm rot}=0.24$). 

In other words, gaseous discs do form in TNG50 dwarfs, but somehow this does not lead to prominent stellar discs as in more massive systems. The next question thus concerns why disc formation is less efficient in dwarfs. Galaxy A hints at an answer. There is very little gas in this galaxy outside the inner clump, and the morphology of that gas suggests that it is disturbed and out of equilibrium. 

Closer examination shows that the outer disc in Galaxy A has indeed been disturbed by feedback energy from young stars in the inner clump, which, because of the decoupled-wind\footnote{Whenever star formation occurs, stellar feedback is implemented as supernova events that create wind particles. These receive an initial wind velocity and are temporarily decoupled from the magnetohydrodynamic equations until one of the following conditions is met: i) a time of $\Delta t = 0.025 ~ t_{\mathrm{Hubble}}$ has elapsed (where $t_{\mathrm{Hubble}}$ is the age of the Universe when the star formation occurs); or ii) the wind particle has reached a gas cell with a density of $\rho_{\mathrm{cell}} \leq 5$\% of the density threshold for star formation. In TNG50, condition ii) is almost always met first \citep[for a more comprehensive analysis of the model, see][]{Vogelsberger2013,Pill2018}.} strategy adopted in TNG50, is dumped outside the clump. This feedback energy is important enough to blow a gap between gas in the inner clump and outer disc in more massive galaxies (see e.g. galaxy B or C), but in dwarfs, where discs tend to be smaller, feedback disrupts incipient discs. 

As a consequence, most stars in dwarfs form in the non-rotating inner clump, leading to compact systems with little rotational support. Discs tend to be more resilient in more massive systems because those discs are more extended and massive, so the 'gaps' blown by feedback from the inner clump have limited, or little, effect overall.

\begin{figure}
    \centering
    \includegraphics[width=0.95\columnwidth]{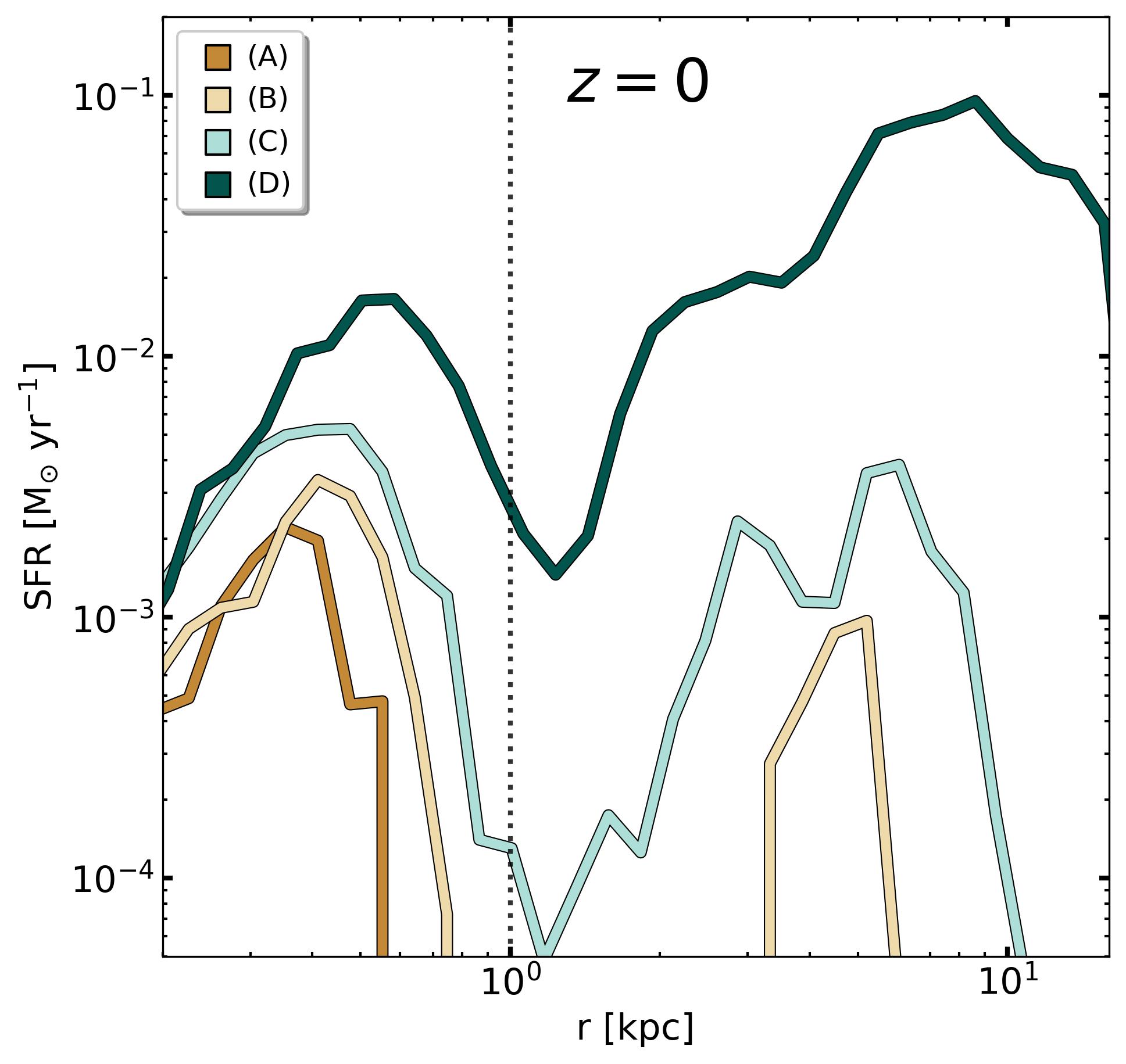}
    \includegraphics[width=0.95\columnwidth]{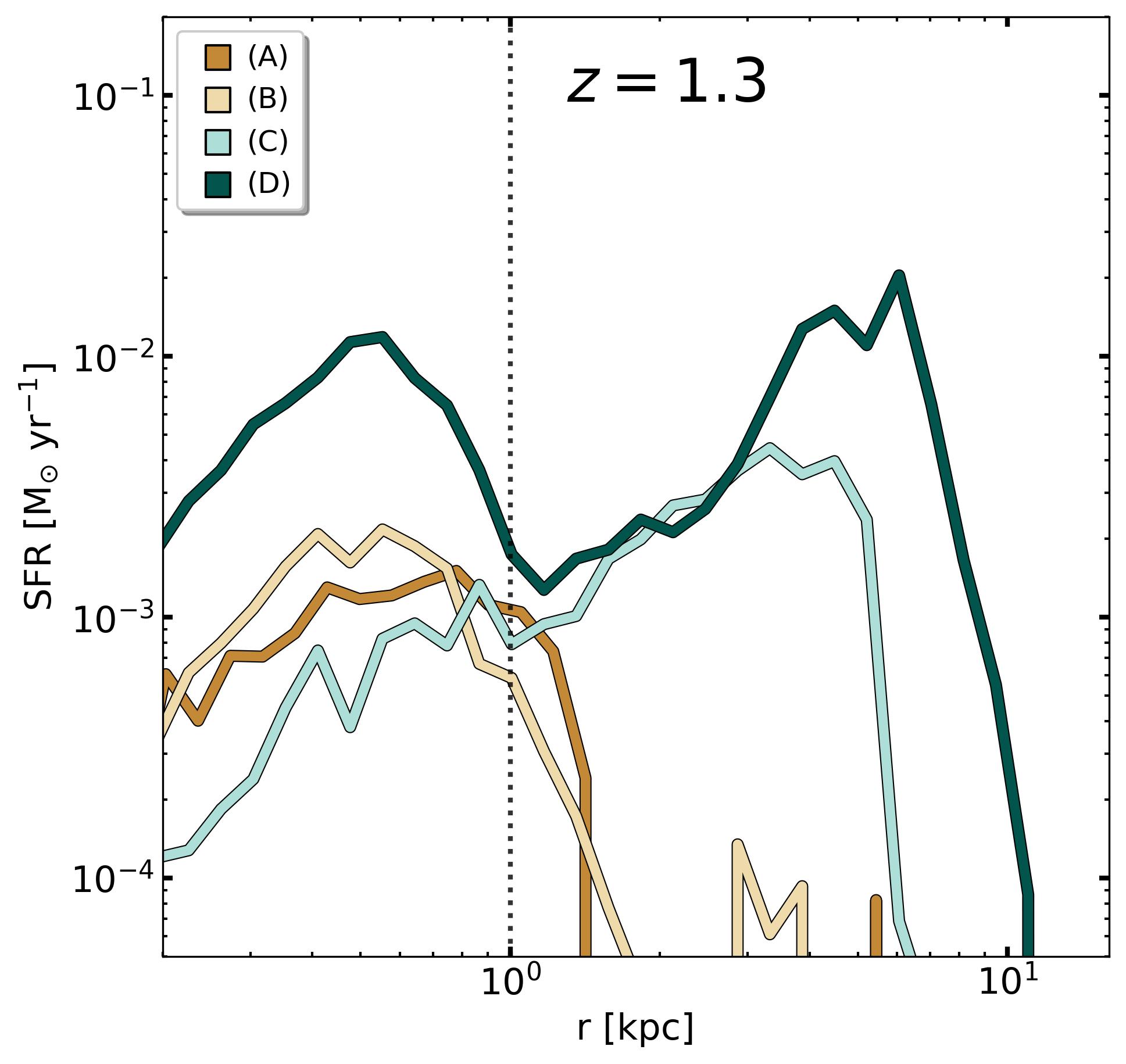}
    
   \caption{Star formation rate (SFR) profiles for the example galaxies (A, B, C, and D) from Section \ref{SecMethods} at $z=0$ (cosmic time $t = 13.8$ Gyr, top panel) and at $z=1.3$ ($t = 4.8$ Gyr, bottom panel). Colours are as for the squares in Figure \ref{FigMstarKrot}: from brown for galaxy A (lowest stellar mass and $\kappa_{\mathrm{rot}}$) to green for galaxy D (highest mass and rotational support). In the presence of a star forming inner clump, the profiles show a relatively quiescent zone between 1 kpc and 2 kpc that separate the central region with the outer gaseous disc. The latter is more extended and more star forming at higher masses, while the size of the unresolved clump evolves slightly. We indicate a galactocentric distance of $r = 1$ kpc, encompassing most inner star formation, with a grey,
dashed vertical line.}
    \label{FigSFRProf}
\end{figure}

Further evidence for this interpretation may be found in the strong correlation between $f_{*,<1}$ and $\kappa_{\rm rot,}$ which can be seen from Figure \ref{FigMstarKrot} and is shown in the bottom panel of Figure \ref{fig:M1kpc_f1kpc_wrt_Mstar}. Clearly, the higher the importance of the inner clump, the lower the rotational support of the galaxy. Since galaxy mass correlates strongly with $f_{*,<1}$, rotational support correlates strongly with $\mathit{M_*}$. Indeed, as shown in the top panel of Figure \ref{fig:M1kpc_f1kpc_wrt_Mstar}, $f_{*,<1}$ correlates inversely with $\mathit{M_*}$, varying from roughly unity at the low-mass end to 0.1-0.2 at the massive end.

The importance of the central clump as a function of $M_*$ is clearly shown in the top panel of Figure \ref{fig:M1kpc_f1kpc_wrt_Mstar}. Simulated galaxies are shown coloured according to the mass of the central black hole ($M_{\rm BH}$), which increases with increasing $M_*$. At fixed stellar mass, galaxies with higher $M_{\rm BH}$ tend to have higher stellar mass enclosed within 1 kpc ($M_{*,<1}$), although the mass of the central black hole makes up typically a negligible fraction ($\sim 0.1$\%) of $M_*$. Most TNG50 dwarfs have no central black hole and are shown with small grey crosses. We conclude that black holes do not play a significant role in the galaxy mass-morphology relation shown in Figure \ref{FigMstarKrot}.

The picture that emerges is thus one where the importance of the inner clump is the key ingredient that determines the rotational support of the stellar component of a TNG50 galaxy. In dwarfs, star formation proceeds mainly or solely in the inner, non-rotating clump, the feedback of which in turn effectively prevents outer gaseous discs from forming. In more massive systems, the importance of the inner clump decreases, and the outer disc component is able to grow and dominate the galaxy's morphology. 

Critical to this interpretation is that the size of the inner clump should be roughly independent of galaxy mass, which would explain why it becomes less and less significant in more extended (massive) systems. We explore this issue next.

\begin{figure*}
    \centering
    \includegraphics[width=0.495\columnwidth]{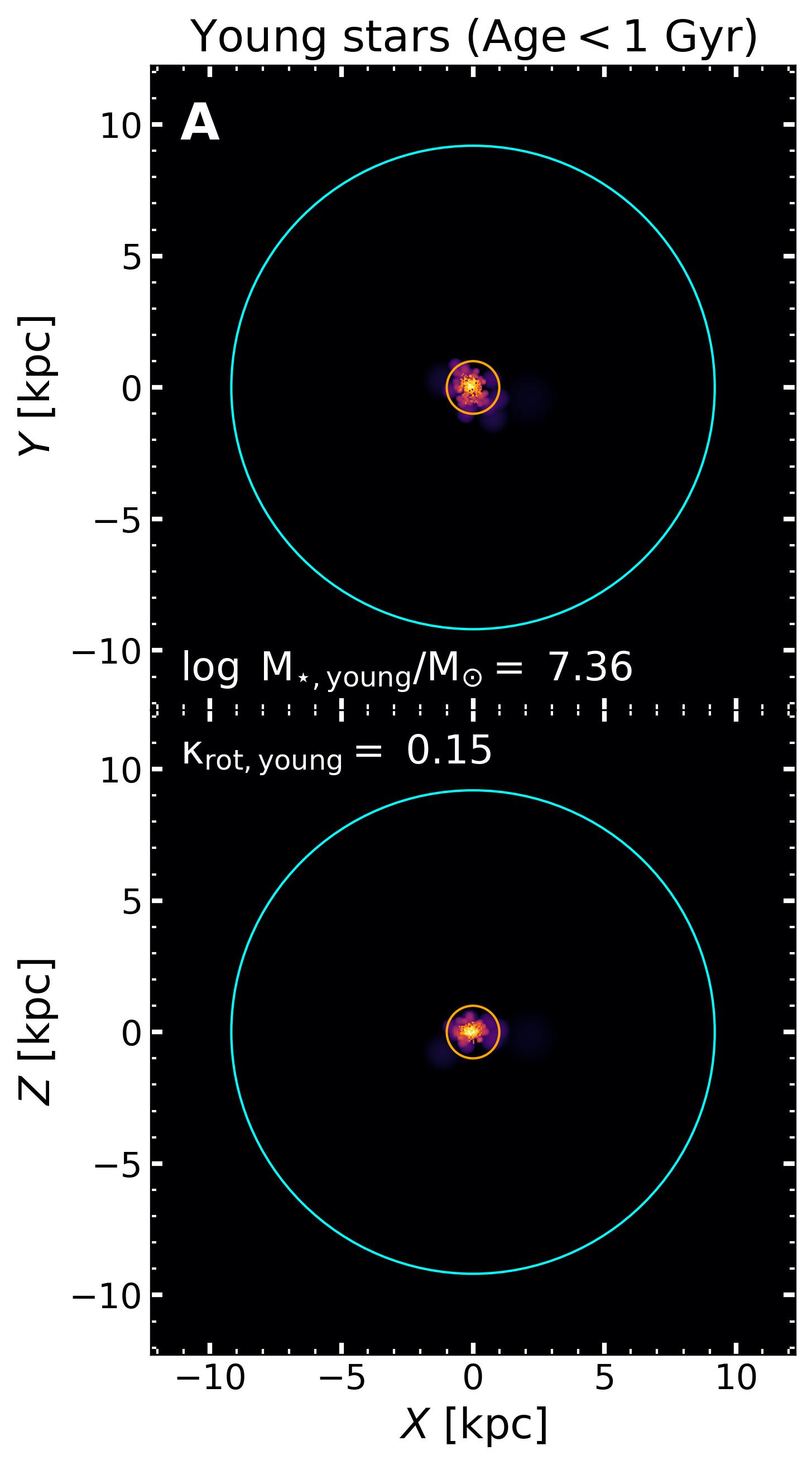}
    \includegraphics[width=0.464\columnwidth]{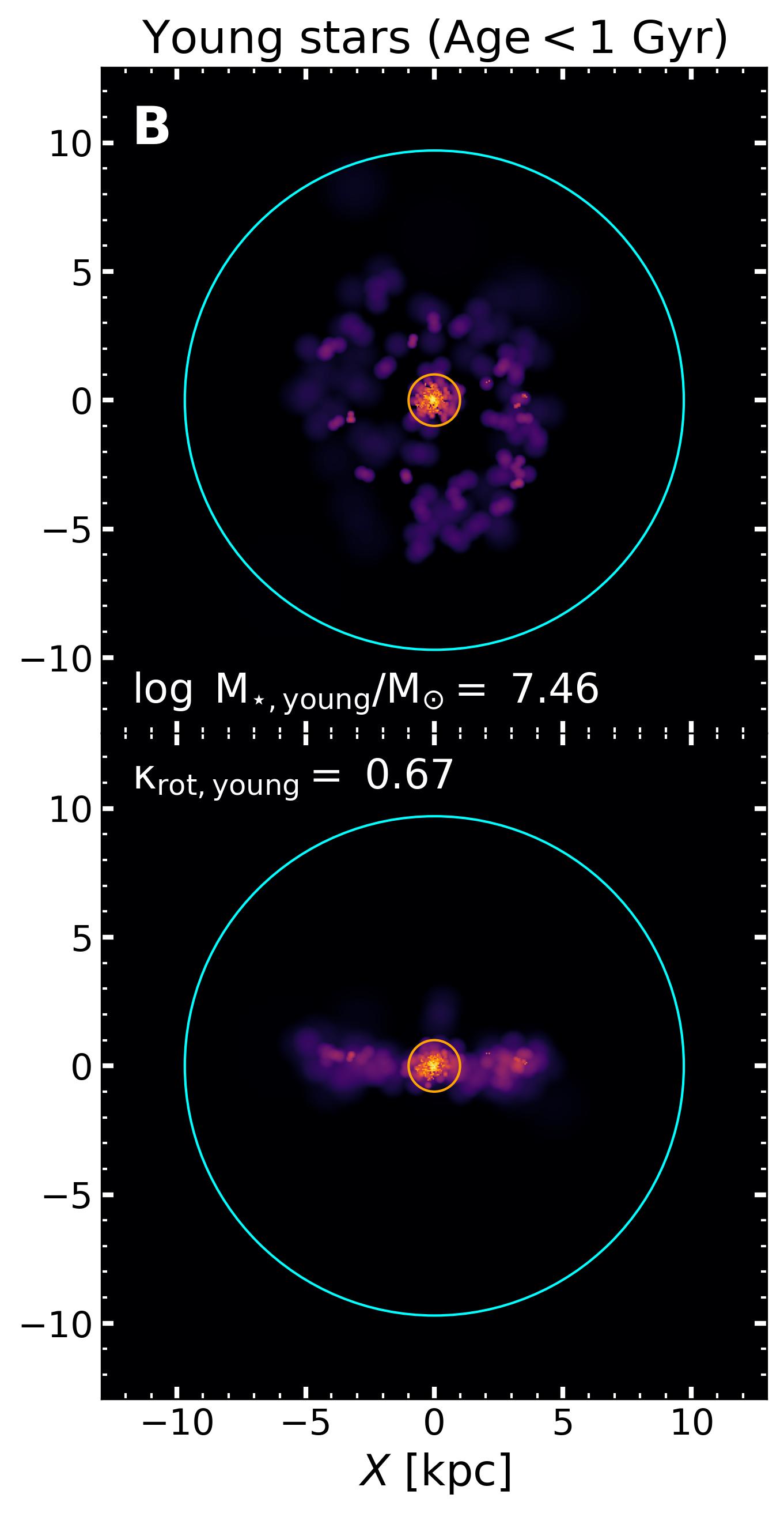}
    \includegraphics[width=0.464\columnwidth]{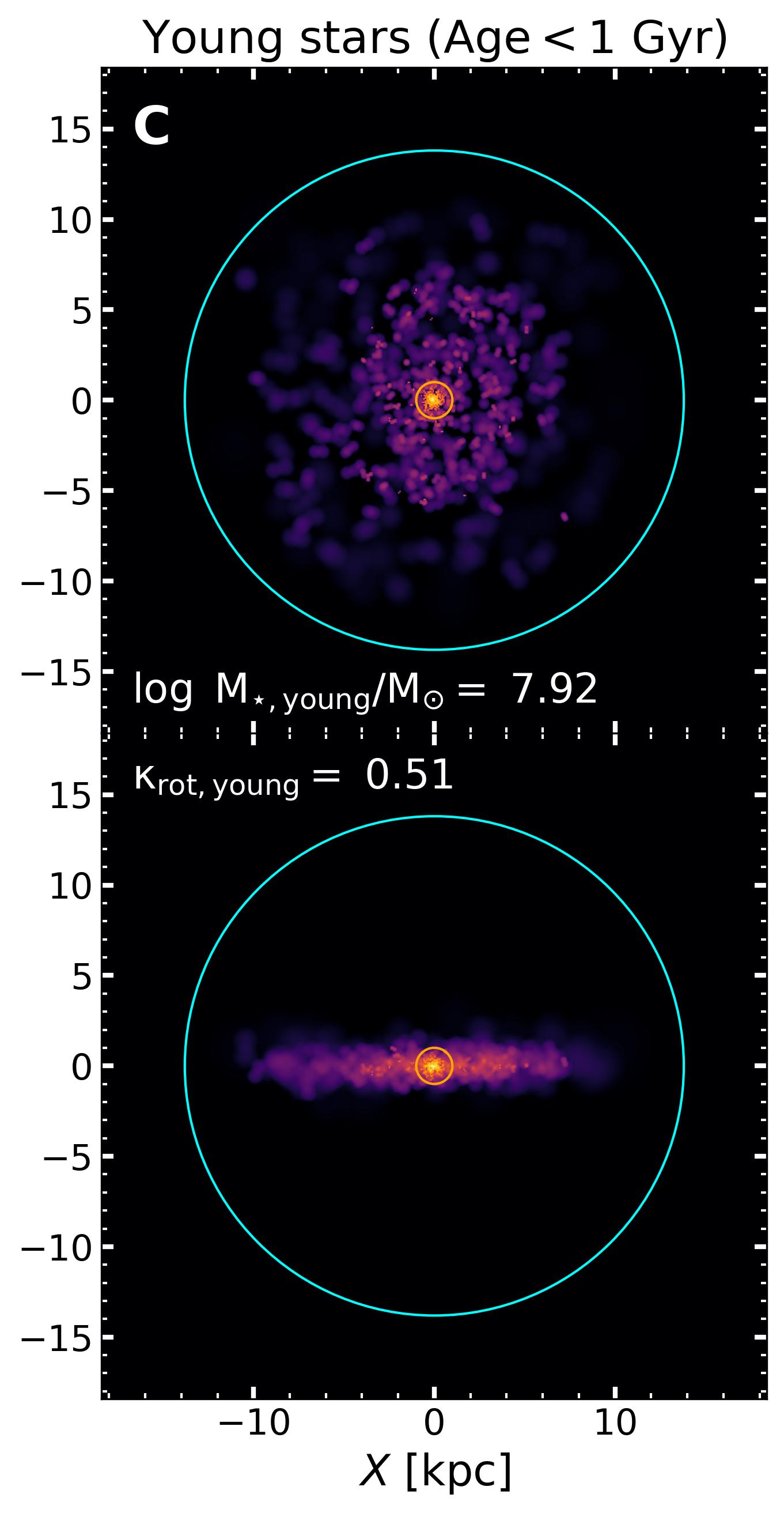}
    \includegraphics[width=0.498\columnwidth]{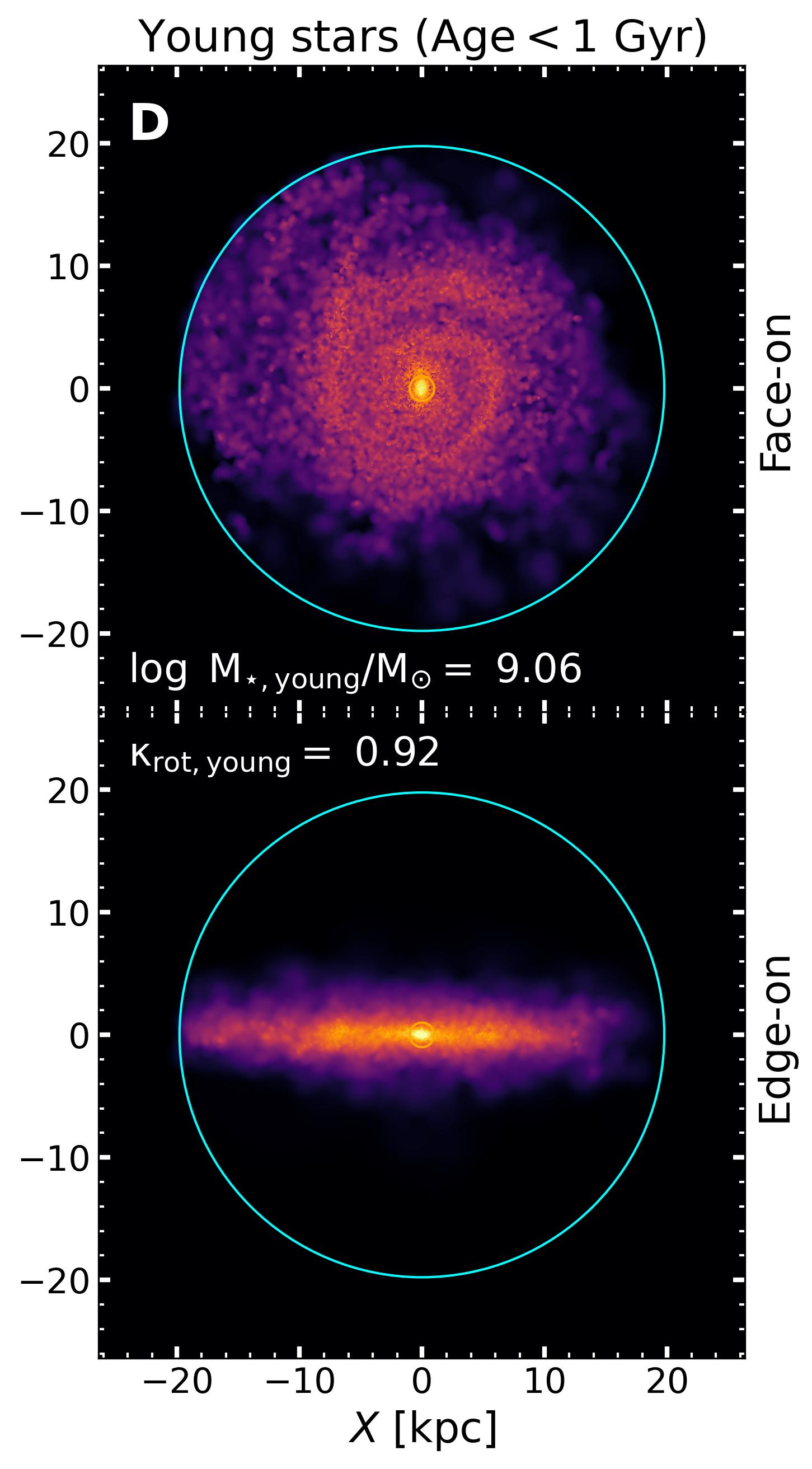}
    
    \caption{Face-on and edge-on projections of young stars (defined as those with an age $< 1$ Gyr at redshift $z=0$) in the four illustrative examples shown in Figure \ref{FigGxExamples} (from left to right, galaxies A, B, C, and D). White labels indicate the mass and rotational support of the young stellar component for each galaxy ($\mathit{M_{*,young}}$ and $\kappa_{\mathrm{rot,young}}$, respectively). Cyan circles show the galaxy size $r = r_{\mathrm{glx}}$, and orange circles mark $r = 1$ kpc. All galaxies exhibit a dense young stellar component at $r < 1$ kpc, with more extended thin discs appearing in the higher mass galaxies. Images were generated with PySPHViewer \citep{BenitezLlambay2017}.}
    \label{FigYoungStars}
\end{figure*}

\begin{figure*}
    \centering
    \includegraphics[width=0.55\columnwidth]{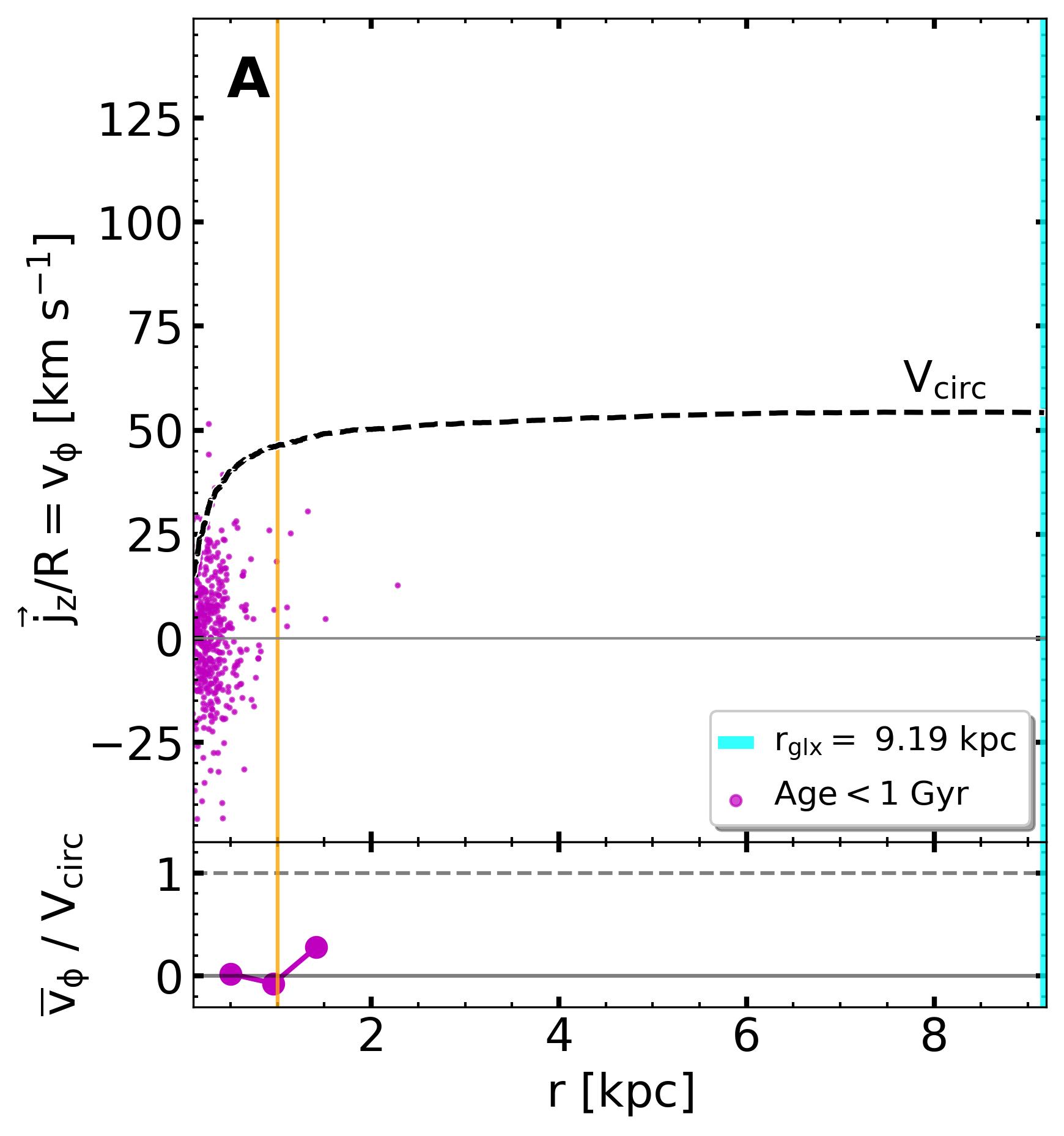}
    \includegraphics[width=0.46\columnwidth]{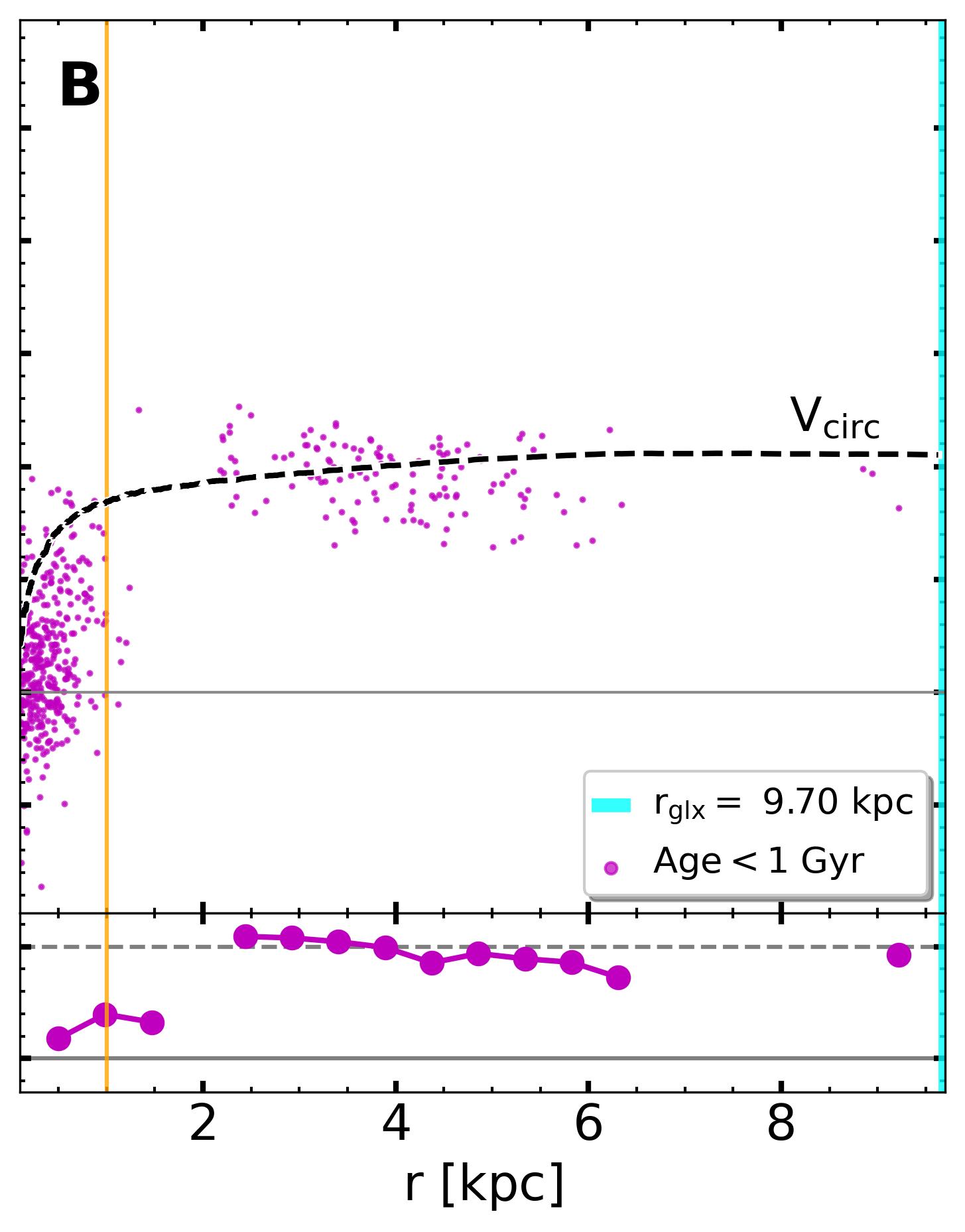}
    \includegraphics[width=0.46\columnwidth]{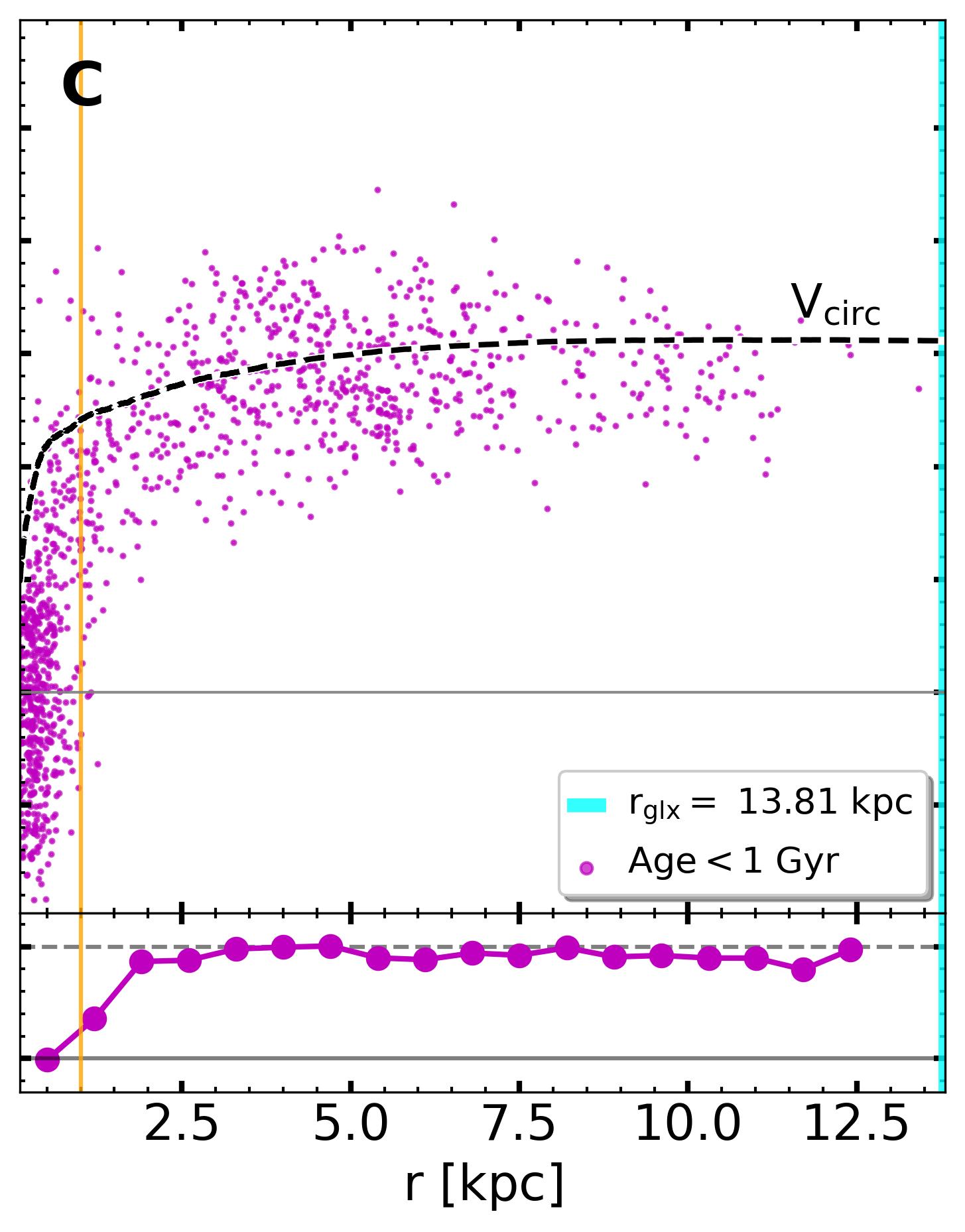}
    \includegraphics[width=0.515\columnwidth]{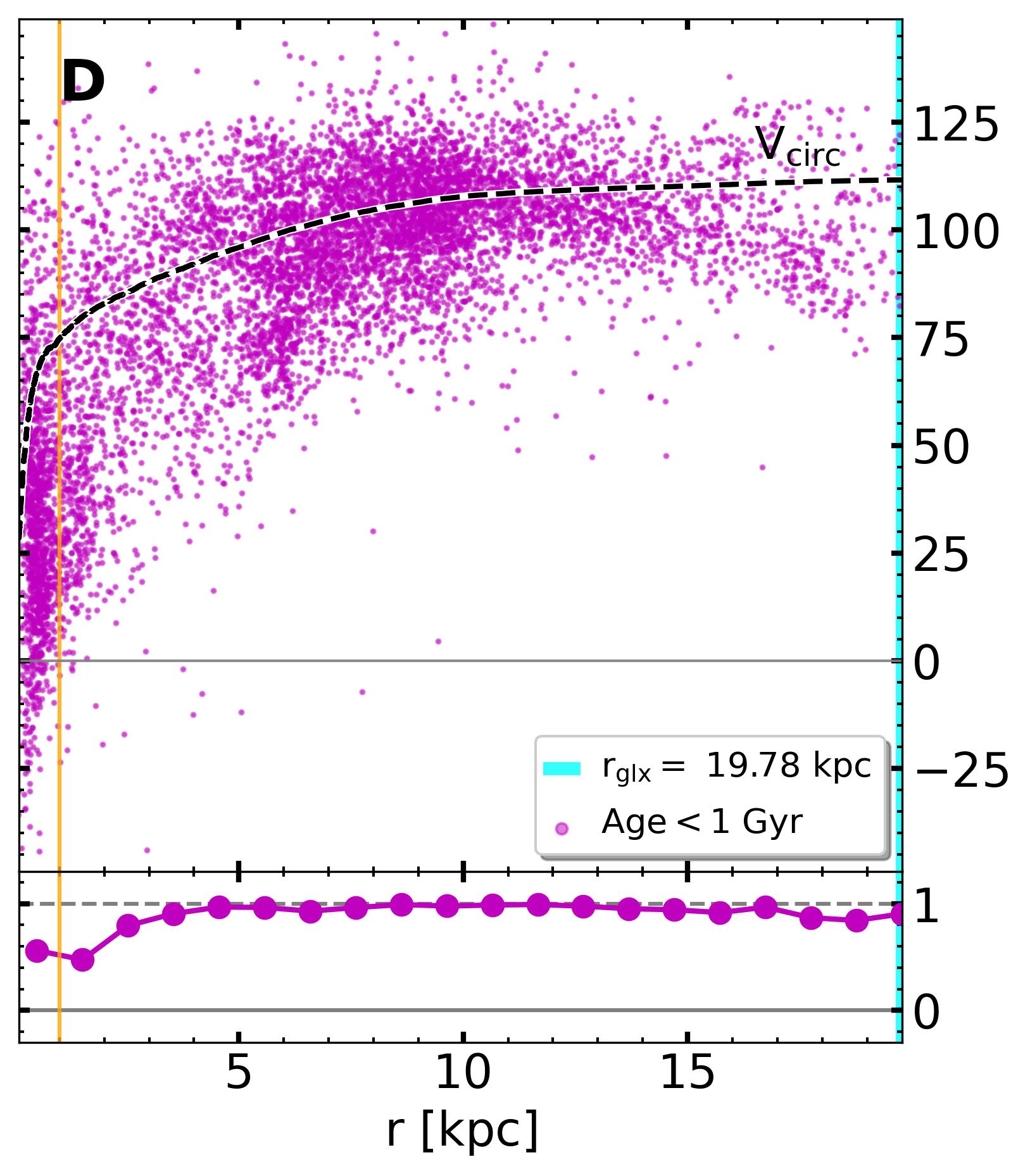}
    
    \caption{Rotation velocity $v_{\phi} = j_z /R$ of young stars (i.e. ages $<1$ Gyr, magenta dots) and circular velocity $V_{\mathrm{circ}} = \sqrt{GM(<r)/r}$ (black dashed lines) for the four example galaxies (A, B, C, and D) from Figure \ref{FigGxExamples}. Bottom panels show the ratio $v_{\phi} / V_{\mathrm{circ}}$ (magenta solid line with circles) as a function of galactocentric distance. Except for the least massive galaxy (A), all have young stars in highly circular orbits at radii of $r \gtrsim 2$ kpc. Conversely, stellar particles with low tangential velocities are found in the inner regions ($r < 1$ kpc). As previously shown, higher mass galaxies have more stars born in the outer regions.}
    \label{FigVphiProf}
\end{figure*}

\subsection{Star formation radial profiles}
\label{SecSFRProf}

Figure \ref{FigSFRProf} shows the star formation rate (SFR) radial profile\footnote{We measured the total SFR in each of 40 log-spaced radial shells between 0.06 and 25 kpc.}
for the four example galaxies presented in Figure \ref{FigGxExamples}. Lines are colour-coded for each galaxy as in Figure \ref{FigMstarKrot}. The top panel shows the four SFR profiles at $z=0$, whereas the bottom panel is analogous to the top, but for $z=1.3$.

Note that the three morphological features of the gas distribution discussed above are clearly reflected here. All four galaxies have an $\sim 1$ kpc-sized star forming inner
clump, independent of galaxy mass, and an outer star forming disc (except for galaxy A, where the outer disc is not forming stars at $z=0$). The clump size does not seem to evolve with time; indeed, it is roughly $1$ kpc in radius in all four galaxies at $z=1.3$ (bottom panel in Figure \ref{FigSFRProf}). Only galaxy B has a less well-defined inner clump at $z=1.3$, suggesting that the clump is a feature that grows more massive and more clearly defined with time.
The inner clump and the outer disc are separated by a gap driven by the feedback energy of young stars in the inner clump, as we discuss below in Section \ref{SecInnerClump}.

\subsection{Rotational support of young stars}

The gas morphology of the four example galaxies shown in Figure \ref{FigGxExamples} suggests that young stars that form outside of the central clump do so in rotationally supported discs. This is shown more clearly in Figure \ref{FigYoungStars}, where we plot face-on and edge-on projections of the young stars (i.e. younger than $1$ Gyr) of the four galaxies at $z=0$.

These examples illustrate that, outside the central clump, stars form in discs, confirming the interpretation that the low values of $\kappa_{\rm rot}$ in dwarf galaxies result not from the fact that gaseous discs cannot form in such systems, but rather because in those systems the majority of stars form in the non-rotating central clump. The case of galaxy A also suggests that outer star forming discs may be short-lived in dwarfs, and easily disrupted by feedback from the inner clump.

This interpretation is consistent with the kinematics of young stars in the four example galaxies, as shown in Figure \ref{FigVphiProf}, where we plot the tangential velocity $v_{\phi} = j_z/R$ of all young stars (purple circles). The circular velocity $V_{\mathrm{circ}} = \sqrt{GM(<r)/r}$ is also shown with a dashed black line. The bottom panels show the radial profile of the median $v_{\phi}$, in units of the circular velocity. This figure makes it clear that young stars born in the inner clump exhibit negligible rotation, while those in the outer regions show tangential velocities generally consistent with the circular velocity, albeit with considerable scatter.

\begin{figure}
    \centering
    \includegraphics[width=0.99\columnwidth]{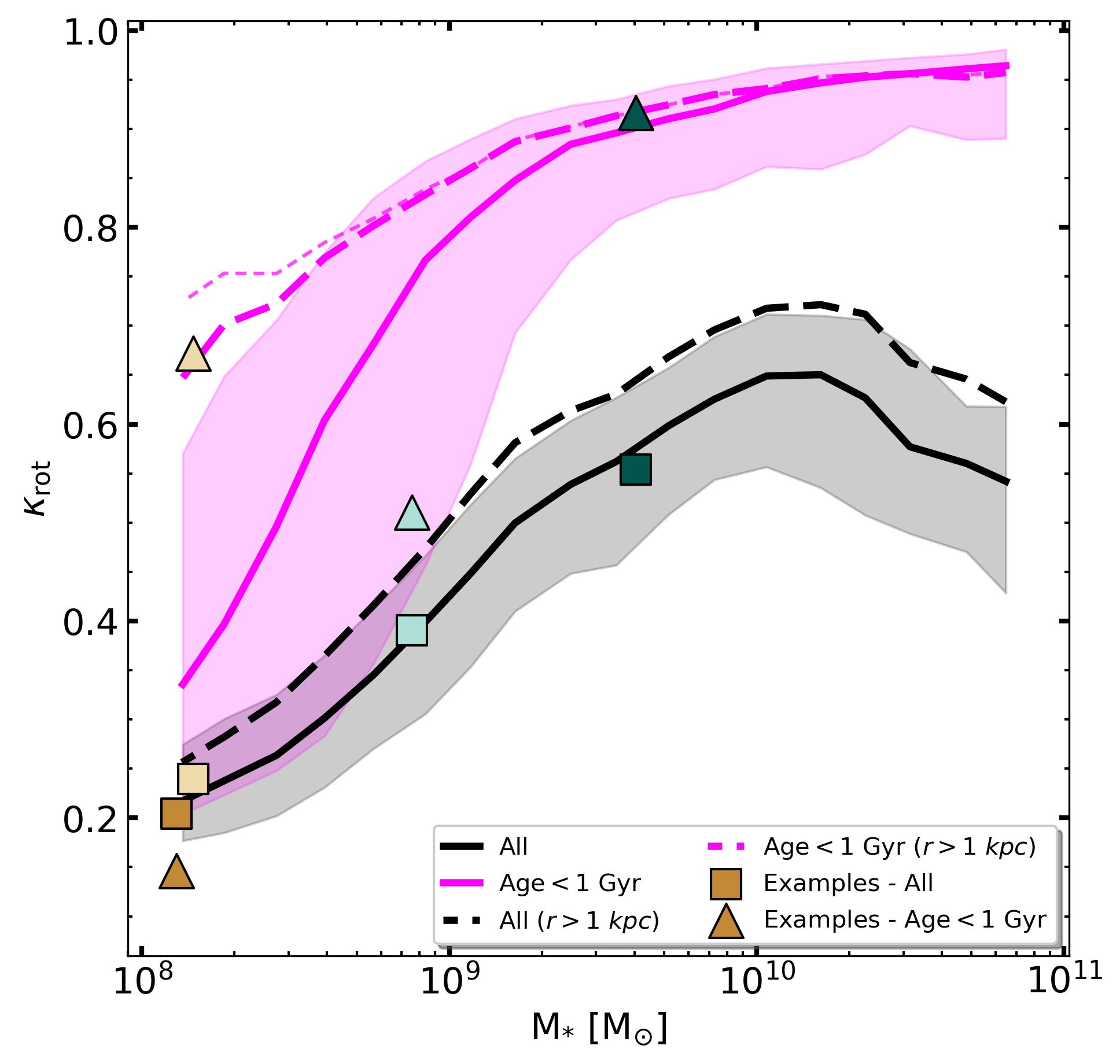}
    \caption{Rotational support parameter $\kappa_{\mathrm{rot}}$ as function of galaxy stellar mass. Solid and dashed lines show results for stars within the galactic radius ($r<r_{\mathrm{glx}}=0.15\, r_{200}$) and in the outer regions ($r > 1~\mathrm{kpc}$), respectively. Black lines represent all stars, while magenta lines represent young stars (i.e. age $< 1$ Gyr). Young stars in the outer regions (magenta dashed line) have $\kappa_{\mathrm{rot}} > 0.7$ and tend to be rotation-supported in all galaxies. Conversely, young stars as a whole (solid magenta line) are non-rotating  at low masses ($\kappa_{\mathrm{rot}} < 0.5$); this is because in these systems most stars form in the central, non-rotating clump. Symbols correspond to galaxies A, B, C, and D, identified according to colour (as in Figure \ref{FigMstarKrot}; squares: $\kappa_{\mathrm{rot}}$; triangles: $\kappa_{\mathrm{rot,young}}$). A thinner magenta line indicates results when only galaxies with at least  $50$ young stars in the outer regions are included in the analysis.}
    \label{FigMstarKrotYoung}
\end{figure}

Figure \ref{FigMstarKrotYoung} shows that the results discussed above for the four example galaxies actually apply to the whole TNG50 galaxy population. This figure shows $\kappa_{\rm rot}$ versus $\mathit{M_*}$ and is analogous to Figure \ref{FigMstarKrot}, but splitting stars into components according to age and radius. Solid lines (and shaded areas) refer to the full TNG50 galaxy population. Black denotes that we used all stars in each galaxy, whereas magenta shows that we restricted the computation to stars younger than $1$ Gyr. Dashed lines correspond to stars outside the inner clump; i.e., $r>1$ kpc. The difference between solid and dashed magenta curves confirms that young stars outside the inner clump form in roughly rotationally supported structures.

Such stars, however, make up a minority of all young stars in dwarfs, which are heavily dominated by the inner clump. As a result, ordered rotation is negligible in most dwarfs. One may worry that there are too few young stars with $r>1$ kpc to compute $\kappa_{\rm rot,young}$ reliably at the faint end. The thin, dashed magenta line of Figure \ref{FigMstarKrotYoung} shows results where we only used galaxies with  a minimum of $50$ young stars in the outer regions; this confirms that young stars in the outer regions are indeed formed in rotationally supported structures.

\begin{figure*}
    \centering

    \includegraphics[width=0.405\textwidth]{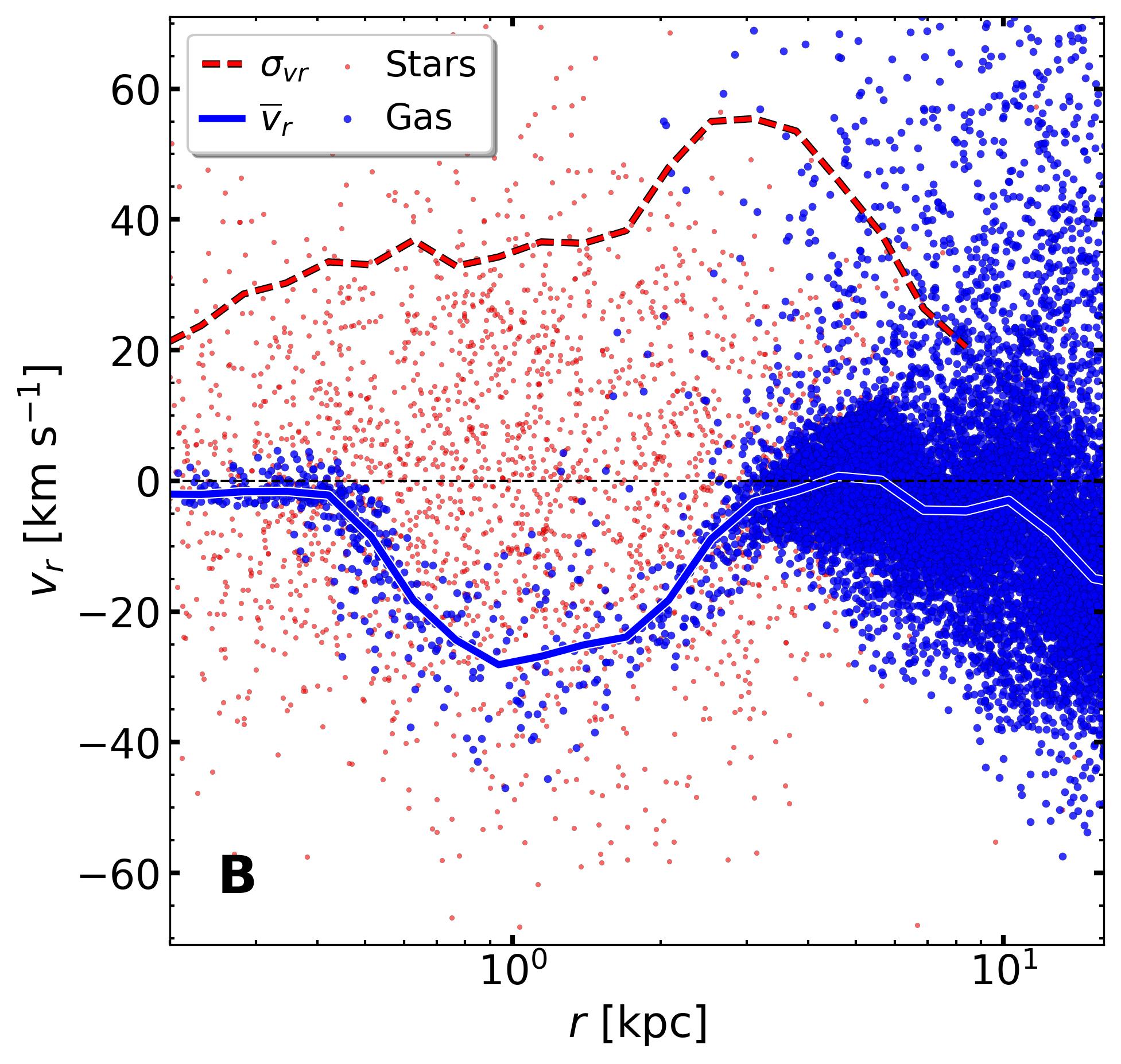}
    \includegraphics[width=0.4\textwidth]{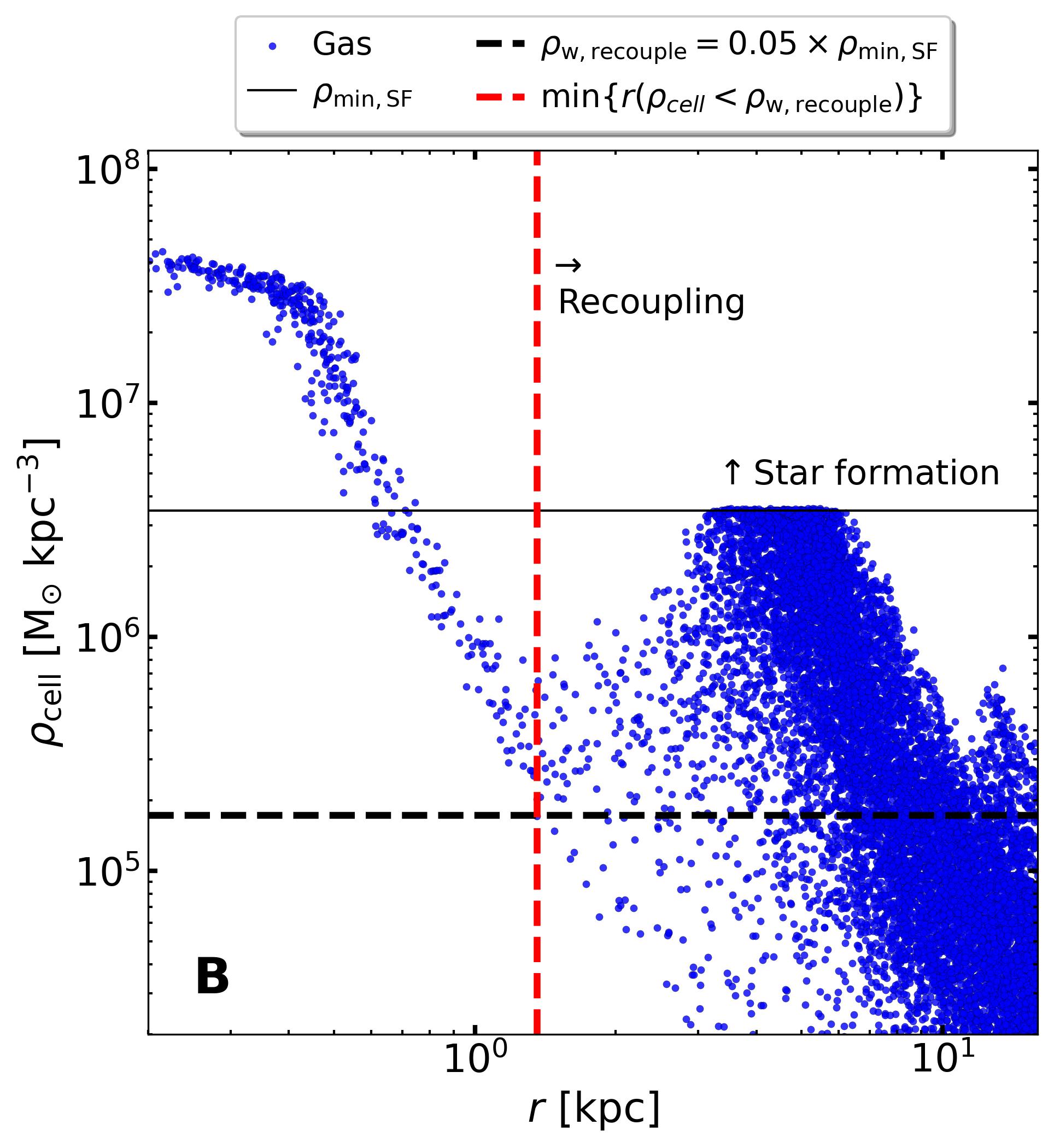}

    \includegraphics[width=0.405\textwidth]{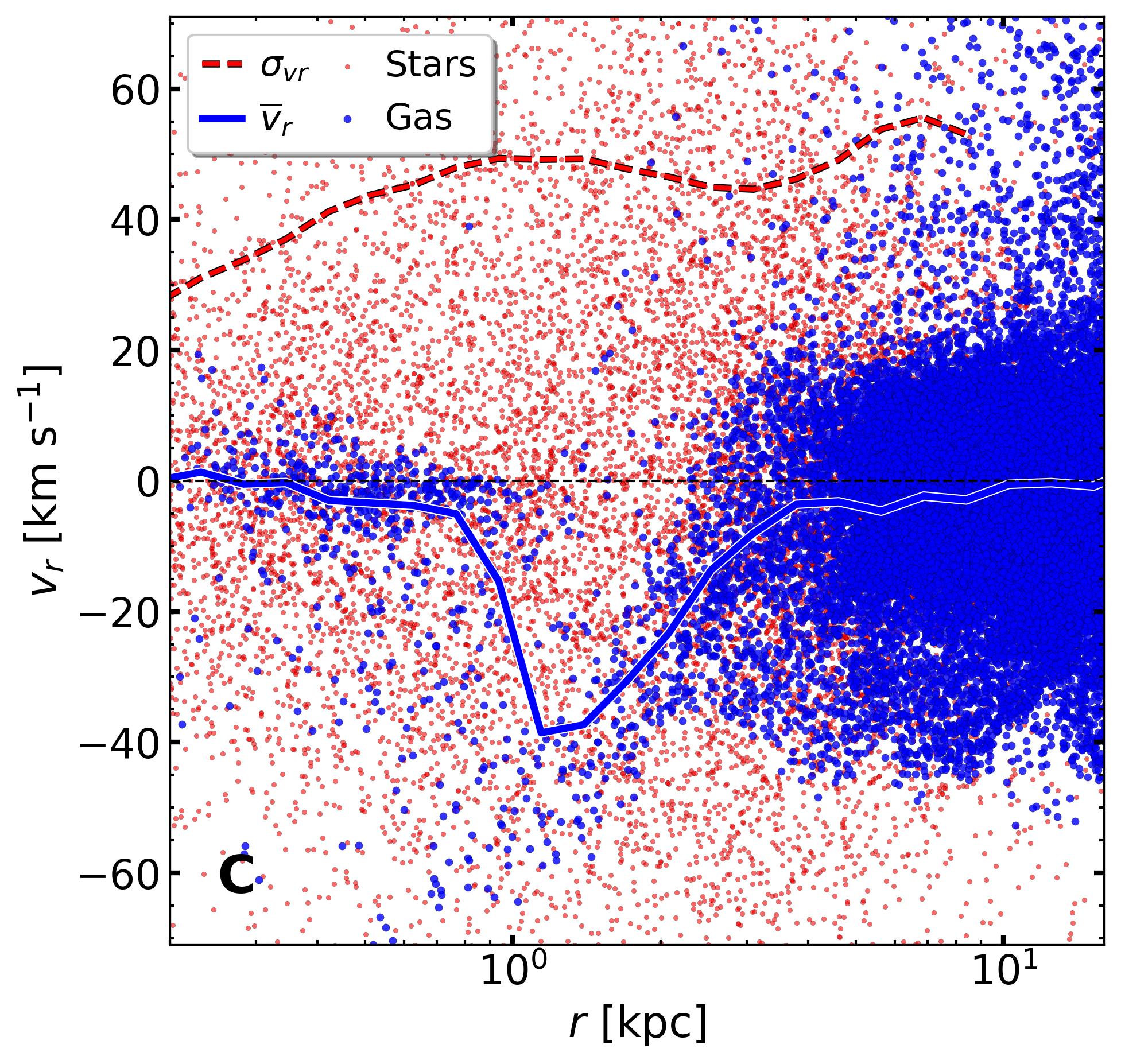}
    \includegraphics[width=0.4\textwidth]{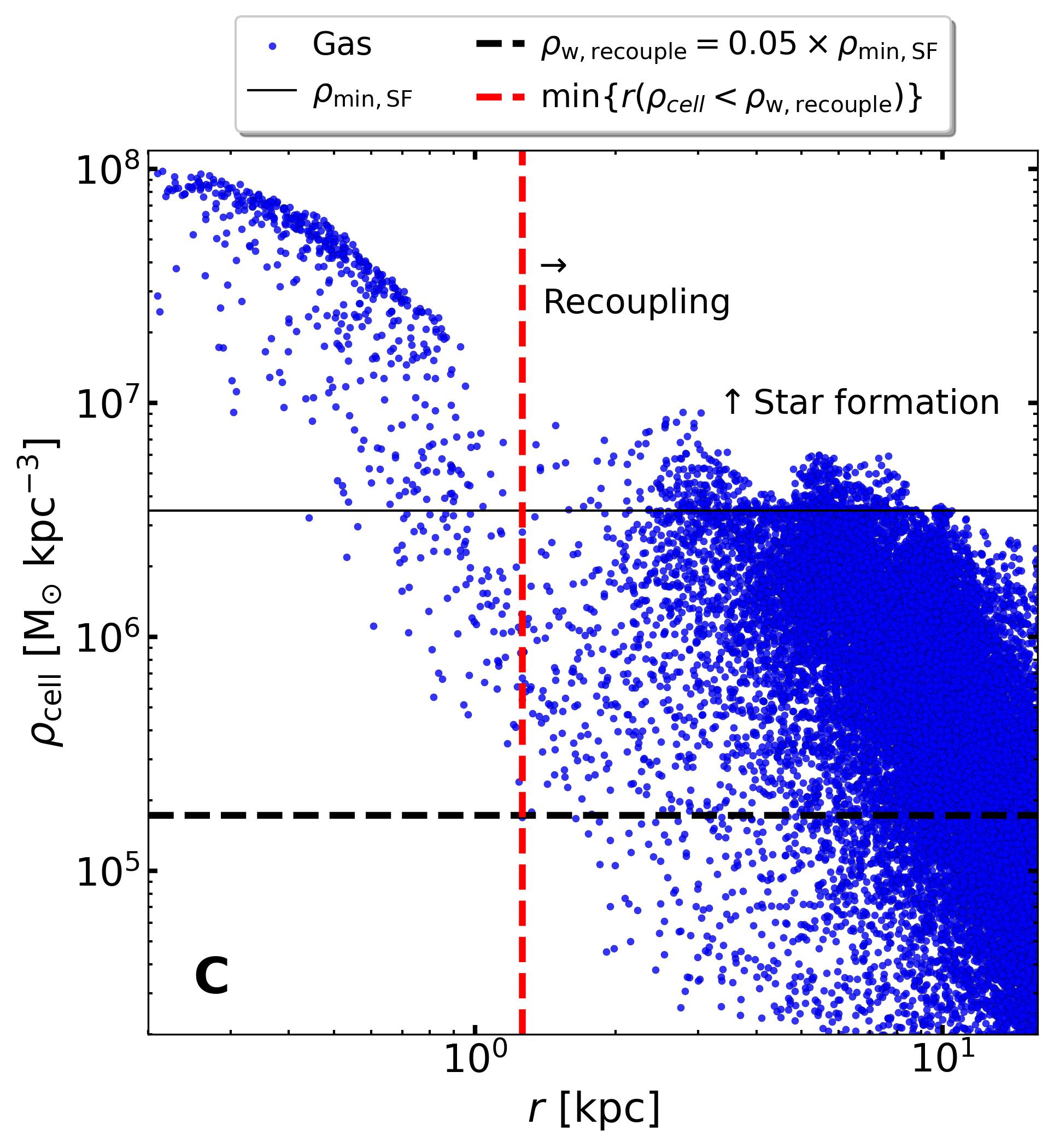}
    
    \caption{Left panels: Radial velocity of stars (red) and gas (blue) for galaxies B and C at redshift $z=0$, which were chosen as illustrative examples. The solid blue curve indicates the median radial velocity of gas cells at a given radius;  the dashed red curve indicates the radial velocity dispersion of star particles. Right panels: Gas cell density versus galactocentric distance. The solid black horizontal line shows the density threshold for star formation ($\rho_{\mathrm{min,SF}} \approx 3.5 \times 10^{6} ~ \mathit{M_{\odot}} ~ \mathrm{kpc^{-3}}$). The dashed horizontal line indicates the minimum wind-recoupling density, defined as $\rho_{\mathrm{w,recouple}} = 0.05 \times \rho_{\mathrm{min,SF}}$. The minimum radius where recoupling may occur (i.e. where a gas cell falls below $\rho_{\mathrm{w,recouple}}$) is indicated with a dashed red vertical line. Note that the quiescent zone between the inner clump and the outer disc roughly coincides with the 'recoupling radius', indicating that the star formation gap is caused by feedback winds.}
    \label{FigClumpProf}
\end{figure*}

\begin{figure}
    \centering
    \includegraphics[width=\columnwidth]{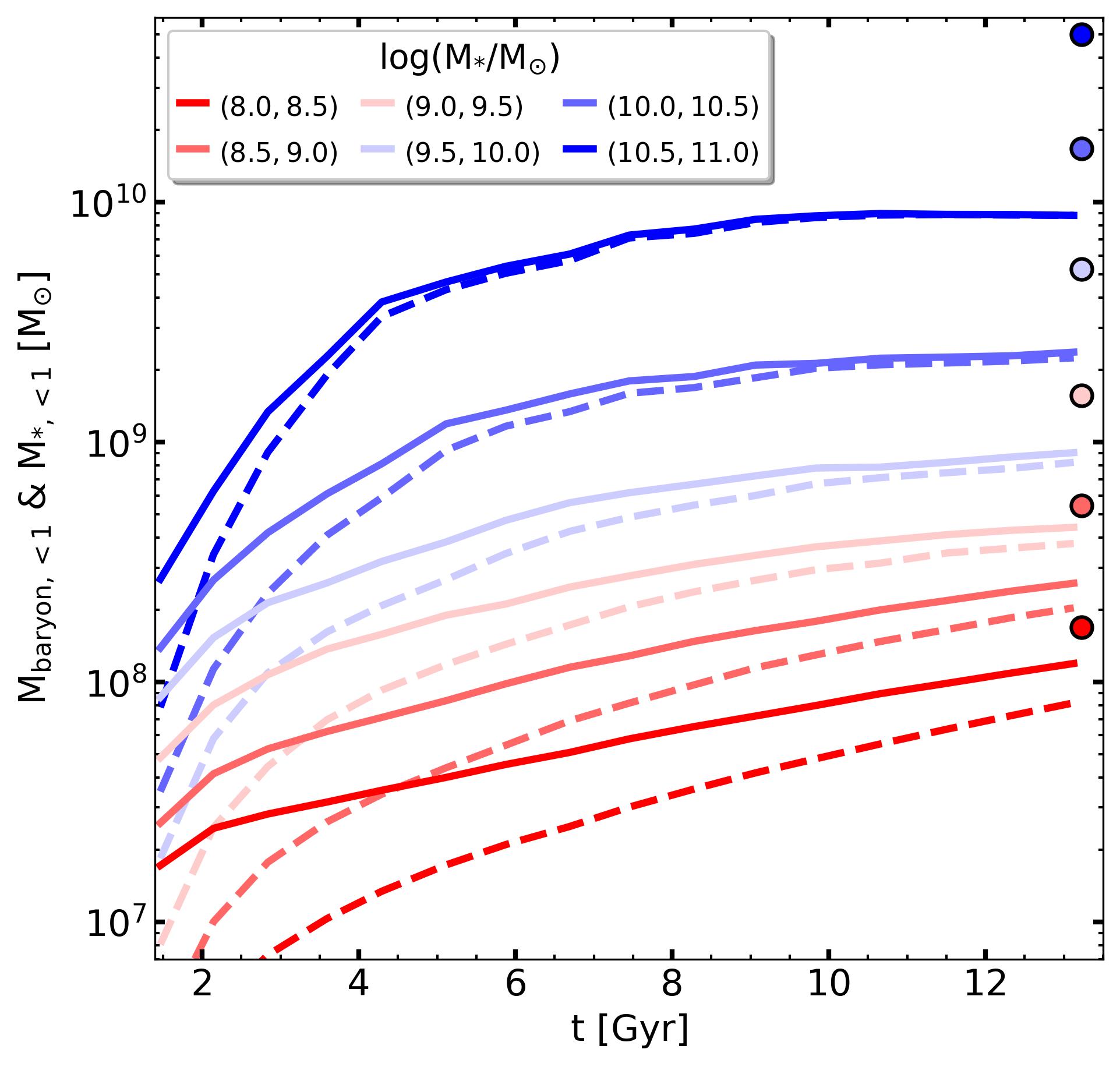}
    \caption{Evolution of median baryonic (solid lines) and stellar mass (dashed lines) in the inner clump (i.e. enclosed within 1 physical kpc) for stacks of galaxies in 0.5 dex $M_*$ bins.  Lines are colour-coded by total stellar mass at redshift $z=0$. The median $M_*$ of each bin is shown by the solid circles plotted at $t \approx 13.5$ Gyr. The inner clump mass grows gradually in lower-mass galaxies (redder colours), indicating significant gas infall into their central regions and ongoing star formation. The inner clumps of more massive galaxies form early and have little gas left at $z=0$.}
    \label{FigEvolInnerClump}
\end{figure}

\begin{figure*}
    \centering
    \includegraphics[width=0.99\linewidth]{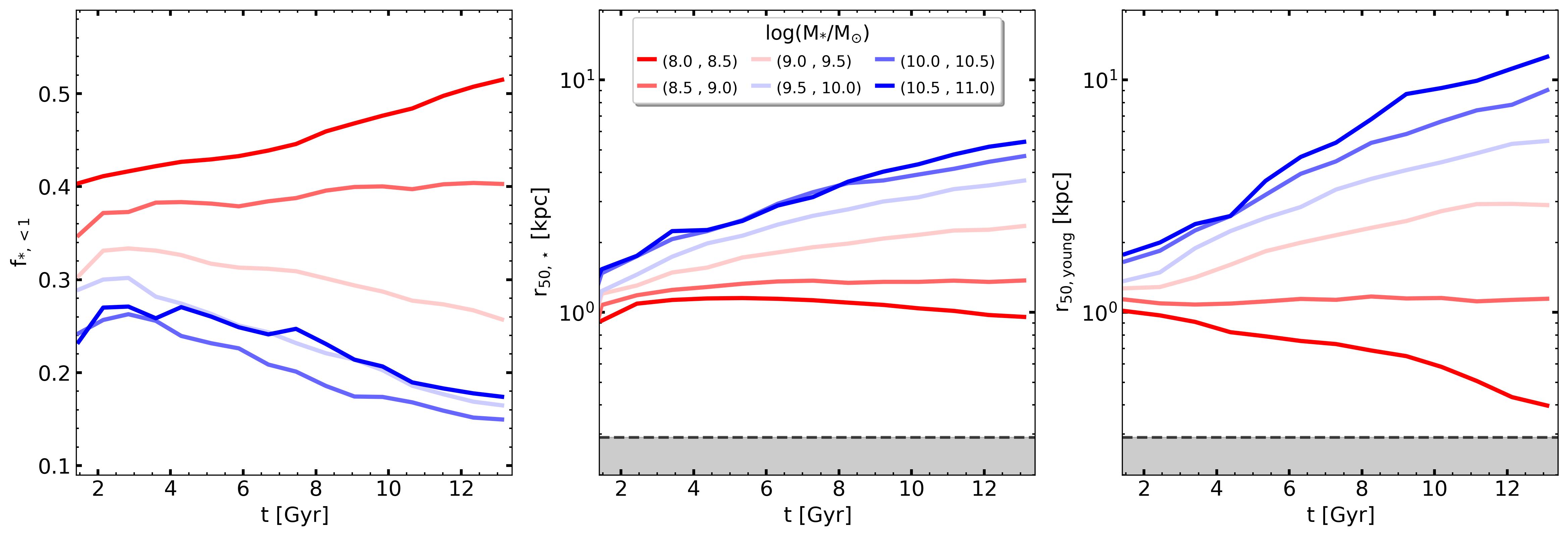}
    
    \caption{Evolution of i) stellar-mass fraction within $r < 1$ kpc, $f_{*,\mathrm{<1}}$ (left panel); ii) median stellar half-mass radius, $r_{50,*}$ (centre panel); and iii) median half-mass radius of young stars (i.e. age $<1$ Gyr), $r_{50,\mathrm{young}}$ (right panel), for the same stacks of galaxies in 0.5 dex $M_*$ bins shown in Figure \ref{FigEvolInnerClump}. Higher mass galaxies (bluer) show a decreasing enclosed stellar-mass fraction as newly formed stars are born at progressively larger radii, which is consistent with inside-out galaxy growth. Conversely, low-mass galaxies (redder) appear to become increasingly centrally concentrated, with an outside-in trend where newly formed stars are found at progressively smaller galactocentric radii.}
    \label{FigSizeEvol}
\end{figure*}

\subsection{Inner clump structure and evolution} \label{SecInnerClump}

It is clear from the above discussion that the present-day morphology of TNG50 galaxies, and dwarfs in particular, is critically dependent on the structure of the inner clump. Therefore, it is important to understand the structure of the clump and its evolution. Figure \ref{FigClumpProf} shows, as a scatter plot, the radial velocity of gas cells and star particles as function of radius for galaxies B and C, which were chosen as illustrative examples. 

In this figure, radial velocity profiles are shown for gas (in blue) and stars (in red) in the left panels, whereas gas-density profiles are shown as scatter plots in the right hand panels. The horizontal lines indicate the threshold density for star formation (thin, solid black line: $\rho_{\rm min,SF}$) as well as the density where feedback-driven winds are allowed to re-couple with the interstellar medium (thick, dashed black line: $\rho_{\rm w,rec} = 0.05 \times \rho_{\rm min, SF}$).

Inside $\sim 1$ kpc, the gas is dense enough to be star forming, and it appears to be near hydrostatic equilibrium (i.e. almost at rest and pressure-supported), except for hints of ongoing accretion at the clump's edge. On the other hand, stars are supported by their random motions. The stellar radial velocity dispersion is shown, as a function of radius, by the dashed red curve in Figure \ref{FigClumpProf}. As shown in the previous subsection for young stars, there is no detectable rotational support for gas in the inner clump.

The inner clump, gap, and outer disc are clearly noticeable in this panel. The coincidence between the gas density just outside $\sim 1$ kpc and the wind recoupling density clearly suggests that it is feedback energy from star formation in the inner clump that carves the gap between the inner clump and outer disc. These winds are responsible for the gap and may also lead to the full disruption of the outer disc in galaxies where the disc has a size comparable to the gap, as is the case for the smallest galaxies in our sample.

We tracked the inner-clump evolution for the whole galaxy population using the simulation's \texttt{SUBLINK} merger trees. We followed the main progenitors of each galaxy through cosmic time in the $1.5 \lesssim t/\mathrm{Gyr} \lesssim 13.8$ range (corresponding to the redshift interval $4 \gtrsim z \gtrsim 0$) and measured the gas and stellar mass of their inner clump, defined at all times as simply the region enclosed within $r = 1$ (physical) kpc.

We then grouped galaxies by their $z=0$ stellar mass, stacking them in 0.5 dex bins. The curves in Figure \ref{FigEvolInnerClump} show the evolution of the median enclosed stellar (dashed) and baryonic (solid) mass of the inner clumps. Lower mass bins are shown by redder colours. Higher mass bins are shown in bluer coloured circles at $t \approx 13.5$ Gyr indicating the median stellar mass $\mathit{M_*}$ of each bin. 

In massive galaxies, the inner clump grows quickly and stagnates relatively early; so, by $z=0$, the clump has transformed essentially all of its baryons into stars. On the other hand, in dwarfs the clump seems to accrete baryonic mass more continuously. These baryons are transformed less efficiently into stars so that, at $z=0$, nearly half of the inner baryonic mass is in the form of gas.

The relative importance of the inner clump also evolves differently depending on galaxy mass, as shown in the left hand panel of Figure \ref{FigSizeEvol}. The clump's relative importance decreases with time in massive galaxies and increases in dwarfs. This has a knock-on effect on the evolution of galaxy sizes (middle panel of the same figure). Massive galaxies grow with time as young stars form progressively at larger radii, whereas star formation in dwarfs moves inward with time as the inner clump becomes more dominant. 

The effect is even more pronounced when tracking only 'young stars' (i.e. those younger than $1$ Gyr at any given time), as shown in the right hand panel of Figure \ref{FigSizeEvol}. Interestingly, in dwarfs, the radius that contains half of young stars moves inwards; as more of the inner clump's gas is transformed into stars, the star forming region inside the clump shrinks, leading to the formation of compact stellar cores, which, in some cases, can dominate the stellar population of a galaxy.

Since star formation occurs only in the central clump or in an outer gaseous disc, the size of the young stellar component is determined by the relative prevalence of these two regions. For example, the seemingly flat evolution shown by the $8.5 < \log(\mathit{M_{*}/M_{\odot}}) < 9.0$ (second reddest line) galaxy bin results because the central clump dominates in half of the systems, whereas the outer disc dominates in the other half, roughly cancelling out the radius evolution. 

The origin of the central clumps warrants further consideration. Their existence implies mechanisms capable of efficiently channelling significant amounts of gas towards the galactic centre, which may arise from a complex interplay of physical factors such as the infall of low-angular-momentum gas through accretion, outward-angular-momentum transport via non-axisymmetric structures such as bars and spiral arms, feedback-driven turbulence, and numerical effects such as artificial viscosity within the gas or  energy deposition of stellar feedback in the interstellar medium. Disentangling the precise contribution of each of these  effects is a challenge that requires a dedicated study, which we defer to a future contribution.

\subsection{Galaxy stellar mass-size relation}

\begin{figure}
    \centering
    \includegraphics[width=0.95\columnwidth]{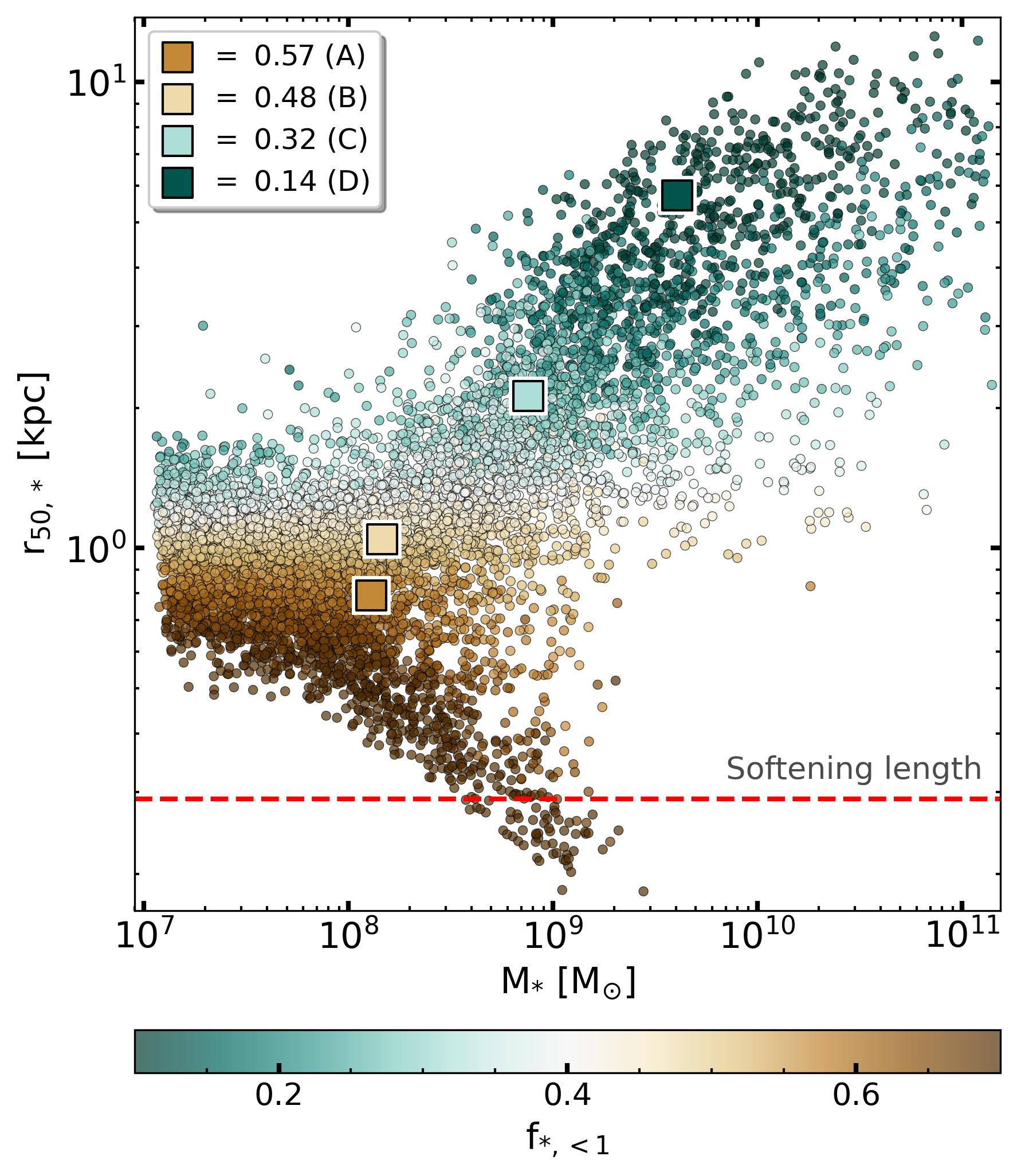}
    
    \caption{Stellar half-mass radius ($r_{50,*}$) as function of stellar mass ($M_{*}$) for our sample of central TNG50 galaxies, extended to $\log(M_{*}/M_{\odot}) = 7$ (virial masses down to $\log(M_{200}/M_{\odot}) \sim 10$). Each galaxy is represented with a circle coloured according to the fraction of stellar mass enclosed within 1 kpc of its centre $f_{*,\mathrm{<1}} = M_{*,\mathrm{<1}}/M_{*}$. Coloured squares highlight galaxies A, B, C, and D from Figure \ref{FigGxExamples}. We note that the stellar-mass-size relation seems to 'converge' to $\sim 1$ kpc in the dwarf regime, with a secondary population of unusually compact galaxies that reach a very small $r_{50,*}$ at $\log(\mathit{M_{*}/M_{\odot}}) \sim 9$. These galaxies reach sizes comparable to the gravitational-force softening length for stars and dark-matter particles ($0.29$ kpc at redshift $z=0$) and correspond to systems where the central clump dominates not only the galaxy stellar mass budget, but also the total mass budget within $1$ kpc. See text for further discussion.}
    \label{FigMstarRh}
\end{figure}

The relative importance of the inner clump discussed above also explains why TNG50 dwarfs have roughly the same stellar half-mass radius, $r_{50,*}$. As Figure \ref{FigMstarRh} shows, the stellar half-mass radius seems to 'converge' to $\sim 1$ kpc at low masses; intriguingly, there also appears to be a 'tongue' of dwarfs with unusually low values of $r_{50,*}$, reaching values as small as $0.2$ kpc (smaller than the gravitational softening length) at $\log(\mathit{M_{*}/M_{\odot}}) \sim 9$. 

This tongue is made out of systems where the central clump not only dominates the  stellar budget, but it is also   massive enough to become self-gravitating. Such systems slip down the tongue as the gas component shrinks and is gradually transformed into stars. The colour-coding in Figure \ref{FigMstarRh} supports this interpretation and shows that the tongue contains almost exclusively clump-dominated systems with large values of $f_{*,<1}$. In addition, we checked that the baryonic mass of central clumps in tongue systems is large enough to be comparable to (or exceed) the dark mass within $1$ kpc, a feature that distinguishes them from other galaxies of comparable $M_*$ (and larger $r_{50,*}$). 

Both the flat dependence of $r_{50,*}$ at low $M_*$ and the presence of the tongue are highly suggestive of an artefact caused by a combination of limited resolution as well as the particular sub-grid implementation of interstellar-medium physics, star formation and feedback in TNG50. A simple confirmation of this is provided by results from TNG100, which is a larger volume, lower resolution simulation that uses the same sub-grid physics implementation as TNG50. The analogous figure for TNG100 is shown in Figure \ref{App_FigMstarRh_TNG50_100}. As expected from a numerically driven feature, TNG100 galaxies show a similar flattening of the radius-mass relation at low $M_*$ and the presence of a similar tongue, but it is shifted to larger values of galaxy mass and radius.
 
Finally, we comment on the idea that the flattening of the $r_{50,*}$-$M_*$ relation may be driven by collisional effects between dark matter and stars due to the finite number of particles in low-mass systems. This was argued by \citet{Ludlow2023} to explain the resolution dependence of low-mass galaxy sizes in the EAGLE simulation series. Although these authors make a compelling case, their interpretation does not seem to apply to TNG50. In particular, the presence of a tongue feature where size decreases with increasing $M_*$ would not be expected in such an interpretation. In TNG50 at least, the flattening of $r_{50,*}$ at low $M_*$ seems to be due to the formation of the central clump, rather than to collisional effects between dark matter and star particles.

\section{Summary and conclusions} \label{SecConc}

We used the TNG50 hydrodynamic cosmological simulation to investigate the rotational support of the stellar component of simulated galaxies. We find that the degree of rotational support (quantified by the parameter $\kappa_{\rm rot}$) correlates strongly with galaxy stellar mass (in the range of $8 < \log(\mathit{M_{*}/M_{\odot}}) < 11$), mainly from dispersion-dominated dwarfs to massive galaxies with prominent centrifugally supported discs. 

This correlation results from the spatial distribution of gas and, consequently, of star formation in simulated galaxies. Star formation profiles are clearly bimodal, with an unresolved, non-rotating dense clump in the inner regions ($r < 1$ kpc) and an extended outer disc ($r > 2$ kpc). These two regions are separated by a relatively quiescent intermediate zone (a `gap'). The relative importance of one region over the other influences the morphology of a galaxy and its rotational support: in massive galaxies the inner clump is relatively unimportant, and, therefore, such galaxies evolve mainly by forming stars in extended discs. On the other hand, the inner clump dominates star formation in dwarf galaxies, which, as a consequence, end up as dispersion-dominated spheroidal systems.

The gaseous inner clump is formed out of low-angular-momentum and pressure-supported gas that becomes dense enough to begin forming stars. These clumps have a typical size of $\sim 1$ kpc, regardless of time or galaxy mass, hinting that it is an artefact of limited resolution and the sub-grid treatment of dense gas in TNG50. The clumps are resilient to disruption, likely because the decoupled-wind feedback scheme adopted in TNG50 dumps the energetic output of young stars in the clump into its surroundings, creating the gap between the inner regions and the outer disc. In dwarfs, this feedback energy can disrupt the outer disc or prevent its assembly altogether, leaving the clump as the sole -- or main -- dwarf-galaxy star forming region.

The relative importance of the inner clumps also has consequences regarding the galaxy mass versus size relation, as well as the radial build-up of galaxies. Nearly all TNG50 dwarfs with $M_*<10^{8.5}\, M_\odot$ have the same size, with stellar half-mass radii comparable to the size of the inner clump (i.e., about $\sim 1$ kpc).  In addition, these dwarfs appear to form from the outside in, with younger stellar populations forming closer to the centre as gas in the inner clump is gradually transformed into stars.

On the other hand, the size of massive galaxies, where the inner clump prevalence is less significant, or negligible, grows with increasing mass. As a consequence, massive TNG50 galaxies appear to form from the inside out, as star formation moves outward in the centrifugally supported outer discs that dominate their morphology. 

Our results suggest caution should be taken when interpreting the evolution of the mass, size, and morphology of the TNG50 galaxy population, especially when using samples that contain a substantial fraction of dwarfs. The highly concentrated, unresolved baryonic clumps located at the centre of TNG50 galaxies are most likely an artefact of limited resolution and of the particular numerical implementation of  star formation  adopted in these simulations, which future improvements should correct. Conclusive answers to questions regarding what sets the morphology and rotational support of a galaxy, and if dwarf in particular, will probably have to wait until such improvements become available.

\begin{acknowledgements}
    This project has received funding from the European Union’s HORIZON-MSCA-2021-SE-01 Research and Innovation programme under the Marie Sklodowska-Curie grant agreement number 101086388 - Project acronym: LACEGAL. This work was partially supported by the Consejo de Investigaciones Científicas y Técnicas de la República Argentina (CONICET) and the Secretaría de Ciencia y Técnica de la Universidad Nacional de Córdoba (SeCyT). The IllustrisTNG simulations were undertaken with compute time awarded by the Gauss Centre for Supercomputing (GCS) under GCS Large-Scale Projects GCS-ILLU and GCS-DWAR on the GCS share of the supercomputer Hazel Hen at the High Performance Computing Center Stuttgart (HLRS), as well as on the machines of the Max Planck Computing and Data Facility (MPCDF) in Garching, Germany. JFN acknowledges the hospitality of the Max-Planck Institute for Astrophysics and of the Donostia International Physics Center during the completion of this manuscript.      
\end{acknowledgements}

\bibliographystyle{aa} % style aa.bst
\bibliography{example} % your references Yourfile.bib

@ARTICLE{Sales2012,
       author = {{Sales}, Laura V. and {Navarro}, Julio F. and {Theuns}, Tom and {Schaye}, Joop and {White}, Simon D.~M. and {Frenk}, Carlos S. and {Crain}, Robert A. and {Dalla Vecchia}, Claudio},
        title = "{The origin of discs and spheroids in simulated galaxies}",
      journal = {\mnras},
         year = 2012,
        month = jun,
       volume = {423},
       number = {2},
        pages = {1544-1555},
          doi = {10.1111/j.1365-2966.2012.20975.x},
archivePrefix = {arXiv},
       eprint = {1112.2220},
 primaryClass = {astro-ph.CO},
       adsurl = {https://ui.adsabs.harvard.edu/abs/2012MNRAS.423.1544S}
}

@ARTICLE{Abadi2003,
       author = {{Abadi}, Mario G. and {Navarro}, Julio F. and {Steinmetz}, Matthias and {Eke}, Vincent R.},
        title = "{Simulations of Galaxy Formation in a {\ensuremath{\Lambda}} Cold Dark Matter Universe. II. The Fine Structure of Simulated Galactic Disks}",
      journal = {\apj},
         year = 2003,
        month = nov,
       volume = {597},
       number = {1},
        pages = {21-34},
          doi = {10.1086/378316},
archivePrefix = {arXiv},
       eprint = {astro-ph/0212282},
 primaryClass = {astro-ph},
       adsurl = {https://ui.adsabs.harvard.edu/abs/2003ApJ...597...21A}
}

@ARTICLE{R-G2017,
       author = {{Rodriguez-Gomez}, Vicente and {Sales}, Laura V. and {Genel}, Shy and {Pillepich}, Annalisa and {Zjupa}, Jolanta and {Nelson}, Dylan and {Griffen}, Brendan and {Torrey}, Paul and {Snyder}, Gregory F. and {Vogelsberger}, Mark and {Springel}, Volker and {Ma}, Chung-Pei and {Hernquist}, Lars},
        title = "{The role of mergers and halo spin in shaping galaxy morphology}",
      journal = {\mnras},
         year = 2017,
        month = may,
       volume = {467},
       number = {3},
        pages = {3083-3098},
          doi = {10.1093/mnras/stx305},
archivePrefix = {arXiv},
       eprint = {1609.09498},
 primaryClass = {astro-ph.GA},
       adsurl = {https://ui.adsabs.harvard.edu/abs/2017MNRAS.467.3083R}
}

@ARTICLE{Nelson2019,
       author = {{Nelson}, Dylan and {Springel}, Volker and {Pillepich}, Annalisa and {Rodriguez-Gomez}, Vicente and {Torrey}, Paul and {Genel}, Shy and {Vogelsberger}, Mark and {Pakmor}, Ruediger and {Marinacci}, Federico and {Weinberger}, Rainer and {Kelley}, Luke and {Lovell}, Mark and {Diemer}, Benedikt and {Hernquist}, Lars},
        title = "{The IllustrisTNG simulations: public data release}",
      journal = {Computational Astrophysics and Cosmology},
         year = 2019,
        month = may,
       volume = {6},
       number = {1},
          eid = {2},
        pages = {2},
          doi = {10.1186/s40668-019-0028-x},
archivePrefix = {arXiv},
       eprint = {1812.05609},
 primaryClass = {astro-ph.GA},
       adsurl = {https://ui.adsabs.harvard.edu/abs/2019ComAC...6....2N}
}

@ARTICLE{Planck2016,
       author = {{Planck Collaboration} and {Ade}, P.~A.~R. and {Aghanim}, N. and {Arnaud}, M. and {Ashdown}, M. and {Aumont}, J. and {Baccigalupi}, C. and {Banday}, A.~J. and {Barreiro}, R.~B. and {Bartlett}, J.~G. and {Bartolo}, N. and {Battaner}, E. and {Battye}, R. and {Benabed}, K. and {Beno{\^\i}t}, A. and {Benoit-L{\'e}vy}, A. and {Bernard}, J. -P. and {Bersanelli}, M. and {Bielewicz}, P. and {Bock}, J.~J. and {Bonaldi}, A. and {Bonavera}, L. and {Bond}, J.~R. and {Borrill}, J. and {Bouchet}, F.~R. and {Boulanger}, F. and {Bucher}, M. and {Burigana}, C. and {Butler}, R.~C. and {Calabrese}, E. and {Cardoso}, J. -F. and {Catalano}, A. and {Challinor}, A. and {Chamballu}, A. and {Chary}, R. -R. and {Chiang}, H.~C. and {Chluba}, J. and {Christensen}, P.~R. and {Church}, S. and {Clements}, D.~L. and {Colombi}, S. and {Colombo}, L.~P.~L. and {Combet}, C. and {Coulais}, A. and {Crill}, B.~P. and {Curto}, A. and {Cuttaia}, F. and {Danese}, L. and {Davies}, R.~D. and {Davis}, R.~J. and {de Bernardis}, P. and {de Rosa}, A. and {de Zotti}, G. and {Delabrouille}, J. and {D{\'e}sert}, F. -X. and {Di Valentino}, E. and {Dickinson}, C. and {Diego}, J.~M. and {Dolag}, K. and {Dole}, H. and {Donzelli}, S. and {Dor{\'e}}, O. and {Douspis}, M. and {Ducout}, A. and {Dunkley}, J. and {Dupac}, X. and {Efstathiou}, G. and {Elsner}, F. and {En{\ss}lin}, T.~A. and {Eriksen}, H.~K. and {Farhang}, M. and {Fergusson}, J. and {Finelli}, F. and {Forni}, O. and {Frailis}, M. and {Fraisse}, A.~A. and {Franceschi}, E. and {Frejsel}, A. and {Galeotta}, S. and {Galli}, S. and {Ganga}, K. and {Gauthier}, C. and {Gerbino}, M. and {Ghosh}, T. and {Giard}, M. and {Giraud-H{\'e}raud}, Y. and {Giusarma}, E. and {Gjerl{\o}w}, E. and {Gonz{\'a}lez-Nuevo}, J. and {G{\'o}rski}, K.~M. and {Gratton}, S. and {Gregorio}, A. and {Gruppuso}, A. and {Gudmundsson}, J.~E. and {Hamann}, J. and {Hansen}, F.~K. and {Hanson}, D. and {Harrison}, D.~L. and {Helou}, G. and {Henrot-Versill{\'e}}, S. and {Hern{\'a}ndez-Monteagudo}, C. and {Herranz}, D. and {Hildebrandt}, S.~R. and {Hivon}, E. and {Hobson}, M. and {Holmes}, W.~A. and {Hornstrup}, A. and {Hovest}, W. and {Huang}, Z. and {Huffenberger}, K.~M. and {Hurier}, G. and {Jaffe}, A.~H. and {Jaffe}, T.~R. and {Jones}, W.~C. and {Juvela}, M. and {Keih{\"a}nen}, E. and {Keskitalo}, R. and {Kisner}, T.~S. and {Kneissl}, R. and {Knoche}, J. and {Knox}, L. and {Kunz}, M. and {Kurki-Suonio}, H. and {Lagache}, G. and {L{\"a}hteenm{\"a}ki}, A. and {Lamarre}, J. -M. and {Lasenby}, A. and {Lattanzi}, M. and {Lawrence}, C.~R. and {Leahy}, J.~P. and {Leonardi}, R. and {Lesgourgues}, J. and {Levrier}, F. and {Lewis}, A. and {Liguori}, M. and {Lilje}, P.~B. and {Linden-V{\o}rnle}, M. and {L{\'o}pez-Caniego}, M. and {Lubin}, P.~M. and {Mac{\'\i}as-P{\'e}rez}, J.~F. and {Maggio}, G. and {Maino}, D. and {Mandolesi}, N. and {Mangilli}, A. and {Marchini}, A. and {Maris}, M. and {Martin}, P.~G. and {Martinelli}, M. and {Mart{\'\i}nez-Gonz{\'a}lez}, E. and {Masi}, S. and {Matarrese}, S. and {McGehee}, P. and {Meinhold}, P.~R. and {Melchiorri}, A. and {Melin}, J. -B. and {Mendes}, L. and {Mennella}, A. and {Migliaccio}, M. and {Millea}, M. and {Mitra}, S. and {Miville-Desch{\^e}nes}, M. -A. and {Moneti}, A. and {Montier}, L. and {Morgante}, G. and {Mortlock}, D. and {Moss}, A. and {Munshi}, D. and {Murphy}, J.~A. and {Naselsky}, P. and {Nati}, F. and {Natoli}, P. and {Netterfield}, C.~B. and {N{\o}rgaard-Nielsen}, H.~U. and {Noviello}, F. and {Novikov}, D. and {Novikov}, I. and {Oxborrow}, C.~A. and {Paci}, F. and {Pagano}, L. and {Pajot}, F. and {Paladini}, R. and {Paoletti}, D. and {Partridge}, B. and {Pasian}, F. and {Patanchon}, G. and {Pearson}, T.~J. and {Perdereau}, O. and {Perotto}, L. and {Perrotta}, F. and {Pettorino}, V. and {Piacentini}, F. and {Piat}, M. and {Pierpaoli}, E. and {Pietrobon}, D. and {Plaszczynski}, S. and {Pointecouteau}, E. and {Polenta}, G. and {Popa}, L. and {Pratt}, G.~W. and {Pr{\'e}zeau}, G. and {Prunet}, S. and {Puget}, J. -L. and {Rachen}, J.~P. and {Reach}, W.~T. and {Rebolo}, R. and {Reinecke}, M. and {Remazeilles}, M. and {Renault}, C. and {Renzi}, A. and {Ristorcelli}, I. and {Rocha}, G. and {Rosset}, C. and {Rossetti}, M. and {Roudier}, G. and {Rouill{\'e} d'Orfeuil}, B. and {Rowan-Robinson}, M. and {Rubi{\~n}o-Mart{\'\i}n}, J.~A. and {Rusholme}, B. and {Said}, N. and {Salvatelli}, V. and {Salvati}, L. and {Sandri}, M. and {Santos}, D. and {Savelainen}, M. and {Savini}, G. and {Scott}, D. and {Seiffert}, M.~D. and {Serra}, P. and {Shellard}, E.~P.~S. and {Spencer}, L.~D. and {Spinelli}, M. and {Stolyarov}, V. and {Stompor}, R. and {Sudiwala}, R. and {Sunyaev}, R. and {Sutton}, D. and {Suur-Uski}, A. -S. and {Sygnet}, J. -F. and {Tauber}, J.~A. and {Terenzi}, L. and {Toffolatti}, L. and {Tomasi}, M. and {Tristram}, M. and {Trombetti}, T. and {Tucci}, M. and {Tuovinen}, J. and {T{\"u}rler}, M. and {Umana}, G. and {Valenziano}, L. and {Valiviita}, J. and {Van Tent}, F. and {Vielva}, P. and {Villa}, F. and {Wade}, L.~A. and {Wandelt}, B.~D. and {Wehus}, I.~K. and {White}, M. and {White}, S.~D.~M. and {Wilkinson}, A. and {Yvon}, D. and {Zacchei}, A. and {Zonca}, A.},
        title = "{Planck 2015 results. XIII. Cosmological parameters}",
      journal = {\aap},
         year = 2016,
        month = sep,
       volume = {594},
          eid = {A13},
        pages = {A13},
          doi = {10.1051/0004-6361/201525830},
archivePrefix = {arXiv},
       eprint = {1502.01589},
 primaryClass = {astro-ph.CO},
       adsurl = {https://ui.adsabs.harvard.edu/abs/2016A&A...594A..13P}
}

@ARTICLE{Pill2018,
       author = {{Pillepich}, Annalisa and {Nelson}, Dylan and {Hernquist}, Lars and {Springel}, Volker and {Pakmor}, R{\"u}diger and {Torrey}, Paul and {Weinberger}, Rainer and {Genel}, Shy and {Naiman}, Jill P. and {Marinacci}, Federico and {Vogelsberger}, Mark},
        title = "{First results from the IllustrisTNG simulations: the stellar mass content of groups and clusters of galaxies}",
      journal = {\mnras},
         year = 2018,
        month = mar,
       volume = {475},
       number = {1},
        pages = {648-675},
          doi = {10.1093/mnras/stx3112},
archivePrefix = {arXiv},
       eprint = {1707.03406},
 primaryClass = {astro-ph.GA},
       adsurl = {https://ui.adsabs.harvard.edu/abs/2018MNRAS.475..648P}
}

@ARTICLE{Springel2018,
       author = {{Springel}, Volker and {Pakmor}, R{\"u}diger and {Pillepich}, Annalisa and {Weinberger}, Rainer and {Nelson}, Dylan and {Hernquist}, Lars and {Vogelsberger}, Mark and {Genel}, Shy and {Torrey}, Paul and {Marinacci}, Federico and {Naiman}, Jill},
        title = "{First results from the IllustrisTNG simulations: matter and galaxy clustering}",
      journal = {\mnras},
         year = 2018,
        month = mar,
       volume = {475},
       number = {1},
        pages = {676-698},
          doi = {10.1093/mnras/stx3304},
archivePrefix = {arXiv},
       eprint = {1707.03397},
 primaryClass = {astro-ph.GA},
       adsurl = {https://ui.adsabs.harvard.edu/abs/2018MNRAS.475..676S}
}

@ARTICLE{Davies1985,
       author = {{Davis}, M. and {Efstathiou}, G. and {Frenk}, C.~S. and {White}, S.~D.~M.},
        title = "{The evolution of large-scale structure in a universe dominated by cold dark matter}",
      journal = {\apj},
         year = 1985,
        month = may,
       volume = {292},
        pages = {371-394},
          doi = {10.1086/163168},
       adsurl = {https://ui.adsabs.harvard.edu/abs/1985ApJ...292..371D}
}

@ARTICLE{Springel2010,
       author = {{Springel}, Volker},
        title = "{E pur si muove: Galilean-invariant cosmological hydrodynamical simulations on a moving mesh}",
      journal = {\mnras},
         year = 2010,
        month = jan,
       volume = {401},
       number = {2},
        pages = {791-851},
          doi = {10.1111/j.1365-2966.2009.15715.x},
archivePrefix = {arXiv},
       eprint = {0901.4107},
 primaryClass = {astro-ph.CO},
       adsurl = {https://ui.adsabs.harvard.edu/abs/2010MNRAS.401..791S}
}

@ARTICLE{Naiman2018,
       author = {{Naiman}, Jill P. and {Pillepich}, Annalisa and {Springel}, Volker and {Ramirez-Ruiz}, Enrico and {Torrey}, Paul and {Vogelsberger}, Mark and {Pakmor}, R{\"u}diger and {Nelson}, Dylan and {Marinacci}, Federico and {Hernquist}, Lars and {Weinberger}, Rainer and {Genel}, Shy},
        title = "{First results from the IllustrisTNG simulations: a tale of two elements - chemical evolution of magnesium and europium}",
      journal = {\mnras},
         year = 2018,
        month = jun,
       volume = {477},
       number = {1},
        pages = {1206-1224},
          doi = {10.1093/mnras/sty618},
archivePrefix = {arXiv},
       eprint = {1707.03401},
 primaryClass = {astro-ph.GA},
       adsurl = {https://ui.adsabs.harvard.edu/abs/2018MNRAS.477.1206N}
}

@ARTICLE{Miranacci2018,
       author = {{Marinacci}, Federico and {Vogelsberger}, Mark and {Pakmor}, R{\"u}diger and {Torrey}, Paul and {Springel}, Volker and {Hernquist}, Lars and {Nelson}, Dylan and {Weinberger}, Rainer and {Pillepich}, Annalisa and {Naiman}, Jill and {Genel}, Shy},
        title = "{First results from the IllustrisTNG simulations: radio haloes and magnetic fields}",
      journal = {\mnras},
         year = 2018,
        month = nov,
       volume = {480},
       number = {4},
        pages = {5113-5139},
          doi = {10.1093/mnras/sty2206},
archivePrefix = {arXiv},
       eprint = {1707.03396},
 primaryClass = {astro-ph.CO},
       adsurl = {https://ui.adsabs.harvard.edu/abs/2018MNRAS.480.5113M}
}

@ARTICLE{York2000,
       author = {{York}, Donald G. and {Adelman}, J. and {Anderson}, John E., Jr. and {Anderson}, Scott F. and {Annis}, James and {Bahcall}, Neta A. and {Bakken}, J.~A. and {Barkhouser}, Robert and {Bastian}, Steven and {Berman}, Eileen and {Boroski}, William N. and {Bracker}, Steve and {Briegel}, Charlie and {Briggs}, John W. and {Brinkmann}, J. and {Brunner}, Robert and {Burles}, Scott and {Carey}, Larry and {Carr}, Michael A. and {Castander}, Francisco J. and {Chen}, Bing and {Colestock}, Patrick L. and {Connolly}, A.~J. and {Crocker}, J.~H. and {Csabai}, Istv{\'a}n and {Czarapata}, Paul C. and {Davis}, John Eric and {Doi}, Mamoru and {Dombeck}, Tom and {Eisenstein}, Daniel and {Ellman}, Nancy and {Elms}, Brian R. and {Evans}, Michael L. and {Fan}, Xiaohui and {Federwitz}, Glenn R. and {Fiscelli}, Larry and {Friedman}, Scott and {Frieman}, Joshua A. and {Fukugita}, Masataka and {Gillespie}, Bruce and {Gunn}, James E. and {Gurbani}, Vijay K. and {de Haas}, Ernst and {Haldeman}, Merle and {Harris}, Frederick H. and {Hayes}, J. and {Heckman}, Timothy M. and {Hennessy}, G.~S. and {Hindsley}, Robert B. and {Holm}, Scott and {Holmgren}, Donald J. and {Huang}, Chi-hao and {Hull}, Charles and {Husby}, Don and {Ichikawa}, Shin-Ichi and {Ichikawa}, Takashi and {Ivezi{\'c}}, {\v{Z}}eljko and {Kent}, Stephen and {Kim}, Rita S.~J. and {Kinney}, E. and {Klaene}, Mark and {Kleinman}, A.~N. and {Kleinman}, S. and {Knapp}, G.~R. and {Korienek}, John and {Kron}, Richard G. and {Kunszt}, Peter Z. and {Lamb}, D.~Q. and {Lee}, B. and {Leger}, R. French and {Limmongkol}, Siriluk and {Lindenmeyer}, Carl and {Long}, Daniel C. and {Loomis}, Craig and {Loveday}, Jon and {Lucinio}, Rich and {Lupton}, Robert H. and {MacKinnon}, Bryan and {Mannery}, Edward J. and {Mantsch}, P.~M. and {Margon}, Bruce and {McGehee}, Peregrine and {McKay}, Timothy A. and {Meiksin}, Avery and {Merelli}, Aronne and {Monet}, David G. and {Munn}, Jeffrey A. and {Narayanan}, Vijay K. and {Nash}, Thomas and {Neilsen}, Eric and {Neswold}, Rich and {Newberg}, Heidi Jo and {Nichol}, R.~C. and {Nicinski}, Tom and {Nonino}, Mario and {Okada}, Norio and {Okamura}, Sadanori and {Ostriker}, Jeremiah P. and {Owen}, Russell and {Pauls}, A. George and {Peoples}, John and {Peterson}, R.~L. and {Petravick}, Donald and {Pier}, Jeffrey R. and {Pope}, Adrian and {Pordes}, Ruth and {Prosapio}, Angela and {Rechenmacher}, Ron and {Quinn}, Thomas R. and {Richards}, Gordon T. and {Richmond}, Michael W. and {Rivetta}, Claudio H. and {Rockosi}, Constance M. and {Ruthmansdorfer}, Kurt and {Sandford}, Dale and {Schlegel}, David J. and {Schneider}, Donald P. and {Sekiguchi}, Maki and {Sergey}, Gary and {Shimasaku}, Kazuhiro and {Siegmund}, Walter A. and {Smee}, Stephen and {Smith}, J. Allyn and {Snedden}, S. and {Stone}, R. and {Stoughton}, Chris and {Strauss}, Michael A. and {Stubbs}, Christopher and {SubbaRao}, Mark and {Szalay}, Alexander S. and {Szapudi}, Istvan and {Szokoly}, Gyula P. and {Thakar}, Anirudda R. and {Tremonti}, Christy and {Tucker}, Douglas L. and {Uomoto}, Alan and {Vanden Berk}, Dan and {Vogeley}, Michael S. and {Waddell}, Patrick and {Wang}, Shu-i. and {Watanabe}, Masaru and {Weinberg}, David H. and {Yanny}, Brian and {Yasuda}, Naoki and {SDSS Collaboration}},
        title = "{The Sloan Digital Sky Survey: Technical Summary}",
      journal = {\aj},
         year = 2000,
        month = sep,
       volume = {120},
       number = {3},
        pages = {1579-1587},
          doi = {10.1086/301513},
archivePrefix = {arXiv},
       eprint = {astro-ph/0006396},
 primaryClass = {astro-ph},
       adsurl = {https://ui.adsabs.harvard.edu/abs/2000AJ....120.1579Y}
}

@ARTICLE{Springel2001,
       author = {{Springel}, Volker and {White}, Simon D.~M. and {Tormen}, Giuseppe and {Kauffmann}, Guinevere},
        title = "{Populating a cluster of galaxies - I. Results at [formmu2]z=0}",
      journal = {\mnras},
         year = 2001,
        month = dec,
       volume = {328},
       number = {3},
        pages = {726-750},
          doi = {10.1046/j.1365-8711.2001.04912.x},
archivePrefix = {arXiv},
       eprint = {astro-ph/0012055},
 primaryClass = {astro-ph},
       adsurl = {https://ui.adsabs.harvard.edu/abs/2001MNRAS.328..726S}
}

@ARTICLE{R-G2015,
       author = {{Rodriguez-Gomez}, Vicente and {Genel}, Shy and {Vogelsberger}, Mark and {Sijacki}, Debora and {Pillepich}, Annalisa and {Sales}, Laura V. and {Torrey}, Paul and {Snyder}, Greg and {Nelson}, Dylan and {Springel}, Volker and {Ma}, Chung-Pei and {Hernquist}, Lars},
        title = "{The merger rate of galaxies in the Illustris simulation: a comparison with observations and semi-empirical models}",
      journal = {\mnras},
         year = 2015,
        month = may,
       volume = {449},
       number = {1},
        pages = {49-64},
          doi = {10.1093/mnras/stv264},
archivePrefix = {arXiv},
       eprint = {1502.01339},
 primaryClass = {astro-ph.GA},
       adsurl = {https://ui.adsabs.harvard.edu/abs/2015MNRAS.449...49R}
}

@MISC{BenitezLlambay2017,
       author = {{Ben{\'\i}tez-Llambay}, Alejandro},
        title = "{Py-SPHViewer: Cosmological simulations using Smoothed Particle Hydrodynamics}",
 howpublished = {Astrophysics Source Code Library, record ascl:1712.003},
         year = 2017,
        month = dec,
          eid = {ascl:1712.003},
        pages = {ascl:1712.003},
archivePrefix = {ascl},
       eprint = {1712.003},
       adsurl = {https://ui.adsabs.harvard.edu/abs/2017ascl.soft12003B}
}

@ARTICLE{SotilloRamos2022,
       author = {{Sotillo-Ramos}, Diego and {Pillepich}, Annalisa and {Donnari}, Martina and {Nelson}, Dylan and {Eisert}, Lukas and {Rodriguez-Gomez}, Vicente and {Joshi}, Gandhali and {Vogelsberger}, Mark and {Hernquist}, Lars},
        title = "{The merger and assembly histories of Milky Way- and M31-like galaxies with TNG50: disc survival through mergers}",
      journal = {\mnras},
         year = 2022,
        month = nov,
       volume = {516},
       number = {4},
        pages = {5404-5427},
          doi = {10.1093/mnras/stac2586},
archivePrefix = {arXiv},
       eprint = {2211.00036},
 primaryClass = {astro-ph.GA},
       adsurl = {https://ui.adsabs.harvard.edu/abs/2022MNRAS.516.5404S}
}

@ARTICLE{McC2012,
       author = {{McConnachie}, Alan W.},
        title = "{The Observed Properties of Dwarf Galaxies in and around the Local Group}",
      journal = {\aj},
         year = 2012,
        month = jul,
       volume = {144},
       number = {1},
          eid = {4},
        pages = {4},
          doi = {10.1088/0004-6256/144/1/4},
archivePrefix = {arXiv},
       eprint = {1204.1562},
 primaryClass = {astro-ph.CO},
       adsurl = {https://ui.adsabs.harvard.edu/abs/2012AJ....144....4M}
}

@ARTICLE{Klypin2015,
       author = {{Klypin}, Anatoly and {Karachentsev}, Igor and {Makarov}, Dmitry and {Nasonova}, Olga},
        title = "{Abundance of field galaxies}",
      journal = {\mnras},
         year = 2015,
        month = dec,
       volume = {454},
       number = {2},
        pages = {1798-1810},
          doi = {10.1093/mnras/stv2040},
archivePrefix = {arXiv},
       eprint = {1405.4523},
 primaryClass = {astro-ph.CO},
       adsurl = {https://ui.adsabs.harvard.edu/abs/2015MNRAS.454.1798K}
}

@ARTICLE{Karachentsev2013,
       author = {{Karachentsev}, Igor D. and {Makarov}, Dmitry I. and {Kaisina}, Elena I.},
        title = "{Updated Nearby Galaxy Catalog}",
      journal = {\aj},
         year = 2013,
        month = apr,
       volume = {145},
       number = {4},
          eid = {101},
        pages = {101},
          doi = {10.1088/0004-6256/145/4/101},
archivePrefix = {arXiv},
       eprint = {1303.5328},
 primaryClass = {astro-ph.CO},
       adsurl = {https://ui.adsabs.harvard.edu/abs/2013AJ....145..101K}
}

@ARTICLE{GAMA2016,
       author = {{Moffett}, Amanda J. and {Lange}, Rebecca and {Driver}, Simon P. and {Robotham}, Aaron S.~G. and {Kelvin}, Lee S. and {Alpaslan}, Mehmet and {Andrews}, Stephen K. and {Bland-Hawthorn}, Joss and {Brough}, Sarah and {Cluver}, Michelle E. and {Colless}, Matthew and {Davies}, Luke J.~M. and {Holwerda}, Benne W. and {Hopkins}, Andrew M. and {Kafle}, Prajwal R. and {Liske}, Jochen and {Meyer}, Martin},
        title = "{Galaxy and Mass Assembly (GAMA): the stellar mass budget of galaxy spheroids and discs}",
      journal = {\mnras},
         year = 2016,
        month = nov,
       volume = {462},
       number = {4},
        pages = {4336-4348},
          doi = {10.1093/mnras/stw1861},
archivePrefix = {arXiv},
       eprint = {1608.05526},
 primaryClass = {astro-ph.GA},
       adsurl = {https://ui.adsabs.harvard.edu/abs/2016MNRAS.462.4336M}
}

@ARTICLE{Naab2014,
       author = {{Naab}, Thorsten and {Oser}, L. and {Emsellem}, E. and {Cappellari}, Michele and {Krajnovi{\'c}}, D. and {McDermid}, R.~M. and {Alatalo}, K. and {Bayet}, E. and {Blitz}, L. and {Bois}, M. and {Bournaud}, F. and {Bureau}, M. and {Crocker}, A. and {Davies}, R.~L. and {Davis}, T.~A. and {de Zeeuw}, P.~T. and {Duc}, P. -A. and {Hirschmann}, M. and {Johansson}, P.~H. and {Khochfar}, S. and {Kuntschner}, H. and {Morganti}, R. and {Oosterloo}, T. and {Sarzi}, M. and {Scott}, N. and {Serra}, P. and {van de Ven}, G. and {Weijmans}, A. and {Young}, L.~M.},
        title = "{The ATLAS$^{3D}$ project - XXV. Two-dimensional kinematic analysis of simulated galaxies and the cosmological origin of fast and slow rotators}",
      journal = {\mnras},
         year = 2014,
        month = nov,
       volume = {444},
       number = {4},
        pages = {3357-3387},
          doi = {10.1093/mnras/stt1919},
archivePrefix = {arXiv},
       eprint = {1311.0284},
 primaryClass = {astro-ph.CO},
       adsurl = {https://ui.adsabs.harvard.edu/abs/2014MNRAS.444.3357N}
}

@ARTICLE{Genel2015,
       author = {{Genel}, Shy and {Fall}, S. Michael and {Hernquist}, Lars and {Vogelsberger}, Mark and {Snyder}, Gregory F. and {Rodriguez-Gomez}, Vicente and {Sijacki}, Debora and {Springel}, Volker},
        title = "{Galactic Angular Momentum in the Illustris Simulation: Feedback and the Hubble Sequence}",
      journal = {\apjl},
         year = 2015,
        month = may,
       volume = {804},
       number = {2},
          eid = {L40},
        pages = {L40},
          doi = {10.1088/2041-8205/804/2/L40},
archivePrefix = {arXiv},
       eprint = {1503.01117},
 primaryClass = {astro-ph.GA},
       adsurl = {https://ui.adsabs.harvard.edu/abs/2015ApJ...804L..40G}
}

@ARTICLE{Blanton2005,
       author = {{Blanton}, Michael R. and {Lupton}, Robert H. and {Schlegel}, David J. and {Strauss}, Michael A. and {Brinkmann}, J. and {Fukugita}, Masataka and {Loveday}, Jon},
        title = "{The Properties and Luminosity Function of Extremely Low Luminosity Galaxies}",
      journal = {\apj},
         year = 2005,
        month = sep,
       volume = {631},
       number = {1},
        pages = {208-230},
          doi = {10.1086/431416},
archivePrefix = {arXiv},
       eprint = {astro-ph/0410164},
 primaryClass = {astro-ph},
       adsurl = {https://ui.adsabs.harvard.edu/abs/2005ApJ...631..208B}
}

@ARTICLE{Swaters2009,
       author = {{Swaters}, R.~A. and {Sancisi}, R. and {van Albada}, T.~S. and {van der Hulst}, J.~M.},
        title = "{The rotation curves shapes of late-type dwarf galaxies}",
      journal = {\aap},
         year = 2009,
        month = jan,
       volume = {493},
       number = {3},
        pages = {871-892},
          doi = {10.1051/0004-6361:200810516},
archivePrefix = {arXiv},
       eprint = {0901.4222},
 primaryClass = {astro-ph.CO},
       adsurl = {https://ui.adsabs.harvard.edu/abs/2009A&A...493..871S}
}

@ARTICLE{Oh2015,
       author = {{Oh}, Se-Heon and {Hunter}, Deidre A. and {Brinks}, Elias and {Elmegreen}, Bruce G. and {Schruba}, Andreas and {Walter}, Fabian and {Rupen}, Michael P. and {Young}, Lisa M. and {Simpson}, Caroline E. and {Johnson}, Megan C. and {Herrmann}, Kimberly A. and {Ficut-Vicas}, Dana and {Cigan}, Phil and {Heesen}, Volker and {Ashley}, Trisha and {Zhang}, Hong-Xin},
        title = "{High-resolution Mass Models of Dwarf Galaxies from LITTLE THINGS}",
      journal = {\aj},
         year = 2015,
        month = jun,
       volume = {149},
       number = {6},
          eid = {180},
        pages = {180},
          doi = {10.1088/0004-6256/149/6/180},
archivePrefix = {arXiv},
       eprint = {1502.01281},
 primaryClass = {astro-ph.GA},
       adsurl = {https://ui.adsabs.harvard.edu/abs/2015AJ....149..180O}
}

@ARTICLE{Mateo1998,
       author = {{Mateo}, Mario L.},
        title = "{Dwarf Galaxies of the Local Group}",
      journal = {\araa},
         year = 1998,
        month = jan,
       volume = {36},
        pages = {435-506},
          doi = {10.1146/annurev.astro.36.1.435},
archivePrefix = {arXiv},
       eprint = {astro-ph/9810070},
 primaryClass = {astro-ph},
       adsurl = {https://ui.adsabs.harvard.edu/abs/1998ARA&A..36..435M}
}

@ARTICLE{Kauffmann2003Mstar,
       author = {{Kauffmann}, Guinevere and {Heckman}, Timothy M. and {White}, Simon D.~M. and {Charlot}, St{\'e}phane and {Tremonti}, Christy and {Brinchmann}, Jarle and {Bruzual}, Gustavo and {Peng}, Eric W. and {Seibert}, Mark and {Bernardi}, Mariangela and {Blanton}, Michael and {Brinkmann}, Jon and {Castander}, Francisco and {Cs{\'a}bai}, Istvan and {Fukugita}, Masataka and {Ivezic}, Zeljko and {Munn}, Jeffrey A. and {Nichol}, Robert C. and {Padmanabhan}, Nikhil and {Thakar}, Aniruddha R. and {Weinberg}, David H. and {York}, Donald},
        title = "{Stellar masses and star formation histories for {}10$^{5}$ galaxies from the Sloan Digital Sky Survey}",
      journal = {\mnras},
         year = 2003,
        month = may,
       volume = {341},
       number = {1},
        pages = {33-53},
          doi = {10.1046/j.1365-8711.2003.06291.x},
archivePrefix = {arXiv},
       eprint = {astro-ph/0204055},
 primaryClass = {astro-ph},
       adsurl = {https://ui.adsabs.harvard.edu/abs/2003MNRAS.341...33K}
}

@ARTICLE{SHIELD2016,
       author = {{McNichols}, Andrew T. and {Teich}, Yaron G. and {Nims}, Elise and {Cannon}, John M. and {Adams}, Elizabeth A.~K. and {Bernstein-Cooper}, Elijah Z. and {Giovanelli}, Riccardo and {Haynes}, Martha P. and {J{\'o}zsa}, Gyula I.~G. and {McQuinn}, Kristen B.~W. and {Salzer}, John J. and {Skillman}, Evan D. and {Warren}, Steven R. and {Dolphin}, Andrew and {Elson}, E.~C. and {Haurberg}, Nathalie and {Ott}, J{\"u}rgen and {Saintonge}, Amelie and {Cave}, Ian and {Hagen}, Cedric and {Huang}, Shan and {Janowiecki}, Steven and {Marshall}, Melissa V. and {Thomann}, Clara M. and {Van Sistine}, Angela},
        title = "{SHIELD: Neutral Gas Kinematics and Dynamics}",
      journal = {\apj},
         year = 2016,
        month = nov,
       volume = {832},
       number = {1},
          eid = {89},
        pages = {89},
          doi = {10.3847/0004-637X/832/1/89},
archivePrefix = {arXiv},
       eprint = {1609.05376},
 primaryClass = {astro-ph.GA},
       adsurl = {https://ui.adsabs.harvard.edu/abs/2016ApJ...832...89M}
}

@ARTICLE{Sales2022,
       author = {{Sales}, Laura V. and {Wetzel}, Andrew and {Fattahi}, Azadeh},
        title = "{Baryonic solutions and challenges for cosmological models of dwarf galaxies}",
      journal = {Nature Astronomy},
         year = 2022,
        month = jun,
       volume = {6},
        pages = {897-910},
          doi = {10.1038/s41550-022-01689-w},
archivePrefix = {arXiv},
       eprint = {2206.05295},
 primaryClass = {astro-ph.GA},
       adsurl = {https://ui.adsabs.harvard.edu/abs/2022NatAs...6..897S}
}

@ARTICLE{LiWhite2009,
       author = {{Li}, Cheng and {White}, Simon D.~M.},
        title = "{The distribution of stellar mass in the low-redshift Universe}",
      journal = {\mnras},
         year = 2009,
        month = oct,
       volume = {398},
       number = {4},
        pages = {2177-2187},
          doi = {10.1111/j.1365-2966.2009.15268.x},
archivePrefix = {arXiv},
       eprint = {0901.0706},
 primaryClass = {astro-ph.CO},
       adsurl = {https://ui.adsabs.harvard.edu/abs/2009MNRAS.398.2177L}
}

@ARTICLE{McQuinn2019,
       author = {{McQuinn}, Kristen. B.~W. and {van Zee}, Liese and {Skillman}, Evan D.},
        title = "{Galactic Winds in Low-mass Galaxies}",
      journal = {\apj},
         year = 2019,
        month = nov,
       volume = {886},
       number = {1},
          eid = {74},
        pages = {74},
          doi = {10.3847/1538-4357/ab4c37},
archivePrefix = {arXiv},
       eprint = {1910.04167},
 primaryClass = {astro-ph.GA},
       adsurl = {https://ui.adsabs.harvard.edu/abs/2019ApJ...886...74M}
}

@ARTICLE{Gutcke2021LYRA,
       author = {{Gutcke}, Thales A. and {Pakmor}, R{\"u}diger and {Naab}, Thorsten and {Springel}, Volker},
        title = "{LYRA - I. Simulating the multiphase ISM of a dwarf galaxy with variable energy supernovae from individual stars}",
      journal = {\mnras},
         year = 2021,
        month = mar,
       volume = {501},
       number = {4},
        pages = {5597-5615},
          doi = {10.1093/mnras/staa3875},
archivePrefix = {arXiv},
       eprint = {2010.07311},
 primaryClass = {astro-ph.GA},
       adsurl = {https://ui.adsabs.harvard.edu/abs/2021MNRAS.501.5597G}
}

@ARTICLE{El-Badry2017,
       author = {{El-Badry}, Kareem and {Wetzel}, Andrew R. and {Geha}, Marla and {Quataert}, Eliot and {Hopkins}, Philip F. and {Kere{\v{s}}}, Dusan and {Chan}, T.~K. and {Faucher-Gigu{\`e}re}, Claude-Andr{\'e}},
        title = "{When the Jeans Do Not Fit: How Stellar Feedback Drives Stellar Kinematics and Complicates Dynamical Modeling in Low-mass Galaxies}",
      journal = {\apj},
         year = 2017,
        month = feb,
       volume = {835},
       number = {2},
          eid = {193},
        pages = {193},
          doi = {10.3847/1538-4357/835/2/193},
archivePrefix = {arXiv},
       eprint = {1610.04232},
 primaryClass = {astro-ph.GA},
       adsurl = {https://ui.adsabs.harvard.edu/abs/2017ApJ...835..193E}
}

@ARTICLE{Ferrarotti2006,
       author = {{Ferrarotti}, A.~S. and {Gail}, H. -P.},
        title = "{Composition and quantities of dust produced by AGB-stars and returned to the interstellar medium}",
      journal = {\aap},
         year = 2006,
        month = feb,
       volume = {447},
       number = {2},
        pages = {553-576},
          doi = {10.1051/0004-6361:20041198},
       adsurl = {https://ui.adsabs.harvard.edu/abs/2006A&A...447..553F}
}

@ARTICLE{Christensen2016,
       author = {{Christensen}, Charlotte R. and {Dav{\'e}}, Romeel and {Governato}, Fabio and {Pontzen}, Andrew and {Brooks}, Alyson and {Munshi}, Ferah and {Quinn}, Thomas and {Wadsley}, James},
        title = "{In-N-Out: The Gas Cycle from Dwarfs to Spiral Galaxies}",
      journal = {\apj},
         year = 2016,
        month = jun,
       volume = {824},
       number = {1},
          eid = {57},
        pages = {57},
          doi = {10.3847/0004-637X/824/1/57},
archivePrefix = {arXiv},
       eprint = {1508.00007},
 primaryClass = {astro-ph.GA},
       adsurl = {https://ui.adsabs.harvard.edu/abs/2016ApJ...824...57C}
}

@ARTICLE{DekelWoo2003,
       author = {{Dekel}, Avishai and {Woo}, Joanna},
        title = "{Feedback and the fundamental line of low-luminosity low-surface-brightness/dwarf galaxies}",
      journal = {\mnras},
         year = 2003,
        month = oct,
       volume = {344},
       number = {4},
        pages = {1131-1144},
          doi = {10.1046/j.1365-8711.2003.06923.x},
archivePrefix = {arXiv},
       eprint = {astro-ph/0210454},
 primaryClass = {astro-ph},
       adsurl = {https://ui.adsabs.harvard.edu/abs/2003MNRAS.344.1131D}
}

@ARTICLE{Cheng2024,
       author = {{Cheng}, Zhuo and {Li}, Cheng and {Li}, Niu and {Yan}, Renbin and {Mo}, Houjun},
        title = "{Post-starburst Galaxies in SDSS-IV MaNGA: Two Broad Categories of Evolutionary Pathways}",
      journal = {\apj},
         year = 2024,
        month = feb,
       volume = {961},
       number = {2},
          eid = {216},
        pages = {216},
          doi = {10.3847/1538-4357/ad1510},
archivePrefix = {arXiv},
       eprint = {2312.03616},
 primaryClass = {astro-ph.GA},
       adsurl = {https://ui.adsabs.harvard.edu/abs/2024ApJ...961..216C}
}

@ARTICLE{B-L2016,
       author = {{Ben{\'\i}tez-Llambay}, A. and {Navarro}, J.~F. and {Abadi}, M.~G. and {Gottl{\"o}ber}, S. and {Yepes}, G. and {Hoffman}, Y. and {Steinmetz}, M.},
        title = "{Mergers and the outside-in formation of dwarf spheroidals}",
      journal = {\mnras},
         year = 2016,
        month = feb,
       volume = {456},
       number = {2},
        pages = {1185-1194},
          doi = {10.1093/mnras/stv2722},
archivePrefix = {arXiv},
       eprint = {1511.06188},
 primaryClass = {astro-ph.GA},
       adsurl = {https://ui.adsabs.harvard.edu/abs/2016MNRAS.456.1185B}
}

@ARTICLE{Riggs2024,
       author = {{Riggs}, Claire L. and {Brooks}, Alyson M. and {Munshi}, Ferah and {Christensen}, Charlotte R. and {Cohen}, Roger E. and {Quinn}, Thomas R. and {Wadsley}, James},
        title = "{Testable Predictions of Outside-in Age Gradients in Dwarf Galaxies of All Types}",
      journal = {\apj},
         year = 2024,
        month = dec,
       volume = {977},
       number = {1},
          eid = {20},
        pages = {20},
          doi = {10.3847/1538-4357/ad8b1e},
archivePrefix = {arXiv},
       eprint = {2408.10379},
 primaryClass = {astro-ph.GA},
       adsurl = {https://ui.adsabs.harvard.edu/abs/2024ApJ...977...20R}
}

@ARTICLE{vdBSwater2001,
       author = {{van den Bosch}, Frank C. and {Swaters}, Rob A.},
        title = "{Dwarf galaxy rotation curves and the core problem of dark matter haloes}",
      journal = {\mnras},
         year = 2001,
        month = aug,
       volume = {325},
       number = {3},
        pages = {1017-1038},
          doi = {10.1046/j.1365-8711.2001.04456.x},
archivePrefix = {arXiv},
       eprint = {astro-ph/0006048},
 primaryClass = {astro-ph},
       adsurl = {https://ui.adsabs.harvard.edu/abs/2001MNRAS.325.1017V}
}

@ARTICLE{Moster2013,
       author = {{Moster}, Benjamin P. and {Naab}, Thorsten and {White}, Simon D.~M.},
        title = "{Galactic star formation and accretion histories from matching galaxies to dark matter haloes}",
      journal = {\mnras},
         year = 2013,
        month = feb,
       volume = {428},
       number = {4},
        pages = {3121-3138},
          doi = {10.1093/mnras/sts261},
archivePrefix = {arXiv},
       eprint = {1205.5807},
 primaryClass = {astro-ph.CO},
       adsurl = {https://ui.adsabs.harvard.edu/abs/2013MNRAS.428.3121M}
}

@ARTICLE{Behroozi2019,
       author = {{Behroozi}, Peter and {Wechsler}, Risa H. and {Hearin}, Andrew P. and {Conroy}, Charlie},
        title = "{UNIVERSEMACHINE: The correlation between galaxy growth and dark matter halo assembly from z = 0-10}",
      journal = {\mnras},
         year = 2019,
        month = sep,
       volume = {488},
       number = {3},
        pages = {3143-3194},
          doi = {10.1093/mnras/stz1182},
archivePrefix = {arXiv},
       eprint = {1806.07893},
 primaryClass = {astro-ph.GA},
       adsurl = {https://ui.adsabs.harvard.edu/abs/2019MNRAS.488.3143B}
}

@ARTICLE{Albers2019,
       author = {{Albers}, Saundra M. and {Weisz}, Daniel R. and {Cole}, Andrew A. and {Dolphin}, Andrew E. and {Skillman}, Evan D. and {Williams}, Benjamin F. and {Boylan-Kolchin}, Michael and {Bullock}, James S. and {Dalcanton}, Julianne J. and {Hopkins}, Philip F. and {Leaman}, Ryan and {McConnachie}, Alan W. and {Vogelsberger}, Mark and {Wetzel}, Andrew},
        title = "{Star formation at the edge of the Local Group: a rising star formation history in the isolated galaxy WLM}",
      journal = {\mnras},
         year = 2019,
        month = dec,
       volume = {490},
       number = {4},
        pages = {5538-5550},
          doi = {10.1093/mnras/stz2903},
archivePrefix = {arXiv},
       eprint = {1909.04040},
 primaryClass = {astro-ph.GA},
       adsurl = {https://ui.adsabs.harvard.edu/abs/2019MNRAS.490.5538A}
}

@ARTICLE{Vogelsberger2013,
       author = {{Vogelsberger}, Mark and {Genel}, Shy and {Sijacki}, Debora and {Torrey}, Paul and {Springel}, Volker and {Hernquist}, Lars},
        title = "{A model for cosmological simulations of galaxy formation physics}",
      journal = {\mnras},
         year = 2013,
        month = dec,
       volume = {436},
       number = {4},
        pages = {3031-3067},
          doi = {10.1093/mnras/stt1789},
archivePrefix = {arXiv},
       eprint = {1305.2913},
 primaryClass = {astro-ph.CO},
       adsurl = {https://ui.adsabs.harvard.edu/abs/2013MNRAS.436.3031V}
}

@ARTICLE{Vogelsberger2014,
       author = {{Vogelsberger}, M. and {Genel}, S. and {Springel}, V. and {Torrey}, P. and {Sijacki}, D. and {Xu}, D. and {Snyder}, G. and {Bird}, S. and {Nelson}, D. and {Hernquist}, L.},
        title = "{Properties of galaxies reproduced by a hydrodynamic simulation}",
      journal = {\nat},
         year = 2014,
        month = may,
       volume = {509},
       number = {7499},
        pages = {177-182},
          doi = {10.1038/nature13316},
archivePrefix = {arXiv},
       eprint = {1405.1418},
 primaryClass = {astro-ph.CO},
       adsurl = {https://ui.adsabs.harvard.edu/abs/2014Natur.509..177V}
}

@ARTICLE{Weinberger2017,
       author = {{Weinberger}, Rainer and {Springel}, Volker and {Hernquist}, Lars and {Pillepich}, Annalisa and {Marinacci}, Federico and {Pakmor}, R{\"u}diger and {Nelson}, Dylan and {Genel}, Shy and {Vogelsberger}, Mark and {Naiman}, Jill and {Torrey}, Paul},
        title = "{Simulating galaxy formation with black hole driven thermal and kinetic feedback}",
      journal = {\mnras},
         year = 2017,
        month = mar,
       volume = {465},
       number = {3},
        pages = {3291-3308},
          doi = {10.1093/mnras/stw2944},
archivePrefix = {arXiv},
       eprint = {1607.03486},
 primaryClass = {astro-ph.GA},
       adsurl = {https://ui.adsabs.harvard.edu/abs/2017MNRAS.465.3291W}
}

@ARTICLE{SpringelHernquist2003,
       author = {{Springel}, Volker and {Hernquist}, Lars},
        title = "{Cosmological smoothed particle hydrodynamics simulations: a hybrid multiphase model for star formation}",
      journal = {\mnras},
         year = 2003,
        month = feb,
       volume = {339},
       number = {2},
        pages = {289-311},
          doi = {10.1046/j.1365-8711.2003.06206.x},
archivePrefix = {arXiv},
       eprint = {astro-ph/0206393},
 primaryClass = {astro-ph},
       adsurl = {https://ui.adsabs.harvard.edu/abs/2003MNRAS.339..289S}
}

@ARTICLE{Chabrier2003,
       author = {{Chabrier}, Gilles},
        title = "{Galactic Stellar and Substellar Initial Mass Function}",
      journal = {\pasp},
         year = 2003,
        month = jul,
       volume = {115},
       number = {809},
        pages = {763-795},
          doi = {10.1086/376392},
archivePrefix = {arXiv},
       eprint = {astro-ph/0304382},
 primaryClass = {astro-ph},
       adsurl = {https://ui.adsabs.harvard.edu/abs/2003PASP..115..763C}
}

@ARTICLE{Walter2008THINGS,
       author = {{Walter}, Fabian and {Brinks}, Elias and {de Blok}, W.~J.~G. and {Bigiel}, Frank and {Kennicutt}, Robert C., Jr. and {Thornley}, Michele D. and {Leroy}, Adam},
        title = "{THINGS: The H I Nearby Galaxy Survey}",
      journal = {\aj},
         year = 2008,
        month = dec,
       volume = {136},
       number = {6},
        pages = {2563-2647},
          doi = {10.1088/0004-6256/136/6/2563},
archivePrefix = {arXiv},
       eprint = {0810.2125},
 primaryClass = {astro-ph},
       adsurl = {https://ui.adsabs.harvard.edu/abs/2008AJ....136.2563W}
}

@ARTICLE{Pakmor2016,
       author = {{Pakmor}, R{\"u}diger and {Springel}, Volker and {Bauer}, Andreas and {Mocz}, Philip and {Munoz}, Diego J. and {Ohlmann}, Sebastian T. and {Schaal}, Kevin and {Zhu}, Chenchong},
        title = "{Improving the convergence properties of the moving-mesh code AREPO}",
      journal = {\mnras},
         year = 2016,
        month = jan,
       volume = {455},
       number = {1},
        pages = {1134-1143},
          doi = {10.1093/mnras/stv2380},
archivePrefix = {arXiv},
       eprint = {1503.00562},
 primaryClass = {astro-ph.GA},
       adsurl = {https://ui.adsabs.harvard.edu/abs/2016MNRAS.455.1134P}
}

@ARTICLE{Dolag2009,
       author = {{Dolag}, K. and {Borgani}, S. and {Murante}, G. and {Springel}, V.},
        title = "{Substructures in hydrodynamical cluster simulations}",
      journal = {\mnras},
         year = 2009,
        month = oct,
       volume = {399},
       number = {2},
        pages = {497-514},
          doi = {10.1111/j.1365-2966.2009.15034.x},
archivePrefix = {arXiv},
       eprint = {0808.3401},
 primaryClass = {astro-ph},
       adsurl = {https://ui.adsabs.harvard.edu/abs/2009MNRAS.399..497D}
}

@ARTICLE{Baldry2012,
       author = {{Baldry}, I.~K. and {Driver}, S.~P. and {Loveday}, J. and {Taylor}, E.~N. and {Kelvin}, L.~S. and {Liske}, J. and {Norberg}, P. and {Robotham}, A.~S.~G. and {Brough}, S. and {Hopkins}, A.~M. and {Bamford}, S.~P. and {Peacock}, J.~A. and {Bland-Hawthorn}, J. and {Conselice}, C.~J. and {Croom}, S.~M. and {Jones}, D.~H. and {Parkinson}, H.~R. and {Popescu}, C.~C. and {Prescott}, M. and {Sharp}, R.~G. and {Tuffs}, R.~J.},
        title = "{Galaxy And Mass Assembly (GAMA): the galaxy stellar mass function at z < 0.06}",
      journal = {\mnras},
         year = 2012,
        month = mar,
       volume = {421},
       number = {1},
        pages = {621-634},
          doi = {10.1111/j.1365-2966.2012.20340.x},
archivePrefix = {arXiv},
       eprint = {1111.5707},
 primaryClass = {astro-ph.CO},
       adsurl = {https://ui.adsabs.harvard.edu/abs/2012MNRAS.421..621B}
}

@ARTICLE{Driver2011,
       author = {{Driver}, S.~P. and {Hill}, D.~T. and {Kelvin}, L.~S. and {Robotham}, A.~S.~G. and {Liske}, J. and {Norberg}, P. and {Baldry}, I.~K. and {Bamford}, S.~P. and {Hopkins}, A.~M. and {Loveday}, J. and {Peacock}, J.~A. and {Andrae}, E. and {Bland-Hawthorn}, J. and {Brough}, S. and {Brown}, M.~J.~I. and {Cameron}, E. and {Ching}, J.~H.~Y. and {Colless}, M. and {Conselice}, C.~J. and {Croom}, S.~M. and {Cross}, N.~J.~G. and {de Propris}, R. and {Dye}, S. and {Drinkwater}, M.~J. and {Ellis}, S. and {Graham}, Alister W. and {Grootes}, M.~W. and {Gunawardhana}, M. and {Jones}, D.~H. and {van Kampen}, E. and {Maraston}, C. and {Nichol}, R.~C. and {Parkinson}, H.~R. and {Phillipps}, S. and {Pimbblet}, K. and {Popescu}, C.~C. and {Prescott}, M. and {Roseboom}, I.~G. and {Sadler}, E.~M. and {Sansom}, A.~E. and {Sharp}, R.~G. and {Smith}, D.~J.~B. and {Taylor}, E. and {Thomas}, D. and {Tuffs}, R.~J. and {Wijesinghe}, D. and {Dunne}, L. and {Frenk}, C.~S. and {Jarvis}, M.~J. and {Madore}, B.~F. and {Meyer}, M.~J. and {Seibert}, M. and {Staveley-Smith}, L. and {Sutherland}, W.~J. and {Warren}, S.~J.},
        title = "{Galaxy and Mass Assembly (GAMA): survey diagnostics and core data release}",
      journal = {\mnras},
         year = 2011,
        month = may,
       volume = {413},
       number = {2},
        pages = {971-995},
          doi = {10.1111/j.1365-2966.2010.18188.x},
archivePrefix = {arXiv},
       eprint = {1009.0614},
 primaryClass = {astro-ph.CO},
       adsurl = {https://ui.adsabs.harvard.edu/abs/2011MNRAS.413..971D}
}

@ARTICLE{Du2021,
       author = {{Du}, Min and {Ho}, Luis C. and {Debattista}, Victor P. and {Pillepich}, Annalisa and {Nelson}, Dylan and {Hernquist}, Lars and {Weinberger}, Rainer},
        title = "{The Evolutionary Pathways of Disk-, Bulge-, and Halo-dominated Galaxies}",
      journal = {\apj},
         year = 2021,
        month = oct,
       volume = {919},
       number = {2},
          eid = {135},
        pages = {135},
          doi = {10.3847/1538-4357/ac0e98},
archivePrefix = {arXiv},
       eprint = {2101.12373},
 primaryClass = {astro-ph.GA},
       adsurl = {https://ui.adsabs.harvard.edu/abs/2021ApJ...919..135D}
}

@ARTICLE{Du2022,
       author = {{Du}, Min and {Ho}, Luis C. and {Yu}, Hao-Ran and {Debattista}, Victor P.},
        title = "{The Origin of the Relation Between Stellar Angular Momentum and Stellar Mass in Nearby Disk-dominated Galaxies}",
      journal = {\apjl},
         year = 2022,
        month = sep,
       volume = {937},
       number = {1},
          eid = {L18},
        pages = {L18},
          doi = {10.3847/2041-8213/ac911e},
archivePrefix = {arXiv},
       eprint = {2201.08579},
 primaryClass = {astro-ph.GA},
       adsurl = {https://ui.adsabs.harvard.edu/abs/2022ApJ...937L..18D}
}

@ARTICLE{OstrikerKim2022,
       author = {{Ostriker}, Eve C. and {Kim}, Chang-Goo},
        title = "{Pressure-regulated, Feedback-modulated Star Formation in Disk Galaxies}",
      journal = {\apj},
         year = 2022,
        month = sep,
       volume = {936},
       number = {2},
          eid = {137},
        pages = {137},
          doi = {10.3847/1538-4357/ac7de2},
archivePrefix = {arXiv},
       eprint = {2206.00681},
 primaryClass = {astro-ph.GA},
       adsurl = {https://ui.adsabs.harvard.edu/abs/2022ApJ...936..137O}
}

@ARTICLE{Adams2014,
       author = {{Adams}, Joshua J. and {Simon}, Joshua D. and {Fabricius}, Maximilian H. and {van den Bosch}, Remco C.~E. and {Barentine}, John C. and {Bender}, Ralf and {Gebhardt}, Karl and {Hill}, Gary J. and {Murphy}, Jeremy D. and {Swaters}, R.~A. and {Thomas}, Jens and {van de Ven}, Glenn},
        title = "{Dwarf Galaxy Dark Matter Density Profiles Inferred from Stellar and Gas Kinematics}",
      journal = {\apj},
         year = 2014,
        month = jul,
       volume = {789},
       number = {1},
          eid = {63},
        pages = {63},
          doi = {10.1088/0004-637X/789/1/63},
archivePrefix = {arXiv},
       eprint = {1405.4854},
 primaryClass = {astro-ph.GA},
       adsurl = {https://ui.adsabs.harvard.edu/abs/2014ApJ...789...63A}
}

@ARTICLE{Oman2019,
       author = {{Oman}, Kyle A. and {Marasco}, Antonino and {Navarro}, Julio F. and {Frenk}, Carlos S. and {Schaye}, Joop and {Ben{\'\i}tez-Llambay}, Alejandro},
        title = "{Non-circular motions and the diversity of dwarf galaxy rotation curves}",
      journal = {\mnras},
         year = 2019,
        month = jan,
       volume = {482},
       number = {1},
        pages = {821-847},
          doi = {10.1093/mnras/sty2687},
archivePrefix = {arXiv},
       eprint = {1706.07478},
 primaryClass = {astro-ph.GA},
       adsurl = {https://ui.adsabs.harvard.edu/abs/2019MNRAS.482..821O}
}

@ARTICLE{Tau2024Obs,
       author = {{Tau}, Elisa A. and {Vivas}, A. Katherina and {Mart{\'\i}nez-V{\'a}zquez}, Clara E.},
        title = "{Extended Stellar Populations in Ultrafaint Dwarf Galaxies}",
      journal = {\aj},
         year = 2024,
        month = feb,
       volume = {167},
       number = {2},
          eid = {57},
        pages = {57},
          doi = {10.3847/1538-3881/ad1509},
archivePrefix = {arXiv},
       eprint = {2312.07279},
 primaryClass = {astro-ph.GA},
       adsurl = {https://ui.adsabs.harvard.edu/abs/2024AJ....167...57T}
}

@ARTICLE{Fu2024And,
       author = {{Fu}, Sal Wanying and {Weisz}, Daniel R. and {Starkenburg}, Else and {Martin}, Nicolas and {Collins}, Michelle L.~M. and {Savino}, Alessandro and {Boylan-Kolchin}, Michael and {C{\^o}t{\'e}}, Patrick and {Dolphin}, Andrew E. and {Longeard}, Nicolas and {Mateo}, Mario L. and {Mercado}, Francisco J. and {Sandford}, Nathan R. and {Skillman}, Evan D.},
        title = "{Stellar Metallicities and Gradients in the Faint M31 Satellites Andromeda XVI and Andromeda XXVIII}",
      journal = {\apj},
         year = 2024,
        month = nov,
       volume = {975},
       number = {1},
          eid = {2},
        pages = {2},
          doi = {10.3847/1538-4357/ad76a2},
archivePrefix = {arXiv},
       eprint = {2407.04698},
 primaryClass = {astro-ph.GA},
       adsurl = {https://ui.adsabs.harvard.edu/abs/2024ApJ...975....2F}
}

@ARTICLE{Pillepich2019,
       author = {{Pillepich}, Annalisa and {Nelson}, Dylan and {Springel}, Volker and {Pakmor}, R{\"u}diger and {Torrey}, Paul and {Weinberger}, Rainer and {Vogelsberger}, Mark and {Marinacci}, Federico and {Genel}, Shy and {van der Wel}, Arjen and {Hernquist}, Lars},
        title = "{First results from the TNG50 simulation: the evolution of stellar and gaseous discs across cosmic time}",
      journal = {\mnras},
         year = 2019,
        month = dec,
       volume = {490},
       number = {3},
        pages = {3196-3233},
          doi = {10.1093/mnras/stz2338},
archivePrefix = {arXiv},
       eprint = {1902.05553},
 primaryClass = {astro-ph.GA},
       adsurl = {https://ui.adsabs.harvard.edu/abs/2019MNRAS.490.3196P}
}

@ARTICLE{Nelson2019TNG50,
       author = {{Nelson}, Dylan and {Pillepich}, Annalisa and {Springel}, Volker and {Pakmor}, R{\"u}diger and {Weinberger}, Rainer and {Genel}, Shy and {Torrey}, Paul and {Vogelsberger}, Mark and {Marinacci}, Federico and {Hernquist}, Lars},
        title = "{First results from the TNG50 simulation: galactic outflows driven by supernovae and black hole feedback}",
      journal = {\mnras},
         year = 2019,
        month = dec,
       volume = {490},
       number = {3},
        pages = {3234-3261},
          doi = {10.1093/mnras/stz2306},
archivePrefix = {arXiv},
       eprint = {1902.05554},
 primaryClass = {astro-ph.GA},
       adsurl = {https://ui.adsabs.harvard.edu/abs/2019MNRAS.490.3234N}
}

@ARTICLE{Ludlow2023,
       author = {{Ludlow}, Aaron D. and {Fall}, S. Michael and {Wilkinson}, Matthew J. and {Schaye}, Joop and {Obreschkow}, Danail},
        title = "{Spurious heating of stellar motions by dark matter particles in cosmological simulations of galaxy formation}",
      journal = {\mnras},
         year = 2023,
        month = nov,
       volume = {525},
       number = {4},
        pages = {5614-5630},
          doi = {10.1093/mnras/stad2615},
archivePrefix = {arXiv},
       eprint = {2306.05753},
 primaryClass = {astro-ph.GA},
       adsurl = {https://ui.adsabs.harvard.edu/abs/2023MNRAS.525.5614L}
}

@ARTICLE{Pillepich2018modelTNG,
       author = {{Pillepich}, Annalisa and {Springel}, Volker and {Nelson}, Dylan and {Genel}, Shy and {Naiman}, Jill and {Pakmor}, R{\"u}diger and {Hernquist}, Lars and {Torrey}, Paul and {Vogelsberger}, Mark and {Weinberger}, Rainer and {Marinacci}, Federico},
        title = "{Simulating galaxy formation with the IllustrisTNG model}",
      journal = {\mnras},
         year = 2018,
        month = jan,
       volume = {473},
       number = {3},
        pages = {4077-4106},
          doi = {10.1093/mnras/stx2656},
archivePrefix = {arXiv},
       eprint = {1703.02970},
 primaryClass = {astro-ph.GA},
       adsurl = {https://ui.adsabs.harvard.edu/abs/2018MNRAS.473.4077P}
}

%%%%%%%%%%%%%%%%% APPENDICES %%%%%%%%%%%%%%%%%%%%%

\begin{appendix}
\label{App1}

\section{Size-mass relation in TNG50 \& TNG100}

In Table \ref{tab:TNGSims} we list numerical parameters of the TNG50-1 and TNG100-1 runs of the Illustris TNG suite \citep[as listed in][]{Nelson2019}. Although TNG50-1 has higher resolution than TNG100-1 (and a smaller volume)  both simulations share the sub-grid physics implementation of e.g. magnetic fields, black hole accretion and feedback, wind directionality, velocity and energy and metal tagging  \citep[see Appendix A][]{Pillepich2018modelTNG}.

\begin{table}[hbt!]
\caption{Numerical parameters for the TNG50-1 and TNG100-1 runs of the IllustrisTNG suite.}
\label{tab:TNGSims}
\centering
\begin{tabular}{c c c c c c}
\hline\hline
Run  & $L_{\rm box}$ & $m_{\rm DM}$ & $m_{\rm baryon}$ & $\epsilon_{\rm DM,*}$ & $\epsilon_{\rm gas,min}$ \\[2pt]
 & [Mpc] & $[\mathrm{M_{\odot}}]$ & $[\mathrm{M_{\odot}}]$ & [kpc] & [kpc] \\[2pt]
\hline\\[-5pt]
TNG50-1 & 51.7 & $4.5 ~ 10^5$ & $8.5 ~ 10^4$ & 0.29 & 0.07 \\[2pt]
TNG100-1 & 110.7 & $7.5 ~ 10^6$ & $1.4 ~ 10^6$ & 0.74 & 0.18 \\[2pt]
\hline
\end{tabular}
\tablefoot{From left to right: Name of the simulation, side of the cosmological box at redshift $z=0$, mass of dark-matter particles, target-mass of stellar particles/gas cells, force softening length for dark matter and stellar particles, minimum force softening length for gas cells.}
\end{table}

\begin{figure}[hbt!]
    \centering
    \includegraphics[width=0.99\columnwidth]{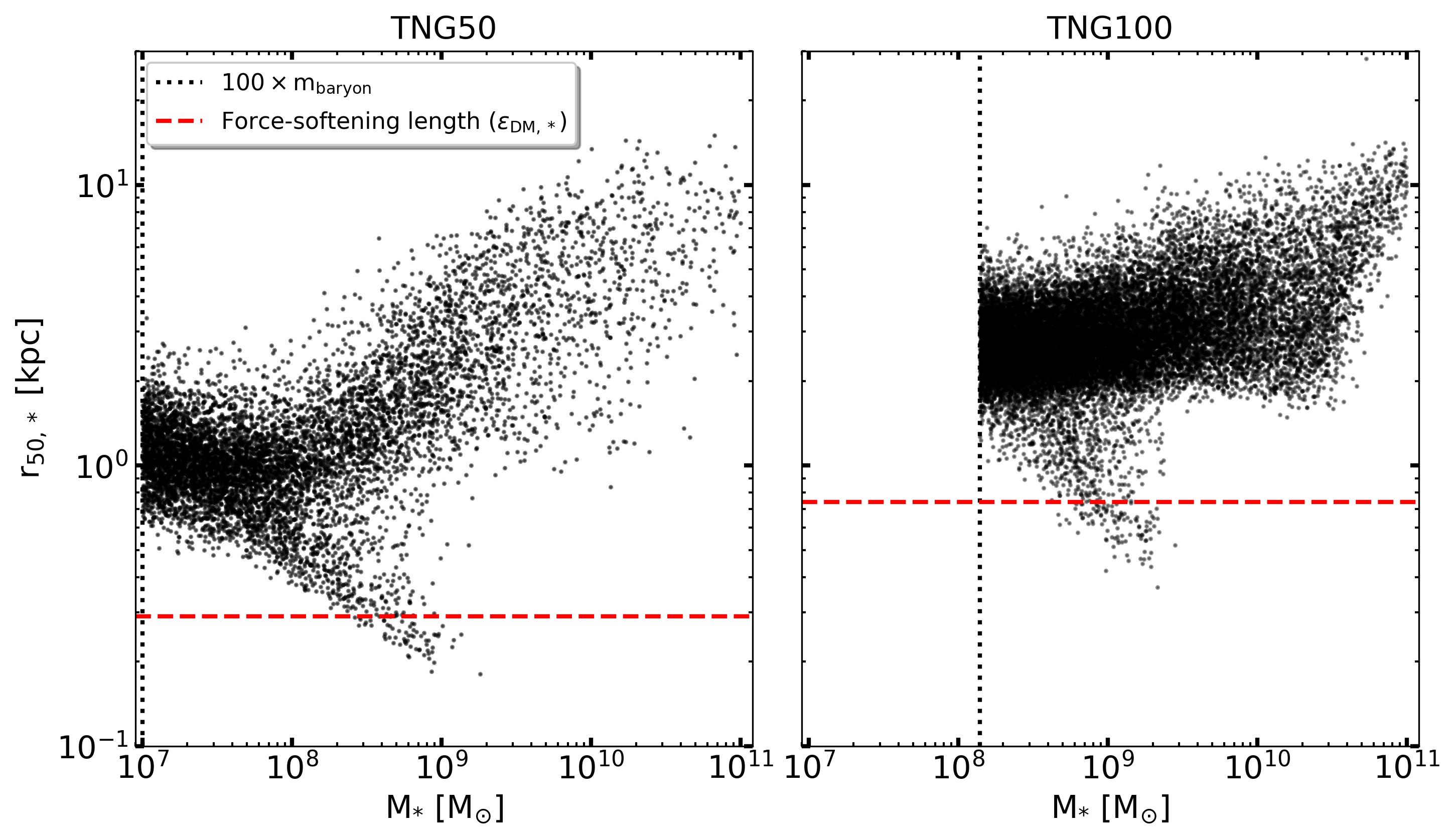}
    
    \caption{Stellar half mass radius ($r_{50,*}$) as a function of stellar mass ($M_{*}$) for all isolated central galaxies in TNG50-1 (left panel, similar to Figure \ref{FigMstarRh}) and in TNG100-1 (right panel). All galaxies shown have more than 100 stellar particles and more than 100 dark matter particles. Red dashed lines indicate the force softening length for dark matter and stellar particles for each simulation. The "tongue" is also present in TNG100, but at slightly higher masses and with larger size, suggesting that this feature is numerical in origin. }
    \label{App_FigMstarRh_TNG50_100}
\end{figure}

The effects of unresolved baryonic clumps at the innermost regions of galaxies are likely due to the sub-grid physics implementation. In Figure \ref{App_FigMstarRh_TNG50_100} we compare the size-mass relation for isolated central galaxies of TNG50-1 (left panel, as in Figure \ref{FigMstarRh}) and TNG100-1 (right panel). A "tongue" of galaxies that reach extremely small size can be identified in both simulations. However, the numerical feature in TNG100-1 occurs at higher stellar masses and larger sizes, a clear indication of numerical artifact. 

\end{appendix}
\end{document}